\documentclass[pre,reprint,superscriptaddress,showpacs,nofootinbib]{revtex4-1}
\usepackage{epsfig}
\usepackage{natbib}
\usepackage{color}
\usepackage{graphics}
\usepackage{amssymb,amsmath}
\usepackage{amsmath}
\usepackage{psfrag}
\usepackage{graphicx}

\usepackage{float}
\usepackage[caption = false]{subfig}

\usepackage{color}
\usepackage{graphics}
\usepackage{amssymb,amsmath}
\usepackage{amsmath}
\usepackage{psfrag}
\usepackage{graphicx}

\newcommand{\be}{\begin{equation}}      
\newcommand{\ee}{\end{equation}}      
      
\newcommand{\bef}{\begin{figure}}      
\newcommand{\eef}{\end{figure}}      
\newcommand{\bea}{\begin{eqnarray}}    
\newcommand{\eea}{\end{eqnarray}}

\begin{document}

\title{Formation of disks with long-lived spiral arms from violent gravitational dynamics}
 \author
 {Francesco Sylos Labini}
 \affiliation{Centro
  Ricerche Enrico Fermi, Via Panisperna 89a, I-00184, Rome, Italia }
\affiliation{Istituto dei Sistemi
  Complessi, Consiglio Nazionale delle Ricerche, Via dei Taurini 19,
   I-00185 Roma, Italia} 
\affiliation{INFN Unit Rome 1, Dipartimento di Fisica, Universit\'a di
  Roma Sapienza, Piazzale Aldo Moro 2, I-00185 Rome, Italia }
  \author{Luis Diego Pinto}
  \affiliation{INAF-IAPS, Istituto di Astrofisica e Planetologia Spaziali,  via del Fosso del Cavaliere, 100, I-00133, Rome, Italia}
  \affiliation{Dipartimento di Fisica, Sapienza, Universit\'a di
  Roma, piazzale Aldo Moro 2, I-00185, Rome, Italia}
  \author{Roberto Capuzzo-Dolcetta}
  \affiliation{Dipartimento di Fisica, Sapienza, Universit\'a di
  Roma, piazzale Aldo Moro 2, I-00185, Rome, Italia}
  
\date{\today}

\begin{abstract}
By means of simple dynamical experiments we study the combined effect of gravitational and gas dynamics in the evolution of an initially out-of-equilibrium, uniform and rotating massive over-density thought of as in isolation.
The rapid variation of the system mean-field potential makes the point like particles (PPs), 
which interact only via Newtonian gravity, 
 form a quasistationary thick disk dominated by rotational motions 
surrounded by  far out-of-equilibrium spiral arms.
On the other side, the gas component is subjected to compression shocks and radiative cooling so as to develop a much flatter disk, where rotational motions are coherent and the  velocity dispersion is smaller than that of PPs.
Around such gaseous disk long-lived, but nonstationary, spiral arms form:
these are made of gaseous particles that move coherently because 
have acquired a specific phase-space correlation during the 
gravitational collapse phase. Such a phase-space 
correlation represents a signature of the violent origin of the arms and implies both 
the motion of  matter and the transfer of energy.
On larger scales, where the radial velocity component is significantly larger than the rotational one, 
the gas follows the same out-of-equilibrium spiral arms traced by PPs
We finally outline the astrophysical and cosmological implications of our results.
\end{abstract}


\maketitle


\section{Introduction}  

Self-gravitating systems, like other ones that interact with a pair potential decaying with 
an exponent smaller than that of the embedding space, i.e.,  with long-range interactions, 
give rise to macroscopic behaviors that are very different from 
the ones arising in short-range interacting systems (SRISs).
Their origin and properties represent an open theoretical problem
because the long-range nature of the interaction displays several 
behaviors that prevent the use of equilibrium statistical mechanics \citep{Padmanabhan_1989,Dauxois_etal_2002,Assisi,Campa_etal_2009,gabrielli_etal_2010a,Campa_etal_2014,chavanis_kEqns_2010,Marcos_2013,Levin_etal_2014,Marcos_etal_2017,diCintio_etal_2018, RCDbook}.
In particular, the relaxation mechanism driving a long-range 
interacting system (LRIS) towards a quasiequilibrium state  is different from that acting in SRISs. 
These latter systems typically tend towards thermal equilibrium through
a rather rapid collisional relaxation process, 
in which  particles exchange energy predominantly by binary encounters.
In this case, to obtain an out-of-equilibrium state it is necessary 
to force the system with an external field.
Instead, LRISs typically show a mean-field (or violent) 
relaxation process in which their global characteristic 
quantities (e.g., size, potential gravitational energy, etc.) 
rapidly vary until they reach a configuration close to a quasistationary state (QSS) \citep{lyndenbell,Padmanabhan_1989,Dauxois:2002pv,Assisi,Campa_etal_2009,Campa_etal_2014}.
This does not correspond to a true 
(i.e., thermodynamical) equilibrium state, but it is such that 
the system is close to be virialized and it is almost time-independent.
The violent  relaxation phase is then followed by a slow adiabatic evolution driven by collisional processes (see, e.g., Ref. \cite{chandra43,Padmanabhan_1989}).
However, this is not the only way LRISs relax; for instance, in the cosmological context, self-gravitating systems often relax through a soft and slow relaxation mechanism.
This occurs when density fluctuations are long-range correlated so that quasistationary 
nonlinear structures of increasing size are formed via a bottom-up hierarchical aggregation process \citep{peebles_1983}. 

%

Gravitational relaxation and the formation of QSSs were studied also in (simpler) one-dimensional (1D) systems 
because in that case one can work out exact solutions
\cite{Wright_etal_1982,Yan_Miller_1997,Tsuchiya_Gouda_2000,Joyce+Worrakitpoonpon_2011,Teles_etal_2011}.
For the full three-dimensional problem, 
given the theoretical difficulties to treat 
out-of-equilibrium LRIS dynamics, an important tool to study their behaviors is represented by numerical experiments.  

While the bottom-up gravitational clustering is usually studied in the context of cosmological simulations, many systematic studies of finite and isolated self-gravitating systems that undergo  a global collapse phase have been reported in the literature \citep{henon_1964,vanalbada_1982,aarseth_etal_1988,aguilar+merritt_1990,theis+spurzem_1999,   boily_etal_2002, roy+perez_2004, boily+athanassoula_2006, barnes_etal_2009, ejection_mjbmfsl, syloslabini_2012, worrakitpoonpon_2014, merritt+aguilar_1985, aguilar+merritt_1990, theis+spurzem_1999,syloslabini_2013, SylosLabini+Benhaiem+Joyce_2015, Benhaiem+SylosLabini_2015, Benhaiem+SylosLabini_2017}.
These experiments  use molecular dynamics and start from simple classes of 
initial conditions (ICs) that do not have the complexity of real 
astrophysical objects and that, of course, do not aim to represent specific realistic systems. 
The motivation for their study is to try to improve the understanding
of the basic physical
mechanisms at play in the evolution of these systems, such as 
the details of the relaxation toward virial equilibria, the dependence of the equilibria properties on the ICs, phenomena such as symmetry breaking, radial orbits instability, etc. 
Although, from the statistical mechanics point of view, a complete understanding is still lacking, these studies have shown that the relaxation dynamics acting during a monolithic collapse is very generically the same for a broad class of ICs and gives rise to
quasivirialized configurations with ellipsoidal shapes.
However, the QSS properties such as the density and velocity profiles depend on the details of the ICs.

It was recently found that a key role in modifying the simple picture outlined above, is played by the system's initial spatial anisotropy.
Indeed, the gravitational collapse phase of initially out-of-equilibrium over-densities
amplifies any initial spatial anisotropy both when they are initially at rest or when have a small isotropic velocity dispersion \citep{Benhaiem+SylosLabini_2015,Benhaiem+SylosLabini_2017}.
Initially spherical systems give rise to almost spherical virialized states with a characteristic density profile that decays as $r^{-4}$
{  \cite{vanalbada_1982,ejection_mjbmfsl,syloslabini_2012}.}
Systems that initially break spherical symmetry form a two-component state whose inner part is virialized and close to spherical, and its outermost regions are out-of-equilibrium and flat. 
This occurs because the initial deviation from spherical symmetry is amplified by the collapse mechanism: in particular,
the system is stretched along the plane identified by the  major and medium
axes of the initial configuration \footnote{The simple numerical experiments considered in Refs.  \cite{Benhaiem+SylosLabini_2015,Benhaiem+SylosLabini_2017,Benhaiem+Joyce+SylosLabini_2017,Benhaiem+SylosLabini+Joyce_2019} 
focused mainly only ellipsoidal ICs, but in a few cases 
  more irregular situations have been considered.}. 
  
Moreover, when the initial over-density has a nonzero angular 
momentum, the gravitational (and dissipationless) collapse gives 
rise to a thick quasiequilibrium disk surrounded by 
out-of-equilibrium 
spiral arms with or without bars and/or 
rings \citep{Benhaiem+Joyce+SylosLabini_2017,Benhaiem+SylosLabini+Joyce_2019}.
These  transient structures involve only a 
fraction of the system's mass and thus 
represent nonequilibrium perturbations of a substantially virialized state.
Such transients, still bound to the system
but dominated by radial motions, 
may continue to evolve for times that are very 
long compared with the intrinsic gravitational 
collapse characteristic time 
$\tau\sim 1/\sqrt{G \rho}$ --- 
with $\rho$ being the system's average mass density.

This fast and violent dynamics is  rather different 
from the slow and soft dynamical mechanisms usually considered in astrophysical contexts.  
Concerning the latter ones,  attention was focused on
 two complementary 
physical systems. 
On the one hand, it is well known since the pioneering work in Refs.  
\cite{Goldreich+Lynden-Bell_1965a,Goldreich+Lynden-Bell_1965b} that galactic disks are remarkably responsive
to small disturbances. For this reason there has been a great effort to 
study the evolution of small-scale disturbances in simplified
models of  rotating self-gravitating disks with and without a 
dissipational gas component 
(see, e.g., Refs.  \cite{Sellwood+Carlberg_1984,Sellwood+Carlberg_2014,Sellwood+Carlberg_2019} and references
therein;  and Refs. \cite{Binney_Tremaine_2008,Dobbs_Baba_2014} for  reviews). 
These models  assume that a disk is already formed and has reached a 
rotational equilibrium (often in the gravitational field of a spherical halo): 
the problem that is considered concerns  how instabilities can give rise to 
spiral arms and/or bars. As such instabilities represent small 
perturbations to the system's gravitational mean field,
 their effects on the global system's conditions are small and 
 the overall  dynamical mechanism is thus soft.

On the other hand, the question of the disk formation is 
studied in the cosmological framework. 
Favored cosmological models, like cold
dark matter (CDM)  type scenarios, assume matter density fluctuations 
that are long-range correlated. Such correlations 
 induce a bottom-up hierarchical clustering: that is, a structure of size $R$ 
 is formed by the aggregation of smaller substructures of size $< R$ rather than   
by the global collapse of an over-density  of  size $R$.
This kind of hierarchical aggregation, being statistically isotropic, gives rise to 
   quasispherical structures with a quasi-isotropic 
velocity dispersion. These are the so-called halo structures \citep{Navarro_etal_1997},
whose formation is ubiquitous in the context of CDM-type cosmological 
simulations.The halos
are not isolated but evolve in a complex gravitational field generated by neighboring structures and thus are subjected to tidal effects and merging.  However, both mechanisms do not violently change  the halos mean field potential 
and thus halos form through a slow  and soft dynamical mechanism that does not involve 
a large  variation of their  mean field. In this scenario it remains open the 
question of a disk formation. 
As first envisaged in Ref. \cite{ELS_1962}, in the cosmological context,
a  disk can be formed by the  dissipational gas collapse: indeed, gas can shock and 
dissipate energy through radiative cooling,
and thus during the gravitational contraction forms a thin disk
if it initially has some angular momentum.
Such a 
disk  is thus embedded in the much larger gravitational field of the 
spherical halo structure {  that is formed by a hierarchical aggregation dynamics:
during such a process the system  mean field potential does not 
substantially vary. }
This situation has motivated the study of simplified ICs 
(see, e.g., Refs. \cite{Katz_1991,Katz+Gunn_1991,Katz_1992})
  in which  structures are formed via a hierarchical 
bottom-up aggregation process driven by gravitational clustering
with the 
inclusion of gas that can cool radiatively.

In this work we  
present several numerical experiments of relatively simple 
ICs
to study the combined effects of gravitational and gas dynamics
during the fast and violent phase occurring in the monolithic collapse
of an isolated overdensity with some initial angular momentum.
Such a process has been overlooked in the literature but, we argue,
can give some interesting insights for the formation 
of real astrophysical structures. 
Indeed, despite the fact that the problem of the 
joint effect of gravity and gas dynamics has been studied through 
numerical experiments of increasing sophistication, 
the focus has been pointed towards a different kind of physical case, i.e. 
when a slow and soft dynamical process takes place.

The paper is organized as follows: 
we start in Sect.\ref{sec:technical} by describing  
the way in which gravitational and gas dynamics are 
implemented in the hydrodynamical code we use to make numerical
simulations and by presenting the properties of the
ICs we have considered.
{Then in  Sect.\ref{sec:dynamics} we
we briefly review the main features of the collapse
dynamics of an isolated overdensity made of purely
self-gravitating particles and  
show  how the  results
change when a gaseous component is added into the system.}
Finally  we draw our main conclusions in
Sect.\ref{sec:concl}.



\section{Models and methods}
\label{sec:technical} 

In this section we summarize the main features of the numerical simulations that we have performed with the aim of  investigating the collapse and the subsequent relaxation to a quasistationary state  of a two-phase (i.e., purely self-gravitating particles and gas) system.
In particular the ICs  chosen 
corresponds 
to an isolated \textit{overdensity} with a simple shape that, in an astrophysical context, can be thought to represent a  proto-galaxy detached from the Hubble flow.

\subsection{Gravitational dynamics of a two-phase system}  

Our numerical experiments consider isolated systems consisting of $N$ point like or \textit{gaseous} particles interacting  by both body force and surface force (i.e., pressure gradients and viscosity).
This particle discretization represents a sampled Lagrangian representation: in particular, we use the smoothed particles hydrodynamics (SPH) approach for the gas dynamics, as described in what follows. 

The representative point particles (PPs) and gas particles (GPs) 
are initially randomly distributed according to a uniform space distribution in the volume corresponding to the initial system.
The inner density fluctuations are small enough that an actual \textit{monolithic} gravitational collapse starts from the initial subvirialized state (see below for more details).

All our simulations have been performed by means of the hydrodynamical SPH code {\tt Gadget-3}. 
This  represents an up-to-date version of the already publicly available (and widely used) code {\tt Gadget-2} \citep{Springel_2005} that has been kindly made available to us by the author.
 {\tt Gadget-3} computes the hydrodynamical evolution of a gas distribution via a
 SPH scheme, by subdividing the fluid into a set of interpolating particles whose 
 spatial distribution is proportional to the  density field.
The  GPs interact via Newtonian force and pressure gradient and by the 
Newtonian force only with the other ensemble of PPs (i.e., point-mass and pressure less objects).
The gravitational interaction is evaluated by direct summation over close neighbors and via a multi-polar expansion on a larger scale. 
In this way, the number of computations is sensibly lower compared to the usual $N^2$ scaling, characteristic of the direct-summation $N$-body algorithms.

The gravitational interaction on the small distance scale is regularized with the so-called ``gravitational softening" $\varepsilon$: the force has its purely Newtonian value at separations greater than $\varepsilon$ ($r\geq \varepsilon$) 
while it is smoothed at shorter separations. 
The assumed functional form of the regularized potential, is a cubic spline interpolating between the exact Newtonian potential at $r = \varepsilon$ and a constant value at $r=0$ where the mutual gravitational force vanishes (the exact expression can be found in Ref. \citep{Springel_2005}). 

A detailed study of the parameter space of the code {\tt Gadget-2},  
for simulations considering only Newtonian gravity, has been reported elsewhere (see
Refs. 
\cite{SylosLabini+Benhaiem+Joyce_2015,Benhaiem+SylosLabini_2015,Benhaiem+SylosLabini_2017, Benhaiem+SylosLabini+Joyce_2019}): here we stress that in purely gravitational simulations performed by using only PPs  without GPs, we always kept energy, momentum and angular momentum conservation at a level of precision better than $1 \%$. 
In this work we consider ICs for which  the initial virial ratio is 
\be
\label{virial-ratio}
 \frac{1}{2} \leq Q_0\equiv \left|\frac{2K(0)}{W(0)}\right| \leq 1
\ee
where  $K(t)$ and $W(t)$ are, respectively, the kinetic and potential energy of the system at time $t$.
In such cases, the maximum system contraction is of about a factor $\sim 2$, i.e., it is
not as extreme as for a purely cold collapse (see, e.g., Refs.
\cite{aarseth_etal_1988,ejection_mjbmfsl,syloslabini_2012,syloslabini_2013}).

\subsection{Gas dynamics} 
 \label{sect:gas_dynamics}
 
The gas component is represented as an inviscid fluid whose time evolution is governed by the set of continuity, Euler (motion) and energy equations.
The Lagrangian form of the continuity equation is
  \be 
  \label{cont}
  \frac{D \rho}{Dt} + \rho
  {\nabla} \cdot \mathbf{v} = 0 \;
  \ee 
(where $\rho$ is the fluid density and $\mathbf{v}$ its velocity) while  Euler's equation of motion is
 \be 
 \label{euler}
 \frac{D \mathbf{v}} {Dt} = -\frac{\nabla P}{\rho} - \nabla \Phi \;,
 \ee 
where $P$ is the pressure and the body force is given by the gradient of the gravitational potential $\Phi(\mathbf{r})$.
The above time derivatives are the usual Lagrangian time derivatives along the flow: 
\be 
\frac{D}{Dt} \equiv
\frac{\partial }{ \partial t} + \mathbf{v}\cdot ~ \nabla.
\ee 
Finally, the thermal energy per unit mass, $u$, evolves according to the first law of thermodynamics, {\it viz.},  
  \be 
 \label{cool}
   \frac{D u}{Dt} = - \frac{P}{\rho} \nabla
  \cdot \mathbf{v} - \frac{\Lambda(u , \rho)}{\rho} \;,
  \ee
where $\Lambda(u,\rho)\geq 0$ represents the radiative \textit{cooling} function per unit volume and we have set the heating function equal to zero. 
 The gas cooling is modeled by adopting the same formalism  discussed in Ref.  \cite{Katz+Gunn_1991}, which considers an optically thin medium in ionization equilibrium, characterized by a primordial cosmological composition.
 {Under the, justified, assumption of optically thin medium no heating term is needed in the energy equation.}
The cooling rate, expressed as a function of density and temperature, plays an important role since it helps the gas component to lose thermal energy and collapse.
Compared to the case in which $\Lambda(u,\rho)$  is neglected, more compact structures can be then formed. The radiative cooling function $\Lambda(u,\rho)$ is evaluated by considering several two-body processes involving both helium and hydrogen atoms: collisional excitation, collisional ionization, recombination, and dielectric recombination.
Moreover, free-free radiation emission processes are taken into account for all the possible ions.

{  The cooling rate function, $\Lambda(u,\rho)$, used in \texttt{Gadget-3} accounts for the various free-bound, bound-bound and free-free processes and is given by the sum of various terms  $\Lambda_i(u,\rho)$ ($i=1,2,...,n)$ each one accounting for the contribution from a specific cooling mechanism.
Basically, each $\Lambda_i(u,\rho)$ term has the form:  
\be
\label{equazione_lambda_i}
\Lambda_i(u,\rho) ~=~ A_i ~h(T) ~n_i~ n_e
\ee
where $A_i$ is a constant, $h(T)$ is a function of the temperature, while $n_i$ and $n_s$ are, respectively, the number density of the particular  `chemical' specie involved and the electron number density.
The species considered are those typical of primordial gas, i.e., neutral hydrogen and helium (H$^0$, He$^0$) and their ions (H$^-$,H$^+$,He$^+$,He$^{++}$).
Furthermore, a proper treatment of the cooling by molecular hydrogen, H$_2$, is implemented in \texttt{Gadget-3} accounting for a nonequilibrium evolution of the abundances of all the ions (see Refs. \cite{katz&al1996,abel&al97,yoshida&al03}).}

As an equation of state we consider the simplest one, i.e.
  \be
  \label{eos}
  P = (\gamma -1 ) \rho u \;,
  \ee
where $\gamma$ is is the adiabatic exponent. 
If we take $\gamma = 5/3$ (mono-atomic ideal gas), the \textit{particle} sound speed is
\be
c_{s} = \sqrt{\frac{5}{3}\frac{P}{\rho}} \;.
\ee   

As we said above, the {\tt Gadget-3} code calculates the evolution of a gas distribution by using an SPH scheme that was introduced  in Refs. \cite{Lucy_1977} and \cite{Gingold+Monaghan_1977} as a Lagrangian method particularly suited to treat the evolution of self-gravitating systems.
An SPH representation of a gas is that of an ensemble of moving particles which sample the fluid density distribution. 
The particles interact   with both surface, small-scale, force (pressure and contact forces) and body, large-scale, forces (gravity).
In the SPH scheme, each particle is characterized by a specific value of density, pressure gradient, and other relevant hydrodynamical quantities, each evaluated by means of a suitable interpolation over a set of neighbor points.
With such a technique the algorithm can work out the quantities useful to solve Eqs.(\ref{cont})-(\ref{eos})  and to find, for the $i$th particle, the density $\rho_i$, the velocity $v_i$, and the internal specific energy $u_i$, which characterize the average status of the system in that specific $i$th point.
For an exhaustive explanation of the formalism see, e.g., Refs. \cite{Monaghan_1992,Monaghan_2005}.

Despite we model the gas as an inviscid fluid, the SPH scheme needs an artificial viscosity to treat properly the fluid evolution during strong compression and to avoid nonphysical oscillations.
The {\tt Gadget-3} code adopts the same form of artificial viscosity as the {\tt Gadget-2} version, i.e., that suggested in Ref. \cite{Mon97}. 
Additional details about the artificial viscosity used in our simulations are found in
Ref. \cite{Springel_2005}. 
Similarly,  {\tt Gadget-3} includes the same numerical scheme indicated in Ref. \cite{Springel_2005} 
for the implementation of the radiative cooling in SPH,
although that was not included in the public release of {\tt Gadget-2}.

\subsection{Initial conditions} 
\label{subsect:mod}

The initial overdensity is characterized by 
its total mass  $M$,  gravitational radius 
\be
\label{rg}
r_g = \frac{GM^2}{|W(0)|}
\ee 
and  total angular momentum $\mathbf{J}$. 
This last is given in the form of a solid body 
rotation and can be quantified by the nondimensional 
\textit{spin} parameter \citep{Peebles_1969,knebe2008} 
\be
\label{lambda}
\lambda= \frac{|\mathbf{J}|}{G\sqrt{M^5/|W|}} \,. 
\ee
We have also examined cases where we gave to the system both 
random motion and solid-body rotation 
(see discussion in Sect.\ref{subsec:discussion}).
In this case  the initial kinetic energy 
has a rotational $K_{rot}$ and a random $K_{ran}$ term such that 
\be
\label{eta}
\eta=\frac{K_{ran}} {K_{rot}} \;. 
\ee
  
We considered prolate, oblate, and triaxial ellipsoids, but hereafter we focus in more detail on the case of a prolate ellipsoid with the three semiaxes such that $b=c$ and $a/c=a/b=3/2$.



Let $N_{GP}$ the number of GPs of mass fixed to $m_{GP}$.
The gas thermal energy per unit mass is
\be
u = \chi T,
\ee
where $T$ is the absolute temperature and 
\be
\chi \equiv \frac{k_B}{\mu m_H (\gamma-1)},
\ee
where $k_B$ is  Boltzmann's constant, $\mu=2.33$  is the mean molecular weight \citep{kra08,liu17}, and $\gamma=5/3$. 
We fix $N_{PP}=N_{GP}$ in the range $\in [10^5,  10^6]$ and 
{  we have taken} 
\be
\psi\equiv \frac{m_{PP}}{m_{GP}}
\ee
to be $\psi \approx 10$ 
that the mass of the gaseous component is $\sim$ 1/10 of the total mass.

For the simulation discussed in more details in 
what follows, and that we consider as a
paradigmatic example of the class 
of systems we explored, PPs and GPs are assumed to have 
the same initial velocity profile, corresponding to a rigid body rotation, and $T_0=40,000$K is the  initial uniform temperature. 
The  total mass is $M_0 = 5 \times 10^{10}$ M$_\odot$, the initial gravitational radius is $R_g \approx 10$ kpc and the spin parameter is $\lambda = 0.3$, corresponding to a virial ratio $Q \approx 3/4$.

In summary the parameters that define a simulation in this class of models 
are 12: $[M, r_g, Q, \lambda, \eta, a,b,c,  T, \psi, \gamma]$.
In order to explore the phase space of this class of systems we note 
the following:

\begin{enumerate}

\item 
By changing the two parameters that determine the timescale of the collapse $\tau$ see Eq.(\ref{tau}) below, i.e., $M, r_g$, 
the typical velocities 
$v \sim \sqrt{GM/r_g}$
of the post-collapse system 
also vary. 
These parameters can be changed if one wants to simulate a specific
astrophysical object: in this work we have considered the case of a medium-size galaxy with
$ r_g \sim 10$ kpc and $v \sim 100$ km/s.

\item  
The three parameters $Q, \lambda, \eta$ define the amount of kinetic energy, 
the angular momentum, and the ratio between rotational and random motions,
respectively.
We have done several tests 
to check their effect by taking all other parameters constant.

\item The three parameters $a,b,c$ determine the shape of the initial ellipsoid.
Depending on the initial shape of the overdensity the type of structures that are formed after the collapse may change.
We stress that the qualitative features of the collapse dynamics do not depend on these parameters, and thus
the results we discuss are quite general for this class of models. 
However, the quantitative characteristics of the post-collapse systems 
finely depend on the properties of the ICs and, in particular, on their shape.

\item 
The three parameters $T, \psi, \gamma$ define the physical properties of the gas component. 
{  If $T$ is high enough then the thermal energy prevents the collapse 
while if $T$ is very low GPs behave initially as PPs.
We have also varied $\psi$ in the range $10-50$ and
we have not observed relevant differences with the case $\psi=0.1$. 
We have fixed $\gamma$ so that the gas is a mono-atomic one.
}

\item {Note that we have performed a series of tests
by both varying the parameters of the code (i.e., 
the softening length, the time step accuracy, etc.) 
and the physical parameters discussed above. We found 
that the results are stable for variation of these 
parameters in a broad range and we refer to Refs.
\cite{syloslabini_2013, SylosLabini+Benhaiem+Joyce_2015, Benhaiem+SylosLabini_2015, Benhaiem+SylosLabini_2017}
for a more extensive discussion of the problem of resolution
for the crucial case of purely self-gravitating simulations.}

\end{enumerate}




\section{Dynamics of the collapse and post-collapse state}  
\label{sec:dynamics}

\subsection{Formation of disks and structures in purely self-gravitating systems} 
\label{sect:formation}

The violent gravitational dynamics of  systems composed by point masses that
{   start } from 
far-out-of  equilibrium configurations generally gives rise to a rich phenomenology, which
 we summarize below (see for more details Refs. \cite{Benhaiem+Joyce+SylosLabini_2017,Benhaiem+SylosLabini+Joyce_2019}).
The ICs consist of a subvirial self-gravitating and isolated system 
with initial uniform mass density, a nonspherical shape (e.g., an ellipsoid) 
and some angular momentum assigned in the form of a solid-body rotational velocity field.  
The overdensity undergoes  a monolithic collapse, driven by its own 
gravitational mean field contrasted by the internal pressure and PP velocity dispersion.  
This occurs whenever the  initial  internal density fluctuations are small. 
Indeed, internal density fluctuations grow during the system's collapse, forming larger and larger substructures.
A simple analytic treatment of the growth of fluctuations, neglecting 
the system's finite size, is based on the linear perturbation analysis 
of the self-gravitating fluid equations in a contracting background \citep{ejection_mjbmfsl}.
This is the same approach used in cosmology but for the case of an 
expanding universe \citep{peebles_1983} (a more detailed approach 
that considers the system's finite size  may be found  in Ref. \cite{aarseth_etal_1988}).
In these conditions the growth of perturbations is controlled by the amplitude of the 
initial fluctuations and by their correlation function. 

The collapse characteristic time-scale for a system with uniform density is of the order of 
\be
\label{tau}
\tau \sim \frac{r_g^{3/2}}{\sqrt{G M}}\;, 
\ee
where $r_g$ is the initial gravitational radius of the system and $M$ is its mass. 
The criterion to define the time $t^*$ when the  collapse is  halted is the following \citep{levin_etal_2008,ejection_mjbmfsl}: $t^*$ corresponds to the time when the size of nonlinear perturbations (defined, e.g., as the scale $\lambda_0$ at which the normalized mass variance is equal to one) becomes of the order of the system's gravitational radius $r_g$.
Thus, if the initial fluctuations have a small enough amplitude and/or they are not strongly correlated,  then  $t^* \approx \tau$ so that the system has had time  to contract by a large factor. 
Otherwise, if density fluctuations have a large amplitude and/or are spatially correlated, bottom-up perturbations grow 
rapidly enough that a large contraction does not occur because of the quicker system fragmentation into many substructures.  
In that case clustering proceeds through a bottom-up aggregation process, i.e., a \textit{slow and soft} dynamics.
If the system is initially  not spherically symmetric, the monolithic collapse eventually leads to the formation of a quasistationary thick disk, in which rotational motions dominate but with a large velocity dispersion. 
In the outermost regions spiral arms are formed, possibly with bars and/or rings, in which particles do not follow steady circular orbits because their velocity has both a rotational and a radial component.
While, globally, the system reaches a quasiequilibrium state close to a virial configuration, its external parts, which contain only a fraction of the system mass, expand (i.e., they are out-of-equilibrium) for  a much longer time than the gravitational collapse timescale $\tau$. 
Let us further consider the origin of the out-of-equilibrium structures. 

{  The variation of the mean gravitational field  during the collapse 
triggers a change of the particle energy distribution, which in turn, 
induces a reassessment of the system's phase-space macroscopic properties. }
This mechanism is both rapid and energetically violent, and it works as follows. 
Although initially all particles are bound, during the collapse a fraction of them can gain some kinetic energy. 
The mechanism of particle energy gain originates from the coupling of the growth of 
inner density fluctuations with the finite size of the system. 
Particles originally placed close to the system boundaries develop a net lag with respect to the bulk 
because the density in the outer regions of the system decreases during the collapse as a consequence 
of the growth of density fluctuations and of the 
corresponding peculiar motions.
While in an inner shell at $R$ the flow of particles from $<R$ or $>R$ is 
statistically symmetrical, in the outermost regions there is an asymmetry because of the system's finite size: 
for this reason, during the collapse, 
 there is a  net outflow of particles in the outer regions so that 
their density becomes 
smaller than that in the inner ones. Correspondingly the collapse time 
becomes larger than that of the others, and thus a time lag is developed.
Thus such particles arrive at the system center when the others are already re-expanding. 
In such a way, these particles move for a short 
time interval in a rapidly varying gravitational field and for this reason they can gain kinetic energy.
In consequence of this mechanism the whole {  particle } energy distribution largely changes.
Given the complex interplay between the growth of density 
fluctuations and the system's finite size, 
both the IC shape and {  the nature of correlations between 
density fluctuations}  determine the details of this process.  
The larger the deviation from spherical symmetry of the ICs, 
the larger the spread of particles' arrival times at the center and thus the larger the particles' energy gain
\citep{Benhaiem+SylosLabini_2015}.

The initial anisotropic distribution is thus amplified by the collapse mechanism 
because the particles that initially lie in the outermost regions, and that are thus 
strongly anisotropically distributed if the system breaks spherical symmetry, get the largest energy increase.
For instance, in the case of a simple prolate ellipsoid with semiaxis $a>b=c$ 
these particles are initially {  located} 
in the region $a<r<b=c$ \citep{Benhaiem+SylosLabini_2015}, 
and they are not spherically symmetric either after the collapse.
Indeed, the collapse amplifies their initial asymmetry, 
because of such  positive energy gain, 
so that 
the final 
distribution becomes 
very anisotropic, with a shape close to a  thick disk whose minor axis coincides with the rotation axis as 
the system is stretched along the plane identified by the major and medium
axes of the initial configuration %
In addition, such a disk is surrounded by  out-of-equilibrium spiral arms that are formed because the most energetic particles, having both a transverse and a radial velocity component, move in an almost central gravitational field and thus conserve angular momentum \citep{Benhaiem+Joyce+SylosLabini_2017}.

In summary, the asymmetric collapse and re-expansion induce a  mass loss from the system, as some particles may gain enough kinetic energy to escape from it, and this leads to a new, marginally stable, configuration of lower energy.
If the initial angular momentum is nonzero, such a state forms a thick disk whose minor axis is oriented parallel to the angular momentum.


\subsection{Evolution of global system quantities} 

For the initial conditions described in Sect.\ref{subsect:mod} the collapse 
and the subsequent relaxation to a QSS, in the system internal region, is characterized by three different time phases.
They can be identified (see Fig.\ref{figure_Rg}) by analyzing the behavior of the system's 
{  dimensionless gravitational radius \cite{Yan_Miller_1997}}:
\begin{figure}
\includegraphics[width = 3.3in]{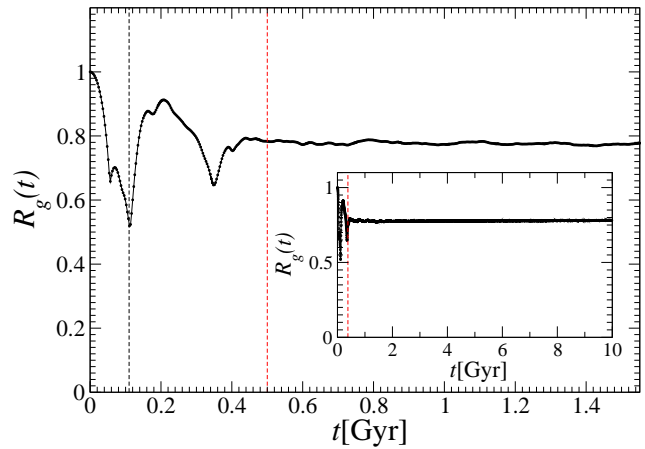}
\caption{
Time evolution of the normalized gravitational radius $R_g(t)$
(see Eq.(\ref{RgN})). 
Results for longer times are illustrated by the inner panel.
{The vertical lines correspond to $t=\tau$ and $t= 5\tau$.}
}
\label{figure_Rg} 
\end{figure}
\be
\label{RgN}
R_g (t) = \frac{W(0)}{W(t)},
\ee
where $W(t)$ is the gravitational potential energy at time $t$ and $W(0)$ that at the initial time.
The first phase corresponds to an initial decrease of $R_g(t)$ up to when it reaches its absolute minimum at $t\approx \tau \approx 0.1$ Gyr where $\tau$ is given
 by Eq.(\ref{tau}). 
This phase is thus driven by an overall contraction of the system and, as we will discuss below, by the dissipation of the gas internal energy. 
{  The gravitationally collapsing nongaseous matter rapidly changes its shape becoming  flat}
--- along the rotation axis ---  to a lesser extent compared to the gas distribution.
Together with the gravitational radius, the mean gravitational potential energy of the system decreases. 
Such a rapid potential variation triggers a large change of the particles' total energy
\footnote{The total energy of a particle, per unit mass, is 
$\epsilon =   (1/2) v_i^2 + \phi_i + u_i$ where $v_i$ is 
the velocity of the $i$th particle, $\phi_i$ its 
gravitational potential per unit mass and $u_i$ 
the specific internal energy (that is zero for a PP).
}
distribution and thus, in turn, of the system's phase-space properties.

This first phase is then followed by a second one, for $\tau < t < 5 \tau \approx 0.5$ Gyr,
characterized by a few damped oscillations of $R_g(t)$ and of the total gravitational potential energy. 
During such oscillations  the system shows rapidly varying transient configurations both in real and in velocity space.
Then, in the third phase, for $r> 5\tau$, the system is relaxed to a QSS in its inner region and $R_g(t)$ has reached its asymptotic value.
However, in the outermost regions there are out-of-equilibrium structures that yet continue to evolve for times $t \gg \tau$.

\begin{figure}
\includegraphics[width = 3.3in]{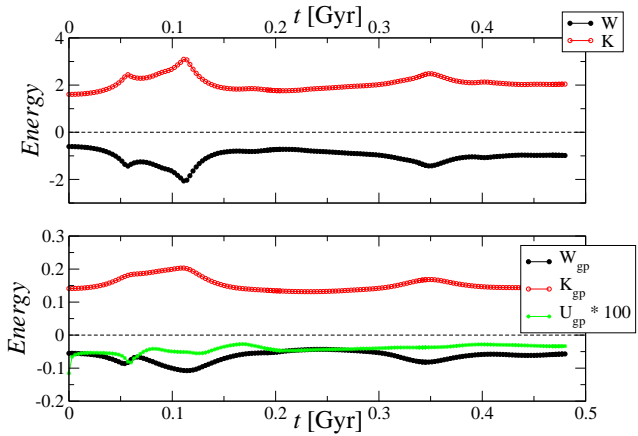}
\caption{ 
Upper panel: time evolution up to $t=5 \tau \approx 0.5$ Gyr of the kinetic 
($K$) and potential energy ($W$),
for all particles.
Note that  both energies are normalized to the initial potential energy, 
Bottom panel: time evolution of the kinetic ($K_{gp}$), potential 
 ($W_{gp}$), and thermal energy ($U_{gp}$ --- multiplied by $10^2$) for the GPs.
}
\label{figure_energy} 
\end{figure}

Figure \ref{figure_energy} 
shows the temporal evolution of the system's  kinetic,
potential energy (top panel) and of the solely gas component (bottom panel) together with its thermal energy. 
The kinetic $K$ and potential energy $W$ of the PPs are initially of the same order of magnitude as the virial ratio is $Q \approx 3/4$. 
Even for the GPs the kinetic energy $K_{gp}$ is of the same order of the 
potential energy $W_{gp}$ amounting to $\sim 10\%$ that of the PPs.
In addition, the total thermal energy $U_{gp}$ of the GPs gives a  negligible 
contribution to the GPs kinetic energy, being $U_{gp}$ about the $1 \%$ of $K_{gp}$
Indeed, given the values of $T,M,r_g$ we have used 
the ratio between the initial internal energy per unit mass and the typical particle's potential energy is 
\be
\label{u_to_e0} 
\frac{u}{|\phi_0|} =  \frac{\chi T}{\frac{GM}{r_g}} \approx 10^{-2} \;.
\ee

\subsection{The three dynamical phases}

\begin{figure}
\includegraphics[width = 3.3in]{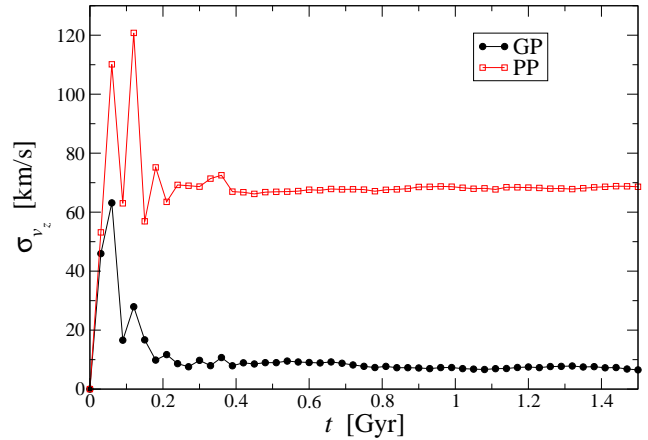}
\caption{Time evolution of the vertical velocity dispersion $\sigma_{v_z}$.
Results for both gas and particle components are shown.} 
\label{figure_Pvz} 
\end{figure}

%
%
During the first phase the GP develop a very flat distribution along the plane orthogonal to the rotation $Z$-axis.
This occurs because  the gas quickly increases its central density during the rapid system's collapse.
When the gas becomes dense enough the radiative cooling function $\Lambda(u,\rho)$ (see Eq.\ref{cool}), according to the schemes adopted in
then literature and used in this work, becomes proportional {  to $\rho^2$} and the cooling processes acquire more and more efficiency ({  see Sect.\ref{sect:gas_dynamics}}).
Thus, thermal energy may be easily dissipated reducing the $Z$
 component of the velocity as a particle crosses the  $XY$ plane. 
In such a way the GP component  develops a much flatter distribution than that of the  PP component.
Such further dissipation lets the GPs  lower their temperature.
%
%
%
%
%
%
For this reason, together with a decrease of the vertical thickness, 
the GPs drastically reduce their vertical velocity dispersion at $ t \approx \tau$ Gyr (see Fig.\ref{figure_Pvz}). 

\begin{figure}
\includegraphics[width = 3.3in]{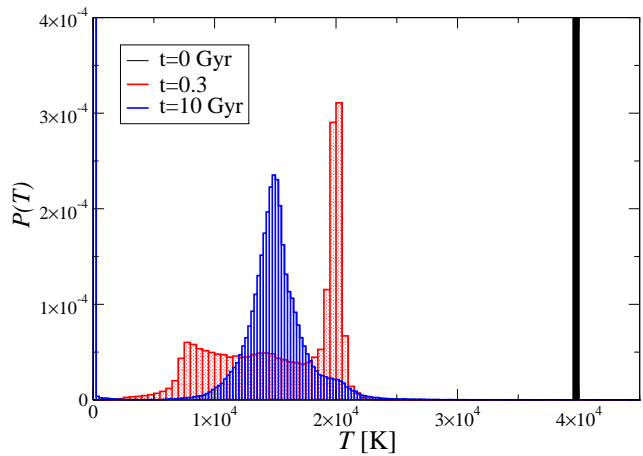}
\caption{PDF of the gas temperature at different times: $t=0,\,0.3,\, 10$ Gyr.
The GP initial temperature distribution is plotted too, and it is a Dirac $\delta$-function
centered at $T=40,000$ K.}
\label{figure_Temp} 
\end{figure}

\begin{figure}
{\includegraphics[width = 3.7in]{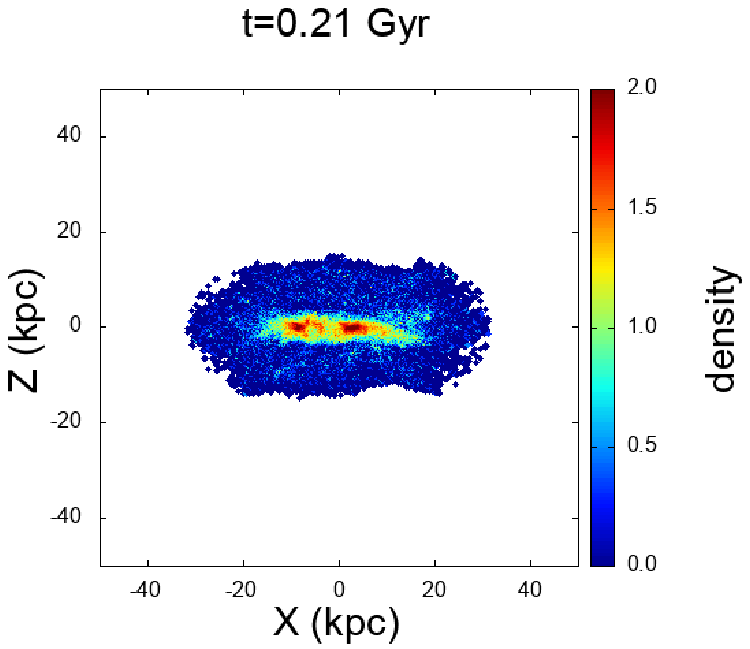}}\\
{\includegraphics[width = 3.7in]{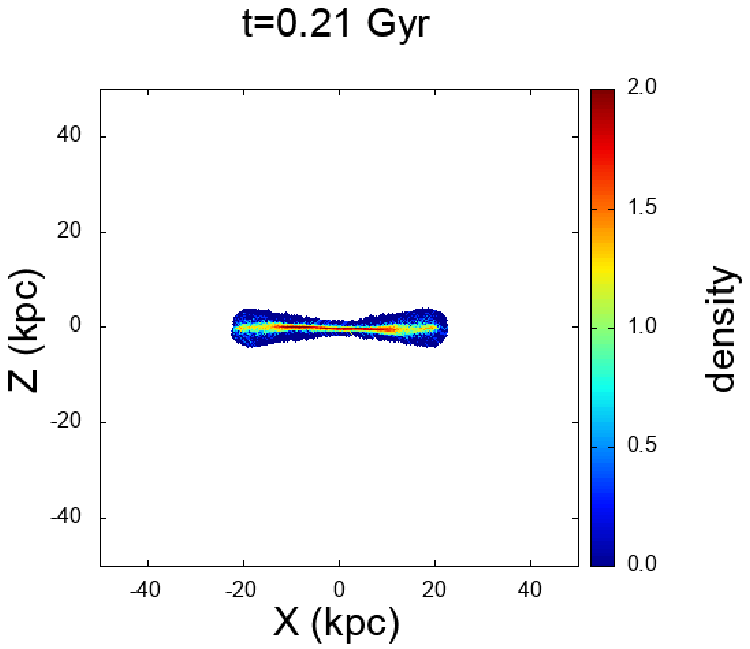}}
\caption{Density map of the PP component (upper panel) and of the GP component (bottom panel) on the $ XZ$  plane at time $t=0.21$ Gyr.
The density is computed in cells in the $XZ$ plane and it is integrated over the $Y$ axis; the color scale  is logarithmic. 
}
\label{figure_GP+SGP_XZ} 
\end{figure}

%
The probability density function (PDF) of the temperature $P(T)$ of the GPs is shown in Fig.\ref{figure_Temp}: 
{  we can observe a progressive} 
redistribution of the SPH particles temperatures towards lower values.
At $t=0.3$ Gyr the distribution $P(T)$ is already 
peaked at $\sim 18,000$ K; then, at longer times, the  GPs 
slowly cool down so that the maximum of  $P(T)$ reaches 
$T \approx  15,000$ K. 
In consequence of the overall system's collapse also the PPs rapidly change their 
spatial distribution by 
contracting along the $Z$ axis, although forming a less flat structure compared to the gaseous disk.
As the PP component is gravitationally dominant, this contraction modifies the system's 
mean gravitational potential (see Fig.\ref{figure_GP+SGP_XZ}).

\begin{figure}
\includegraphics[width = 3.in]{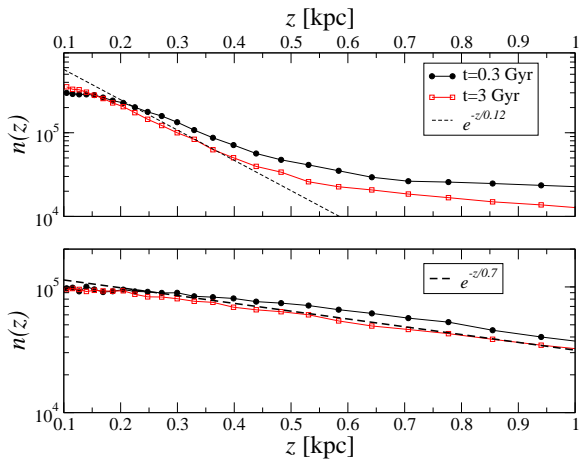}
\caption{Time evolution of the vertical number density profile of the GPs (upper panel) and PPs (bottom panel) at $t=0.3$ Gyr and $t=3$ Gyr.
As a reference we report (dashed lines) an exponential decay $n(z) \sim \exp(-z/z_0)$
where $z_0=0.12$ kpc for the gas component and 
$z=0.7$ kpc for the nongaseous component.
 }
\label{figure_nz} 
\end{figure}

The differences acquired by the vertical configurations of the density distributions of the two components are clear in Fig.\ref{figure_nz}, which shows the vertical density profile of both the PPs and the GPs: they both display an exponential decay 
\[
n(z) \sim \exp(-z/z_0)
\]
with $z_0 \approx 0.7$ kpc for the PPs 
and  $z_0 \approx 0.12$ kpc for the GPs.
Thus, at $t \ge 0.2 $ Gyr the GPs form an extremely flat disk and the PPs a thicker disk: when the density increases such that the cooling becomes very efficient, the gas component decouples from the PPs component and it starts to have a different time-evolution. 
{  Note that
 PPs and GPs have a different velocity dispersion and this is the reason why the 
GPs do not follow the same trend of the PPs in their vertical density profile (see Fig.\ref{figure_nz}).}

\begin{figure}
\includegraphics[width = 3.5in]{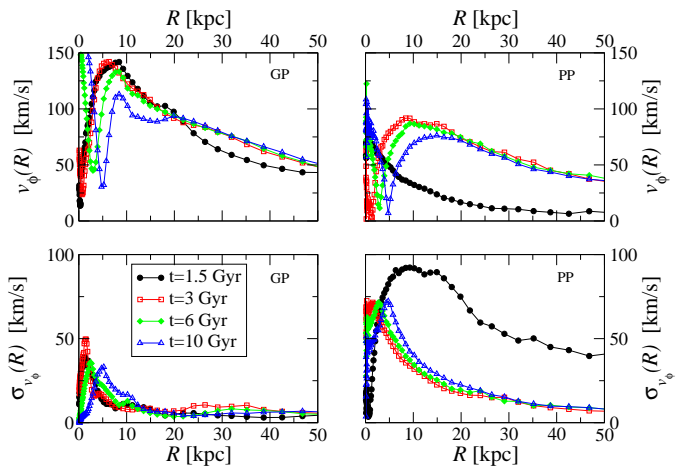}
\caption{Azimuthal velocity profile (upper panels) and its dispersion profile  (bottom panels) 
for the GPs (left panels) and PPs (right panels)
at different times (see upper left labels).
}
\label{figure-Vphi} 
\end{figure}

\begin{figure}
\includegraphics[width = 3.5in]{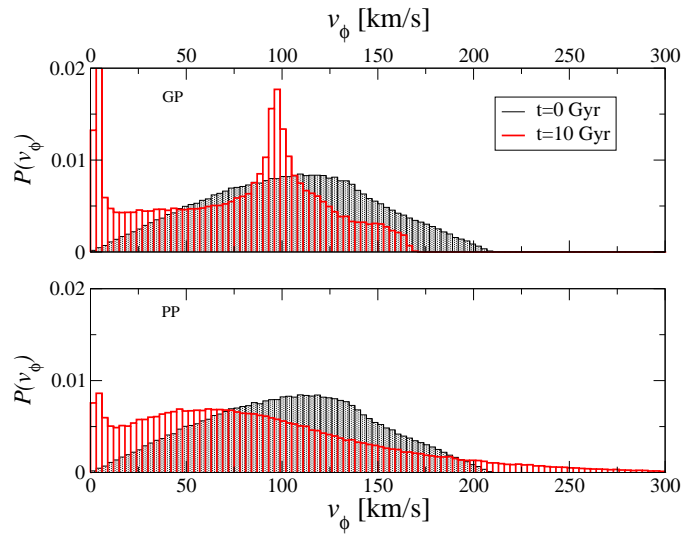}
\caption{
PDF of the azimuthal velocity for GPs (upper panel) and PPs (bottom panel) at $t=0$ and $t=10$ Gyr. }
\label{figure-Pvphi} 
\end{figure}

%
The difference in the motion of GPs and PPs after the collapse can  be noticed  
by looking at the profile of the azimuthal velocity  $v_\phi(R)$, i.e.,  the mean azimuthal 
velocity evaluated in concentric circular coronas in the disk as a function of the 
two-dimensional
(2D) disk radius $R$, and of  its dispersion profile $\sigma_{v_\phi}(R)$
\footnote{Unless differently specified  we adopt a cylindrical coordinates system $(R,\phi,z)$. }.
In particular, the amplitude of $v_\phi(R)$ of the GPs component is larger (by about  a 
factor $\sim 2$) than that of PPs, while the dispersion is  smaller by 
 a factor $\sim 6-8$ (see Fig.\ref{figure-Vphi}). 
%
%
%
%
Such a noticeable difference is due to the fact that the motion of the  gas component, 
being confined on a thin disk, is much more coherent than that of nongaseous matter.
The signature of such a coherence is shown by the PDF
$P(v_\phi)$ of the azimuthal velocity: this is more peaked for GPs than for PPs
 (see Fig.\ref{figure-Pvphi}). 
{  Note that the maximum radial anisotropy}
\be 
\beta = 1 - \frac{\langle v_t^2 \rangle}{2 \langle v_R^2 \rangle} \rightarrow 1 \;,
\ee
{  is reached in the outermost regions of the system corresponding to 
the peak of $P(v_\phi)$ for $v_\phi \rightarrow 0$.} 
Indeed, when a  particle increases its distance from the system's center, 
it decreases its tangential velocity because it approximately moves in a central potential conserving its angular momentum.

\begin{figure}
\includegraphics[width = 3.5in]{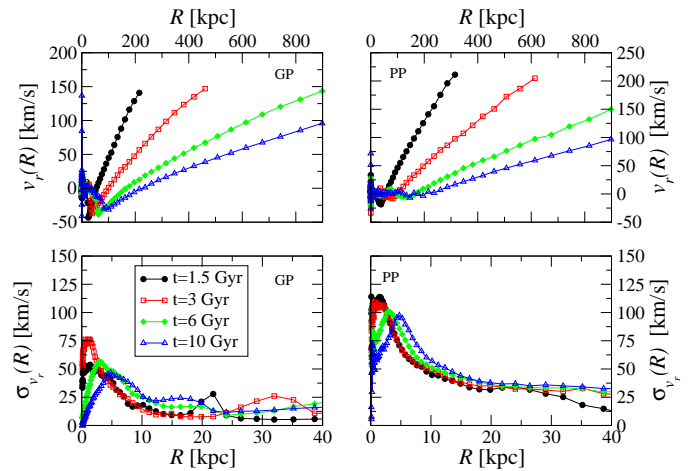}
\caption{ 
Radial velocity profile (upper panels) and its dispersion profile  (bottom panels) 
for the GPs (left panels) and PPs (right panels)
at different times (see labels).}
\label{figure-Vr} 
\end{figure}

\begin{figure}
\includegraphics[width = 3.5in]{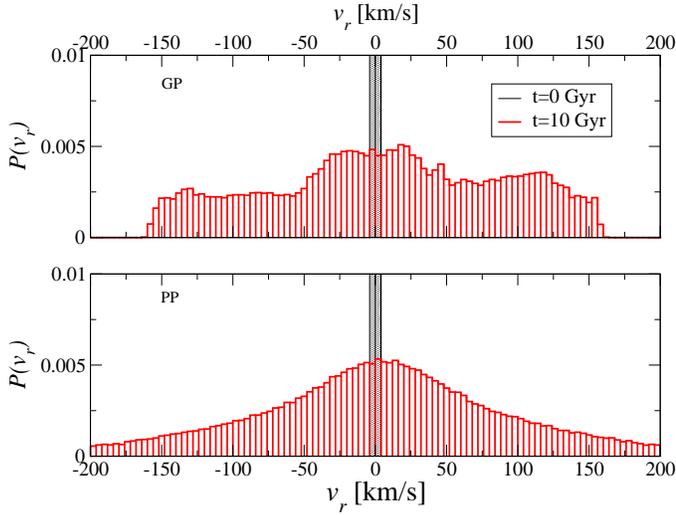}
\caption{
PDF of the radial velocity  for GPs (upper  panel) and PPs (bottom panel) at the initial time and $t=10$ Gyr.}
\label{figure-Pvr} 
\end{figure}

%
%
%
The radial velocity profile $v_R(R)$  (computed in circular coronas) 
 and the relative dispersion profile $\sigma_{v_R}(R)$ 
of both the PPs and GPs show a large-distance time-dependent tail (see Fig.\ref{figure-Vr}): 
this is due to the out-of-equilibrium particles that have
gained the largest amount of energy during the collapse phase.
At small distances the dispersion $\sigma_{v_R}(R)$
is also larger, by a factor 2-3, for PPs than for GPs.
%
%
%
%
%
%
The PDF of the radial velocity is shown in Fig.\ref{figure-Pvr}: one may note that 
while PP have an approximately symmetric PDF the GPs show an asymmetrical one with a persistent tail both at positive values.
%

\begin{figure}
\includegraphics[width = 3.5in]{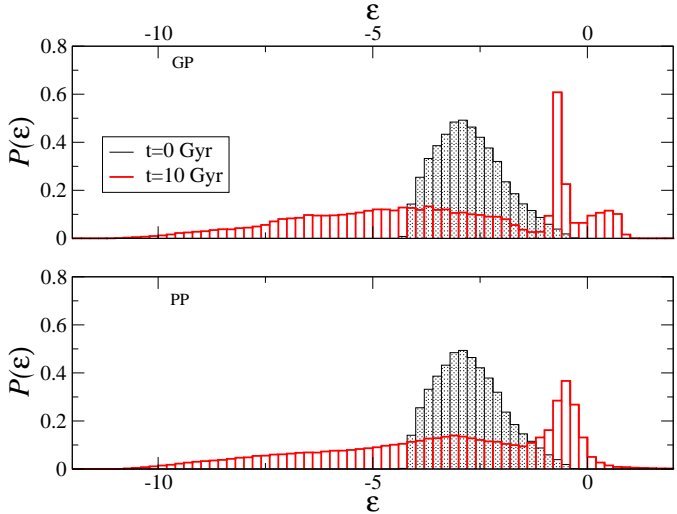}
\caption{
  Energy distribution  at $t=0$ and $t=9$ Gyr for the GPs (upper panel) and PPs (bottom panel). 
}
\label{figure_Pe} 
\end{figure}

%
%
%
At the end of the second phase, at $t \approx 5 \tau \approx 0.5$ Gyr, the system has almost reached its asymptotic state. 
In particular, the particle energy PDF $P(\epsilon)$ quickly relaxes to an 
almost time-independent shape that determines the properties of the QSS (see Fig.\ref{figure_Pe}).
This distribution has undergone to a substantial change at $t \sim \tau$ in consequence 
of the rapid variation of the system's mean gravitational field.
Indeed, particles moving in a rapidly varying potential field do not conserve energy 
and the variation of the mean-field potential triggers the change in the system's
macroscopic properties as we discussed above for the PP case. 
The energy change is, however, different for the PPs and the GPs, reflecting their different dynamical evolution.
In particular, because GPs can dissipate energy they have an energy distribution 
with a negative tail that is more extended for  than for PP. Note that the fraction of the GPs mass for $r>10$ kpc is
less than $10\%$ of the total GPs mass.
{  The boundary conditions are open and 
thus the escaping particles increase their distance 
indefinitely.}
%
%
%
%
%
%

\begin{figure}
\includegraphics[width = 3.5in]{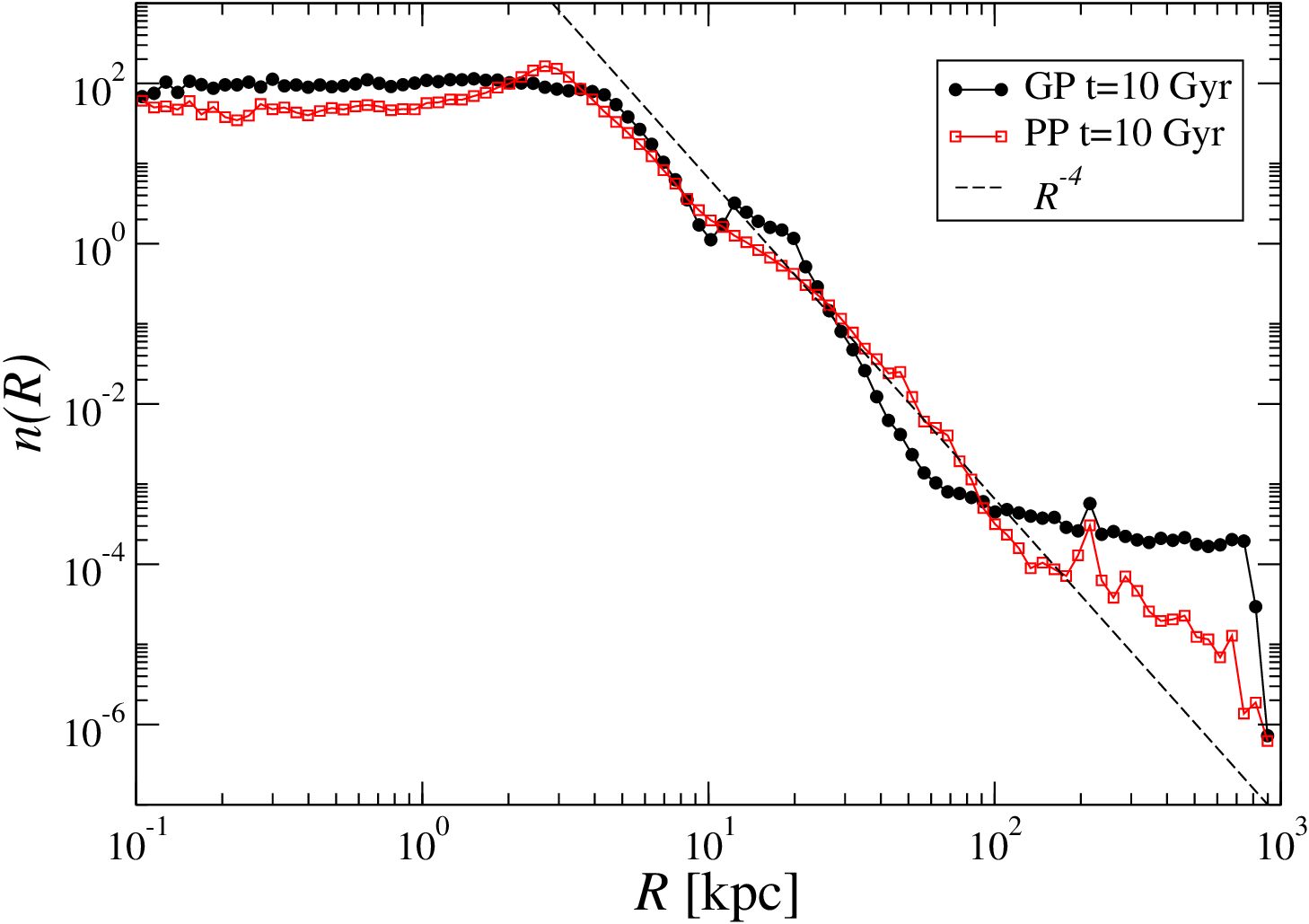}
\caption{
Number density profile for both the PPs and GPs 
(arbitrarily re-scaled on the $Y$-axis) at different times (see labels).
}
\label{figure-densityprofile} 
\end{figure}

\begin{figure}
\includegraphics[width = 3.5in]{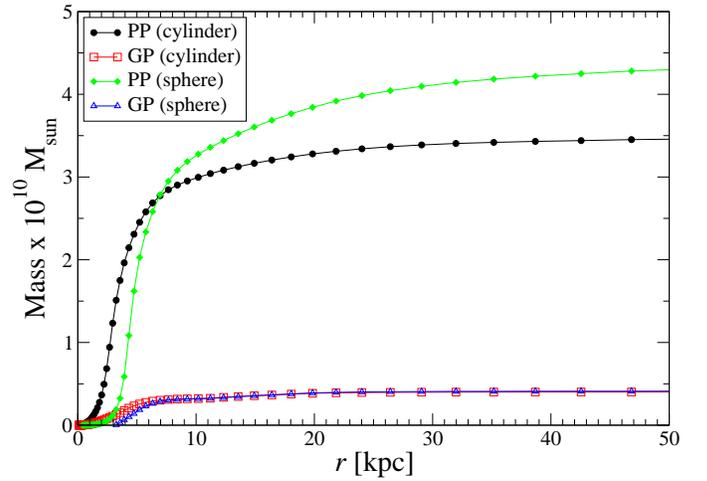}
\caption{Integrated mass as function of the radius of a cylinder of thickness $\Delta Z =2$ kpc and in spheres.
{  Note that the mass is in units of $10^{10} M_\odot$.}
}
\label{integrated-mass} 
\end{figure}

The density profile (see Fig.\ref{figure-densityprofile}) shows a flat core and an approximate $n(R) \sim R^{-4}$ 
decay at large distances: both behaviors are typically formed after a violent enough collapse \citep{syloslabini_2012}.
Note that the large-distance tail continues to evolve for long times, 
due to the particles with energy close to or larger than zero.
%
%
%
%
%
%
Finally, the {behavior of integrated mass versus radius } 
is reported in Fig.\ref{integrated-mass}: we show both the  integrated mass $M(R)$ computed in a tiny cylinder with thickness $\Delta Z=2$ kpc (in cylindrical coordinates) and the integrated mass $M(r)$ computed in spheres. 
The difference between the two is due to the fact that the PP is not confined on the thin disk as its mass is distributed in a larger volume around it.
%
%
%
%
%
%
%
%
%
%
%
%

\subsection{The inhomogeneous gas velocity field} 

Given the complex dynamical mechanism at work, the velocity field of the system formed after the collapse is rather heterogeneous: not only the PPs and the GPs have a different velocity field but, in both cases, its properties depend on scales.
Let us now focus on the GP component given that the PP component shows the evolution we described above for the case of a purely self-gravitating collapse: GPs represent indeed a small perturbation of the system mass and thus the evolution of the PPs is unperturbed by the presence of the GPs. 

\begin{figure*}
{\includegraphics[width = 2in]{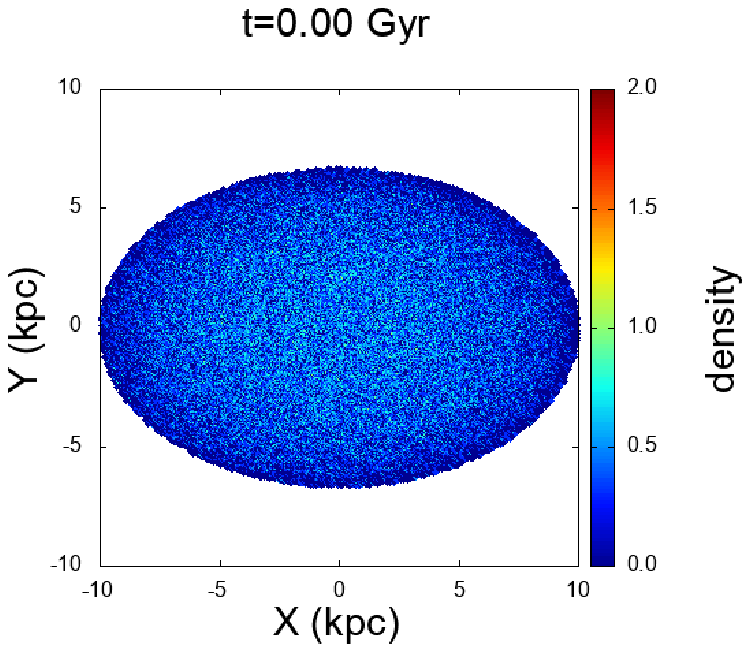}}
{\includegraphics[width = 2in]{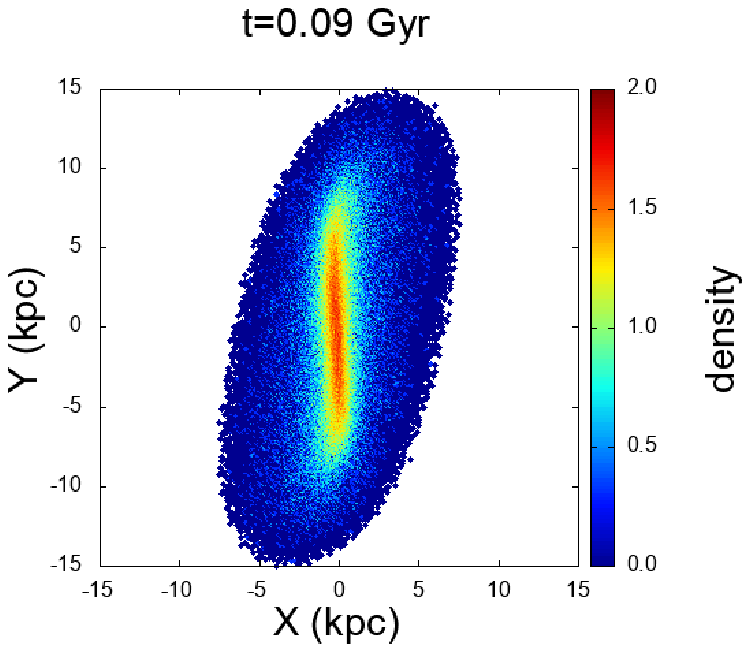}}
{\includegraphics[width = 2in]{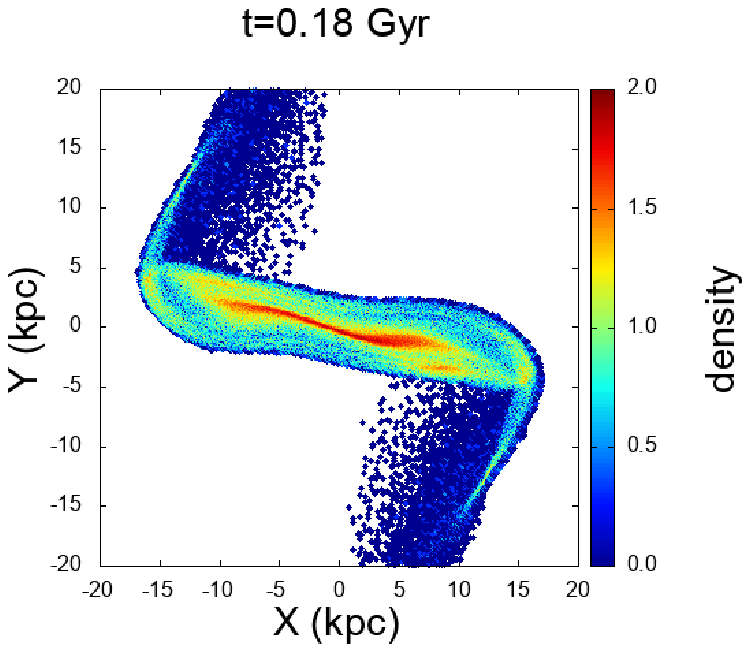}}\\
{\includegraphics[width = 2in]{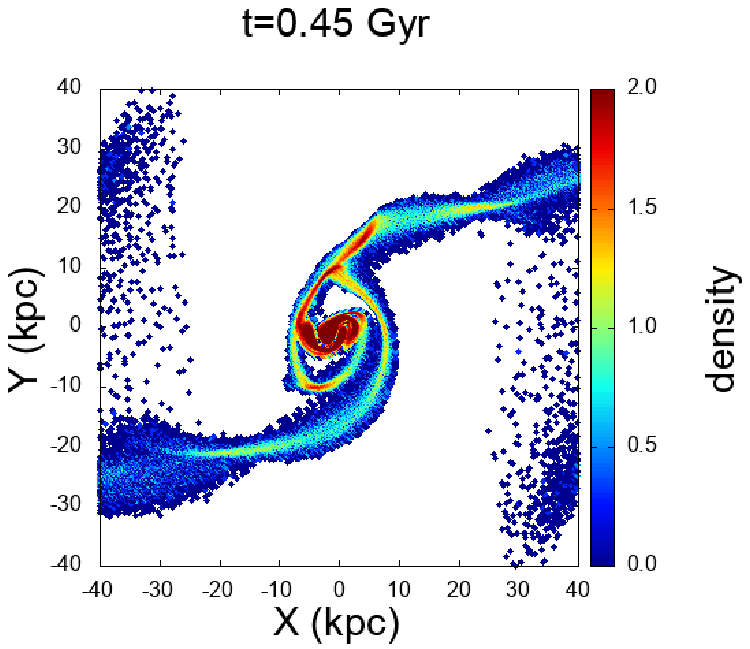}}
{\includegraphics[width = 2in]{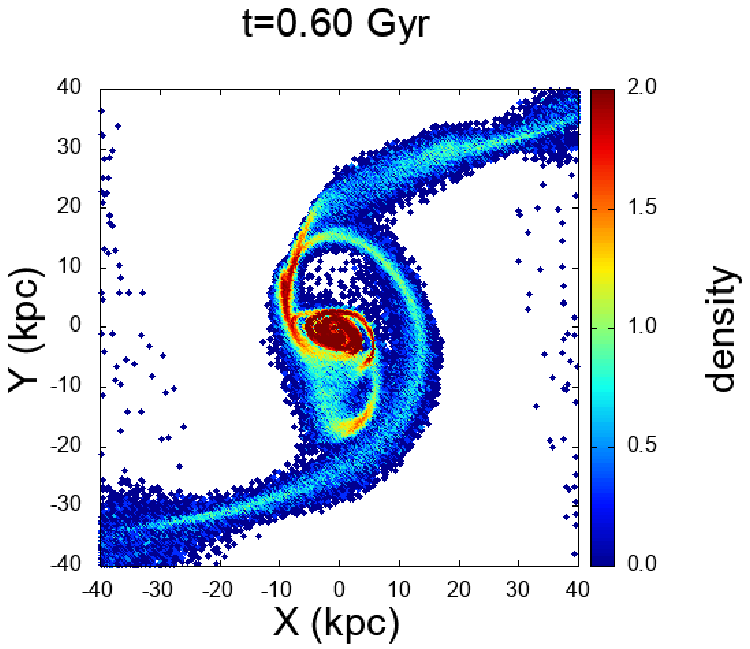}}
{\includegraphics[width = 2in]{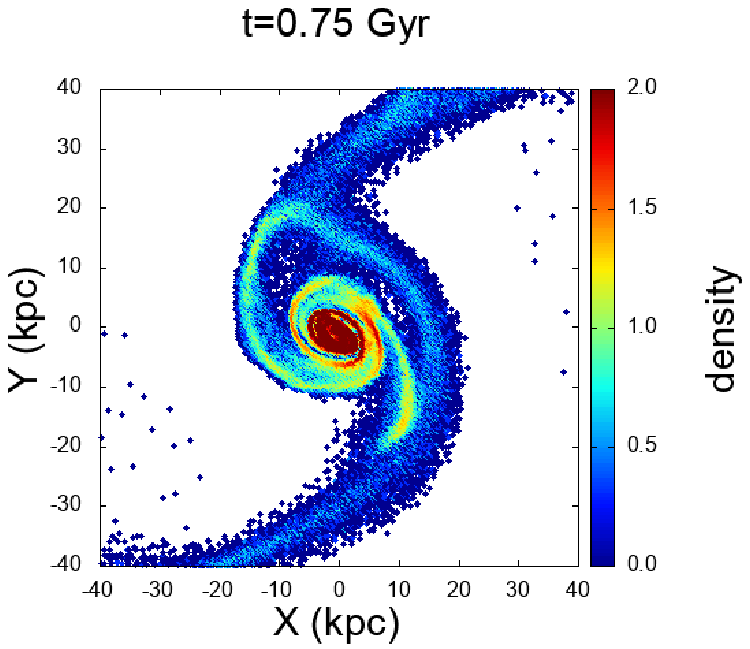}}\\
{\includegraphics[width = 2in]{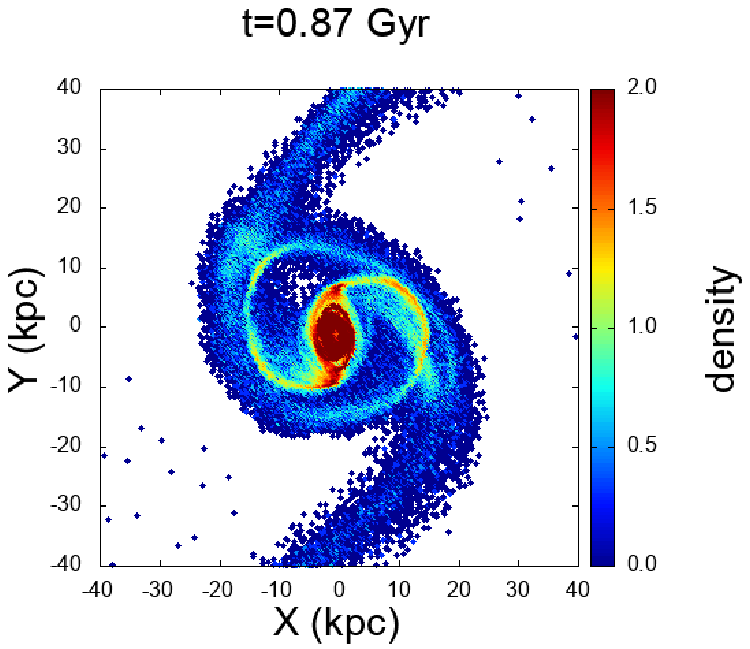}}
{\includegraphics[width = 2in]{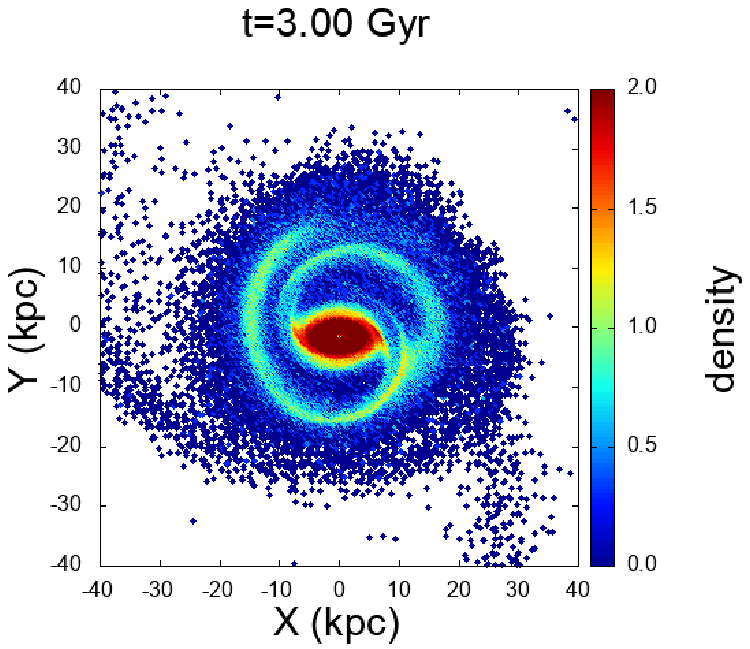}}
{\includegraphics[width = 2in]{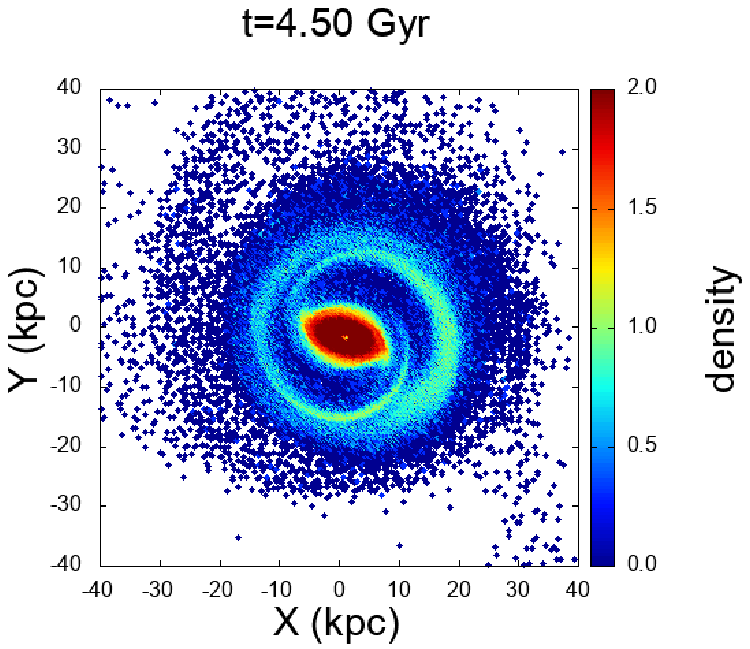}}\\
{\includegraphics[width = 2in]{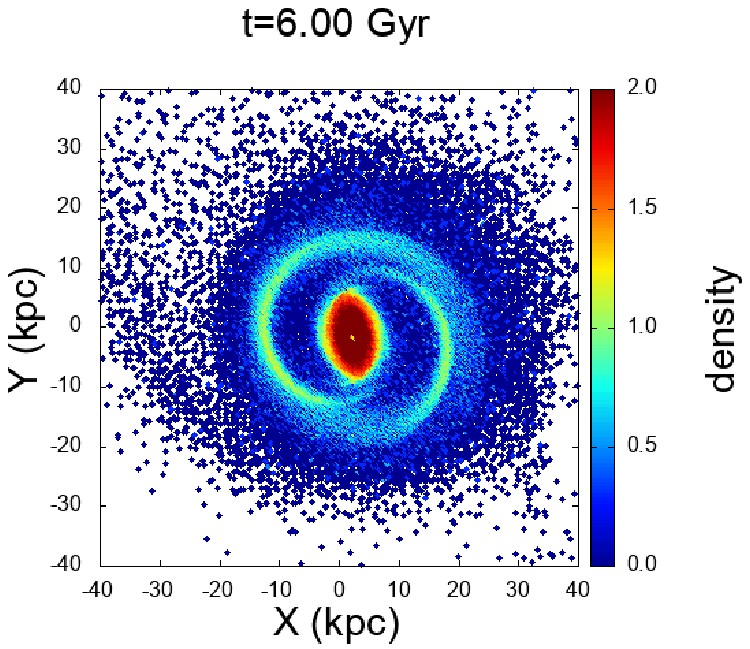}}
{\includegraphics[width = 2in]{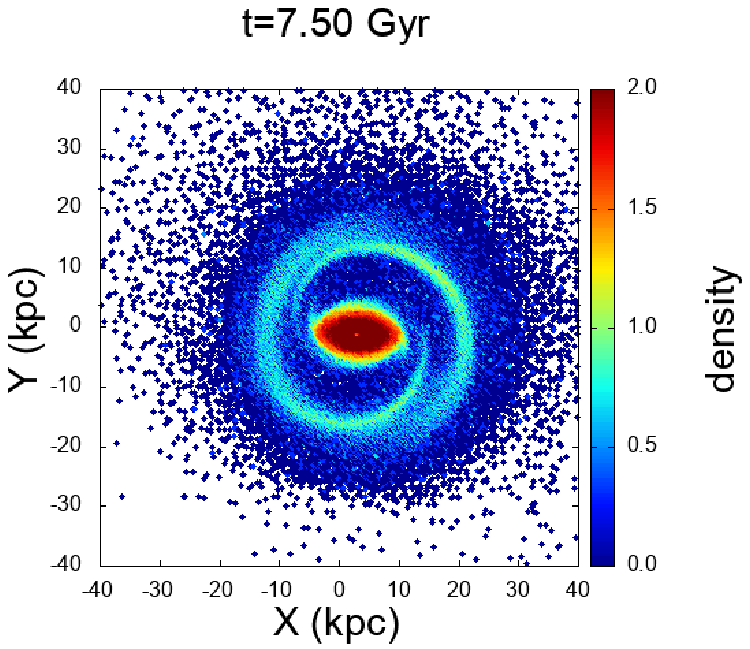}}
{\includegraphics[width = 2in]{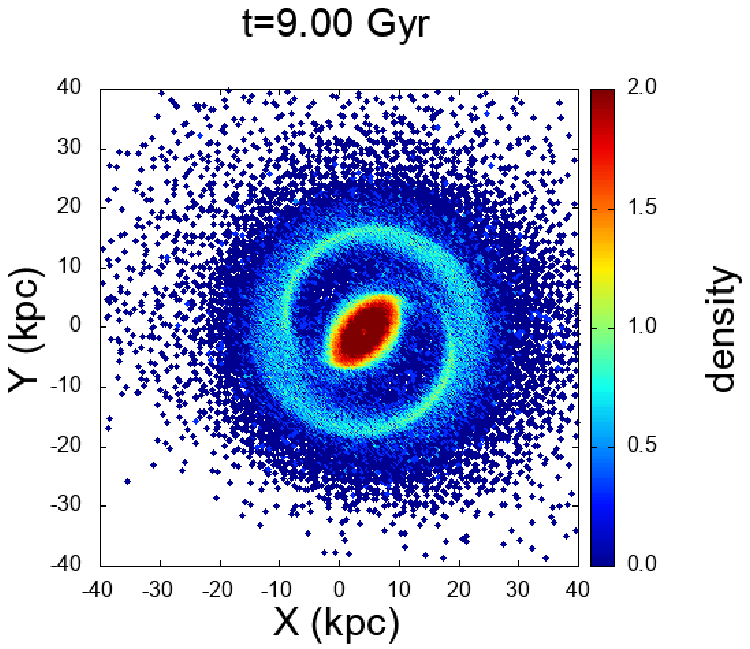}}
\caption{
Projection of various times slices of the GP component on the $XY$ plane: the color code corresponds to the logarithm of the density integrated over the $Z$ axis.  
The corresponding time is reported at the top of each each panel. 
}
\label{GP-XY} 
\end{figure*}

\begin{figure*}
{\includegraphics[width = 2in]{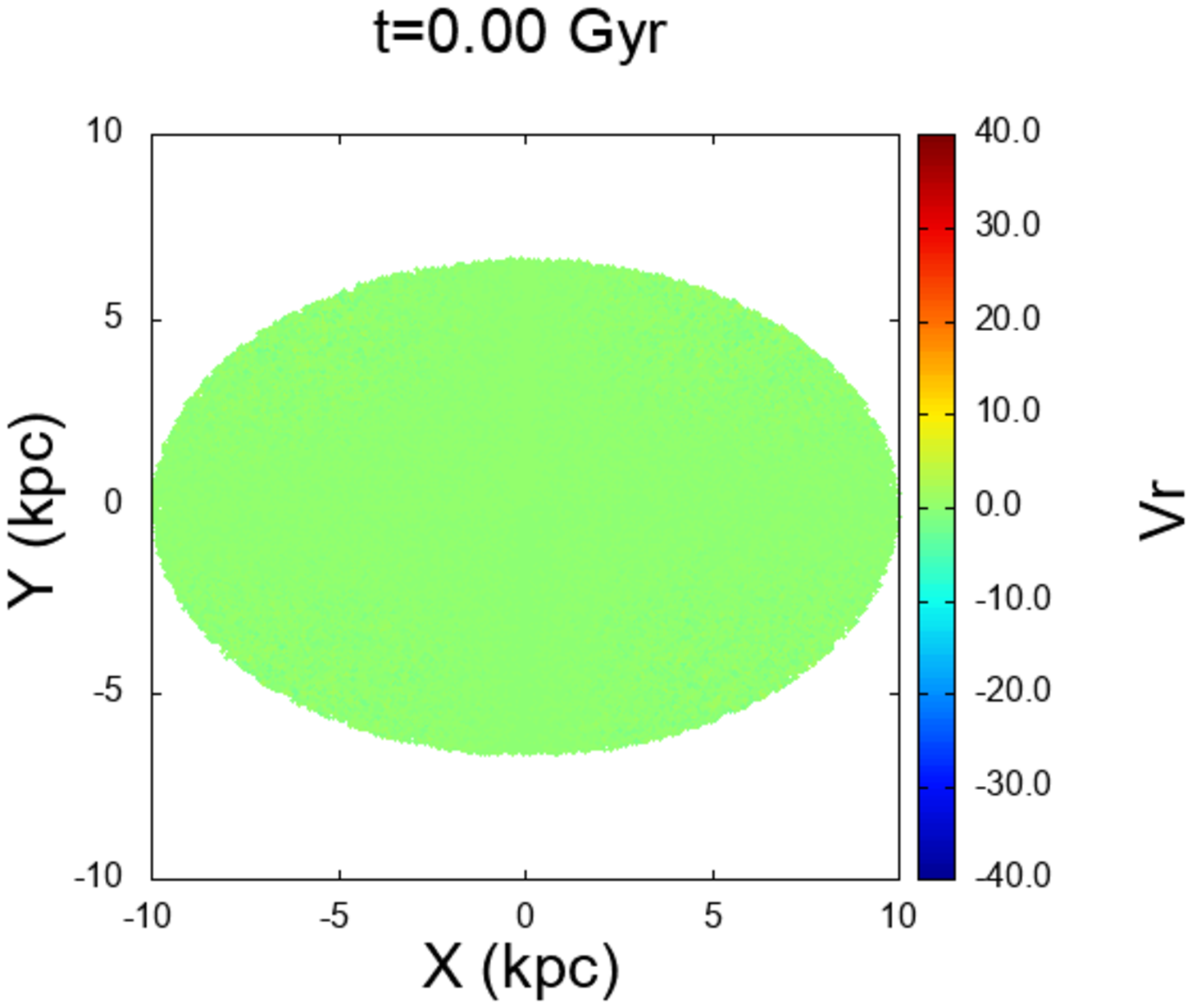}}
{\includegraphics[width = 2in]{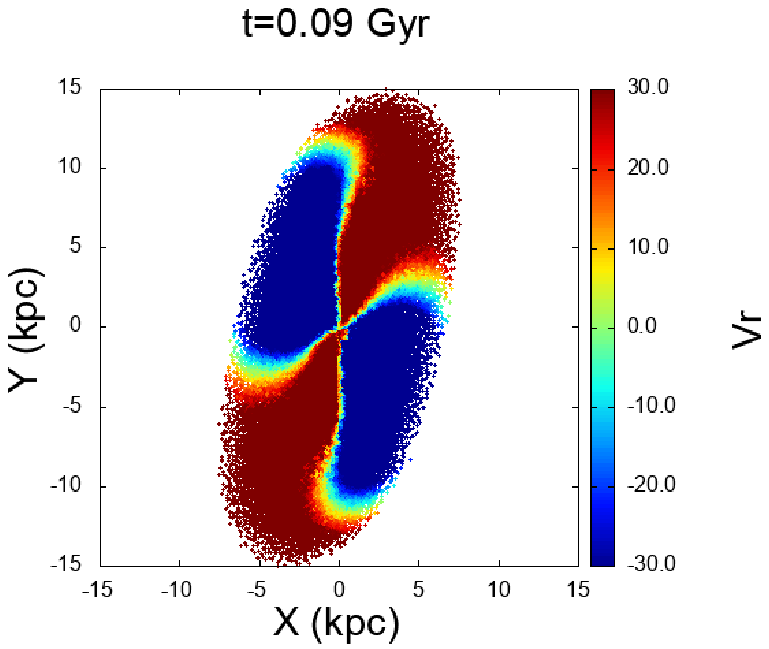}}
{\includegraphics[width = 2in]{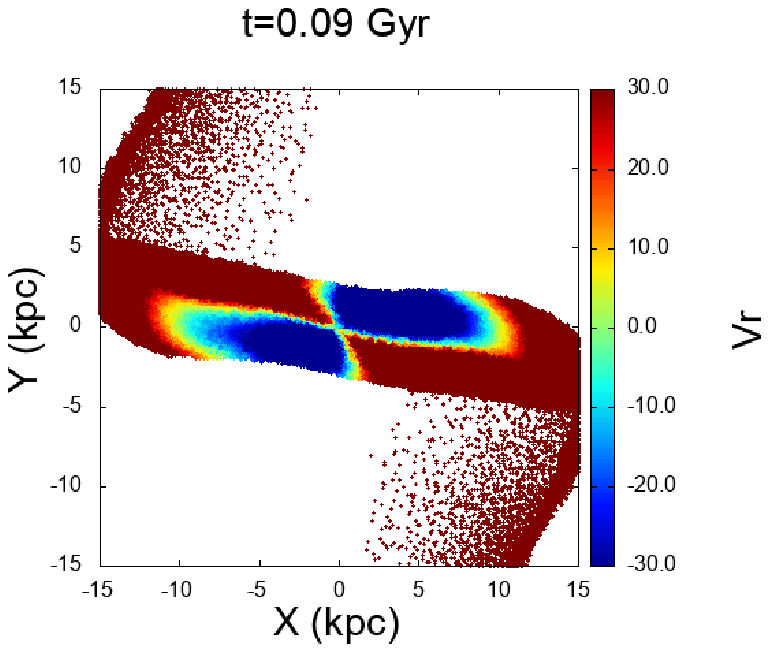}}\\
{\includegraphics[width = 2in]{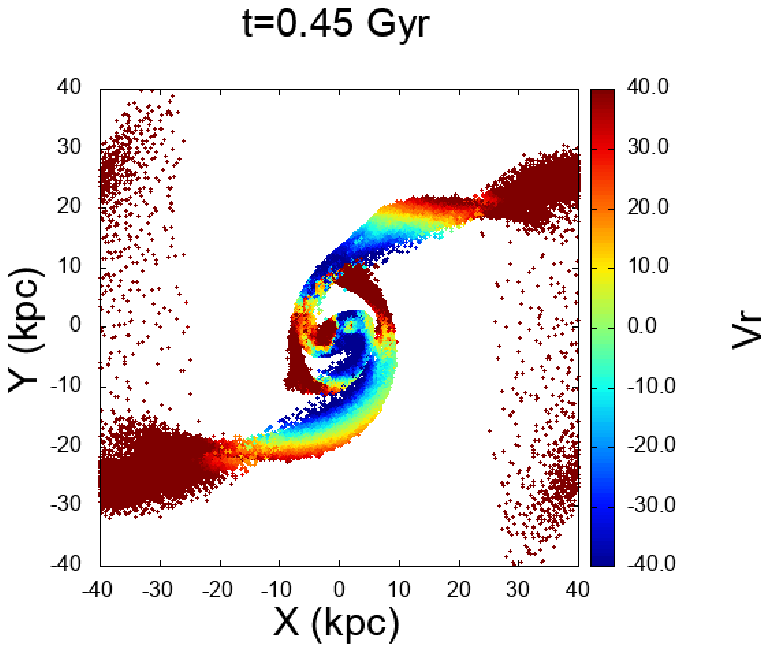}}
{\includegraphics[width = 2in]{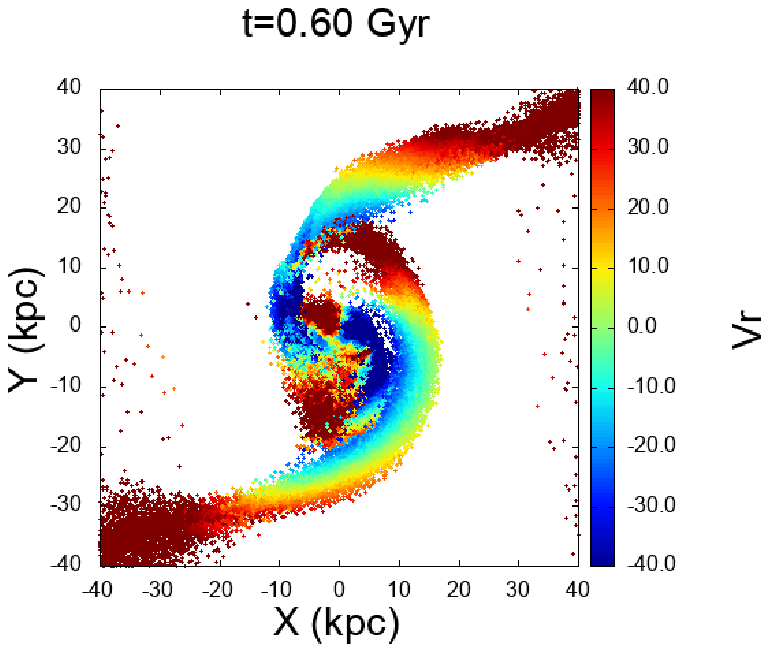}}
{\includegraphics[width = 2in]{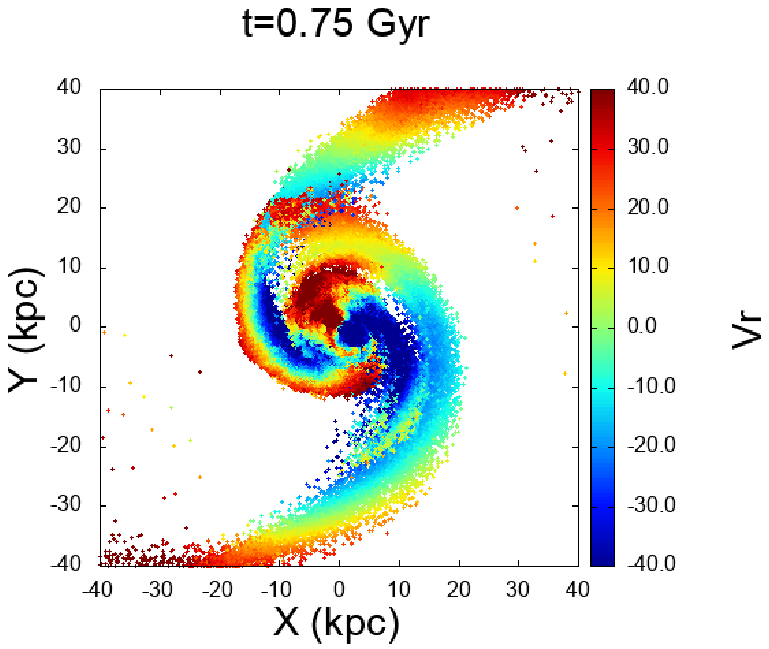}}\\
{\includegraphics[width = 2in]{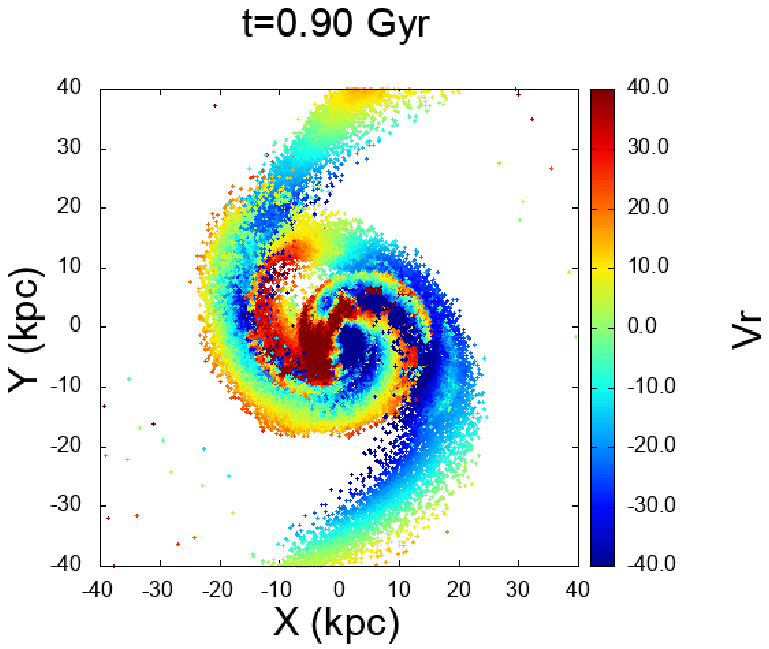}}
{\includegraphics[width = 2in]{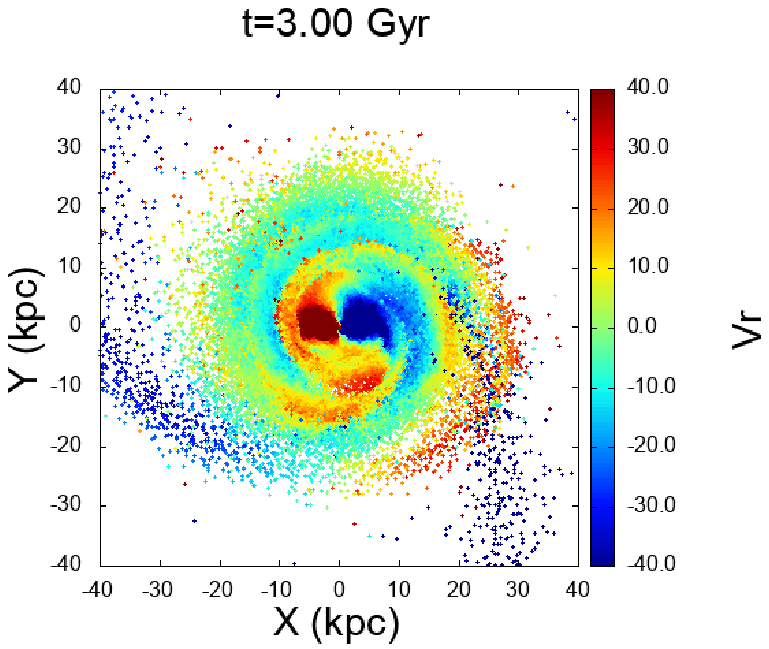}}
{\includegraphics[width = 2in]{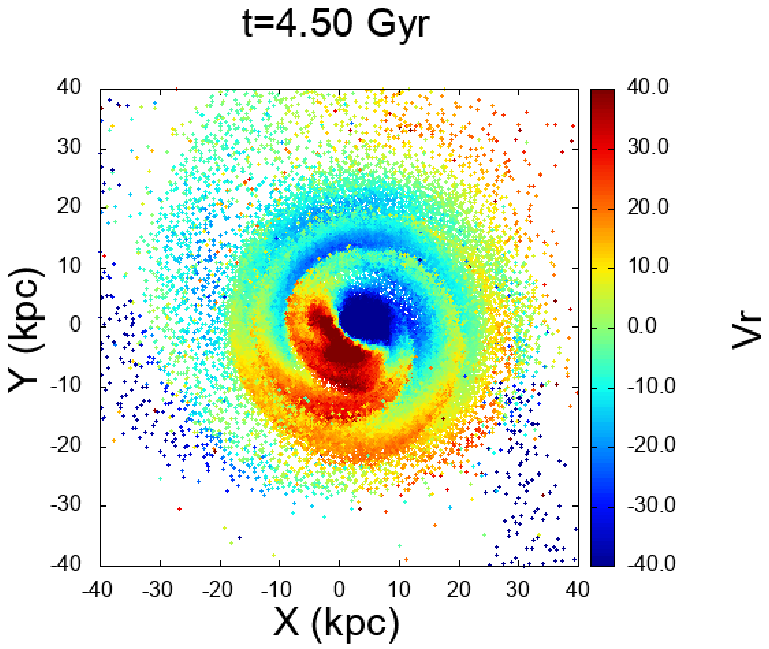}}\\
{\includegraphics[width = 2in]{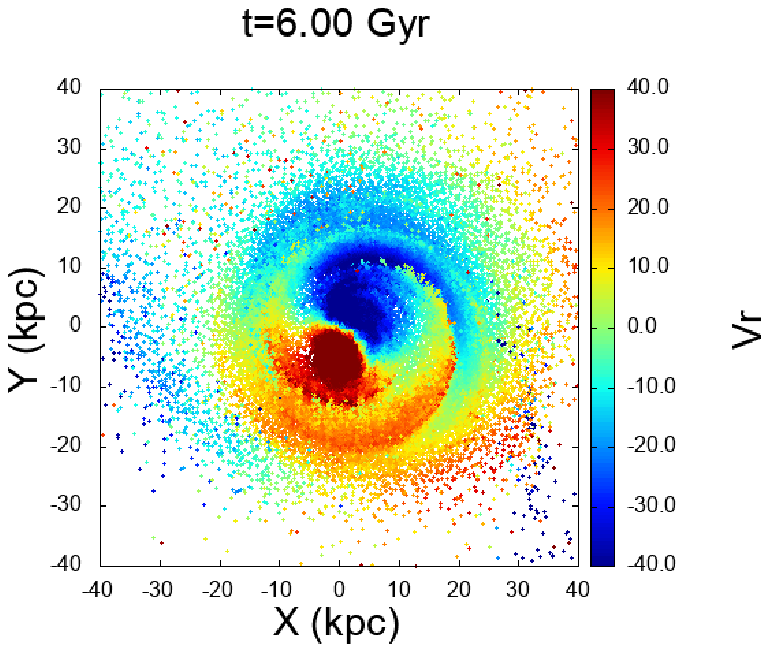}}
{\includegraphics[width = 2in]{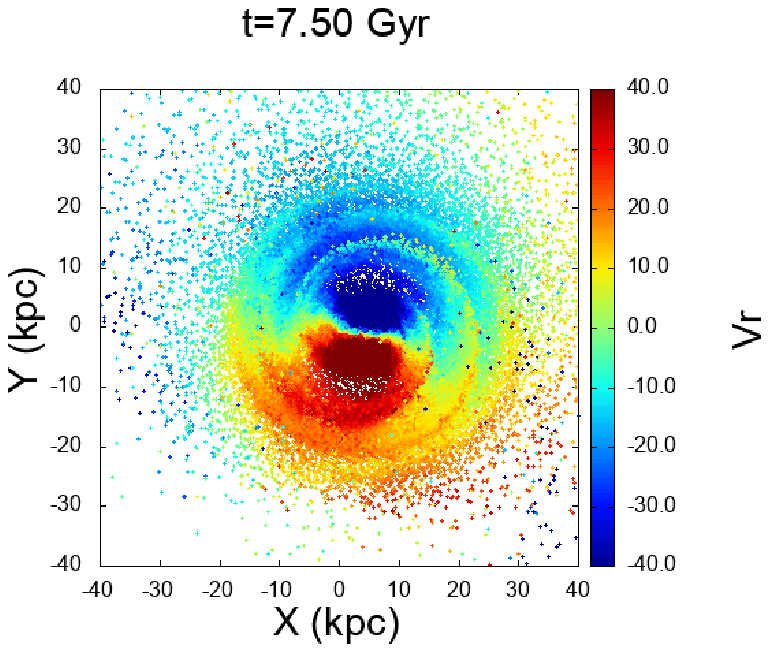}}
{\includegraphics[width = 2in]{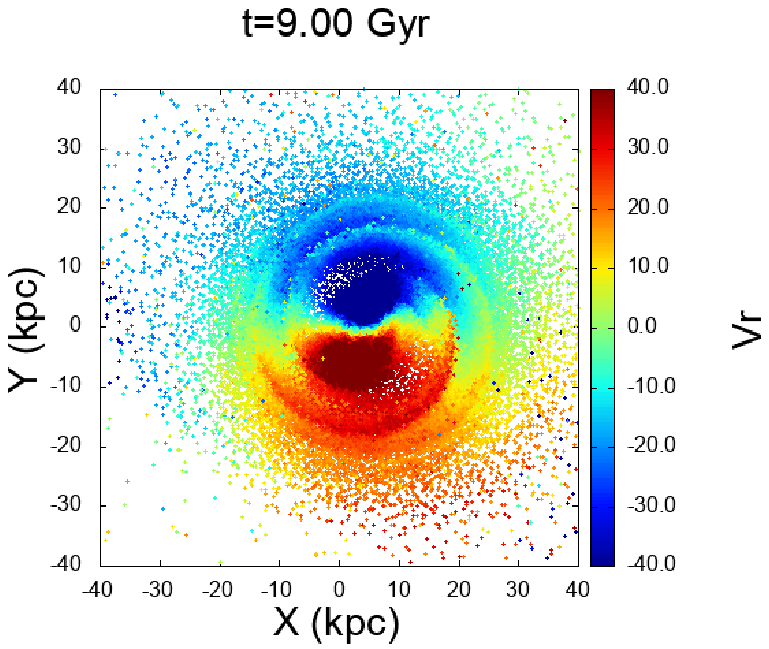}}
\caption{
Projection of various time slices of the GP component on the $XY$ plane: the color code corresponds to the modulus of the radial  velocity. 
The corresponding time is reported at the top of each panel. 
}
\label{GP-VR} 
\end{figure*}

\begin{figure*}

{\includegraphics[width = 2in]{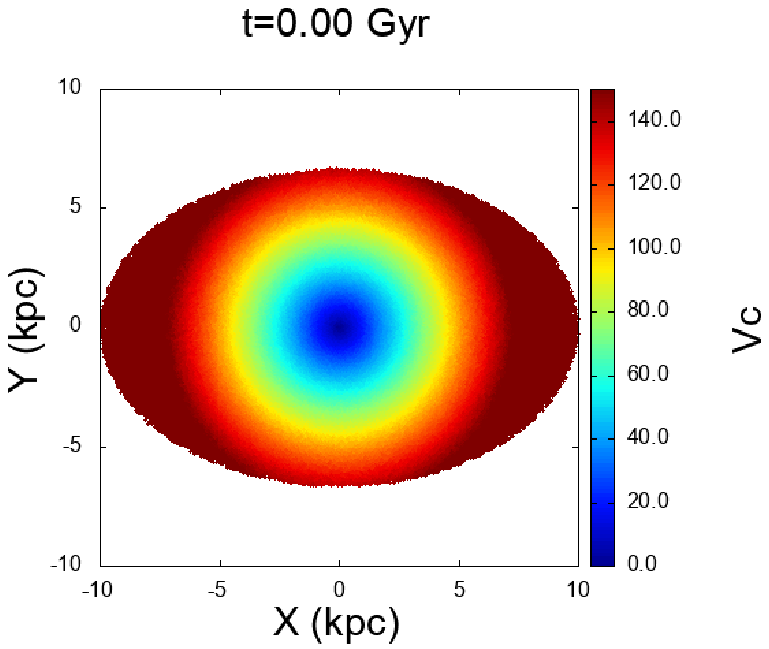}}
{\includegraphics[width = 2in]{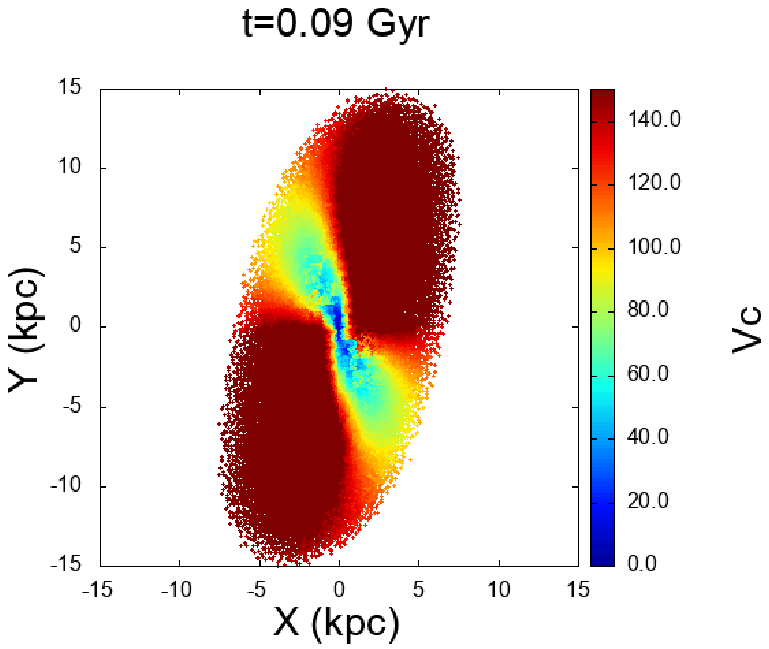}}
{\includegraphics[width = 2in]{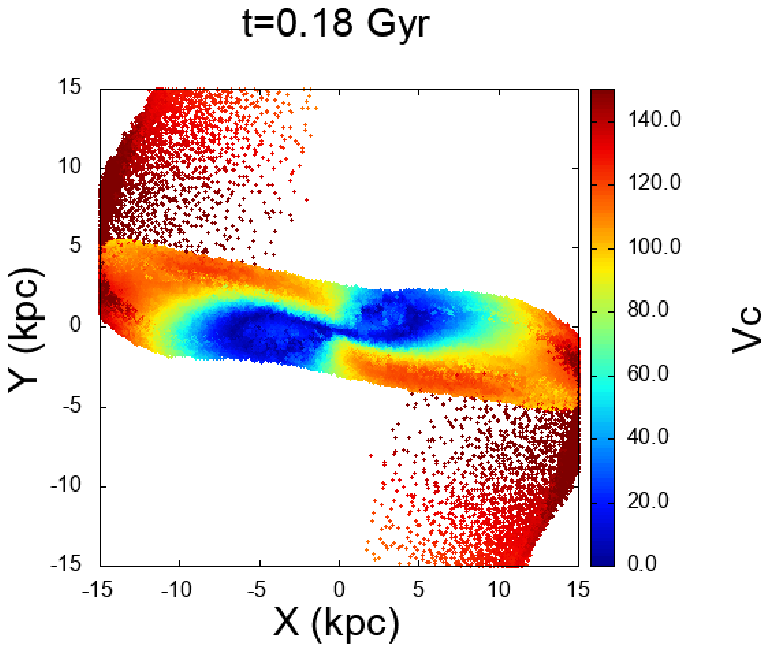}}\\
{\includegraphics[width = 2in]{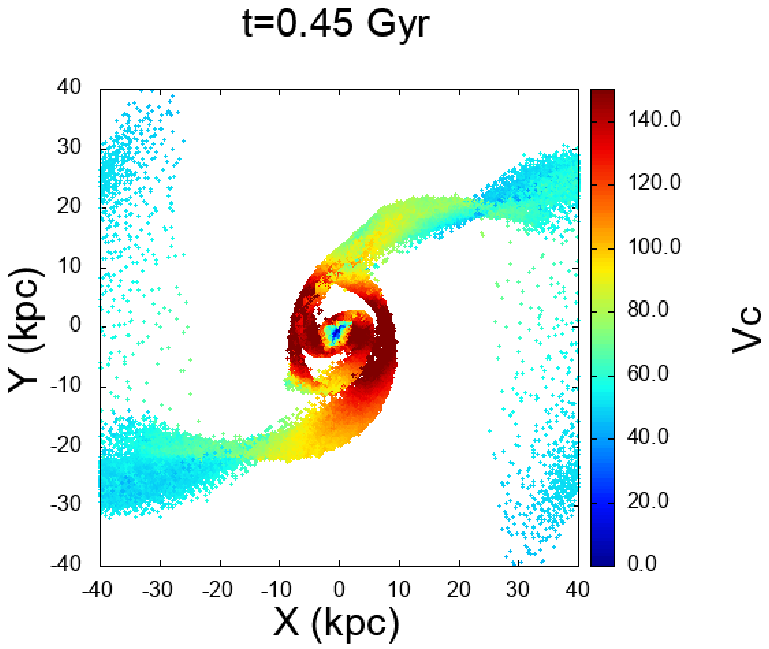}}
{\includegraphics[width = 2in]{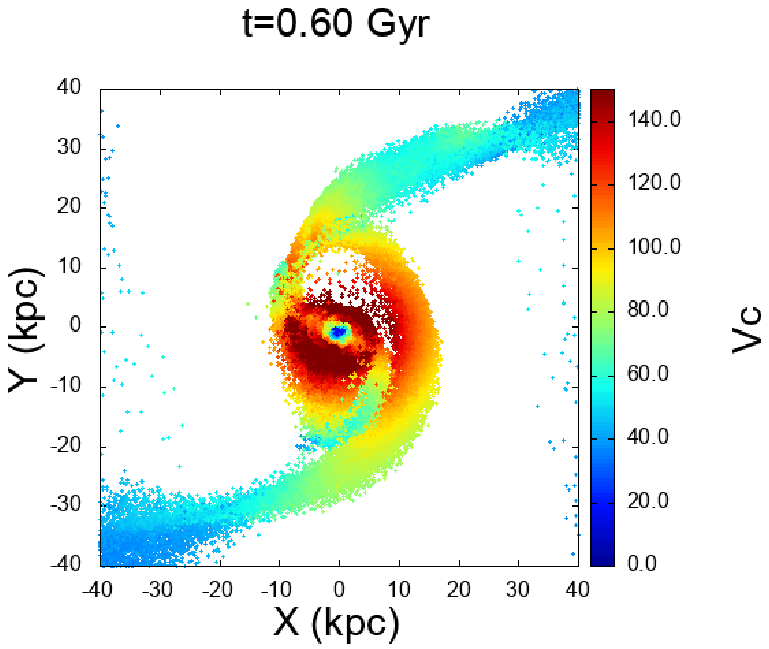}}
{\includegraphics[width = 2in]{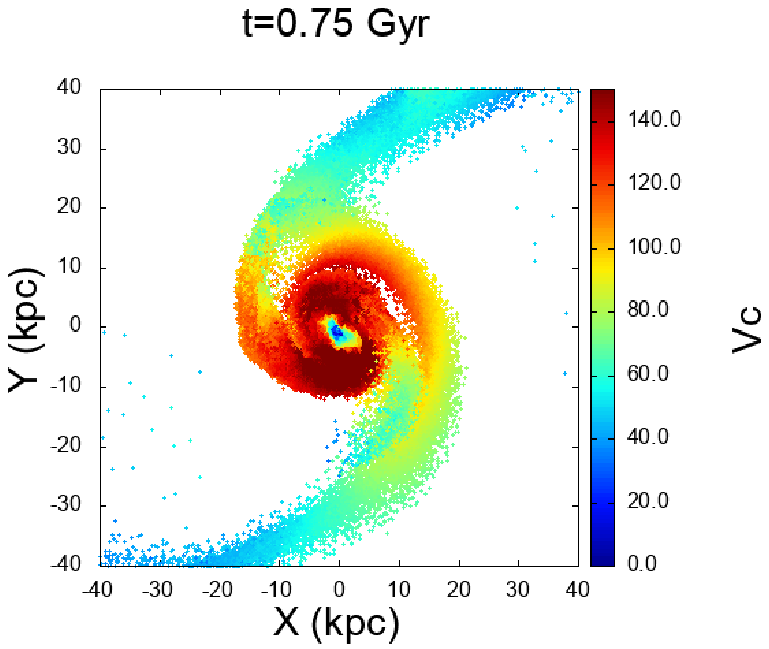}}\\
{\includegraphics[width = 2in]{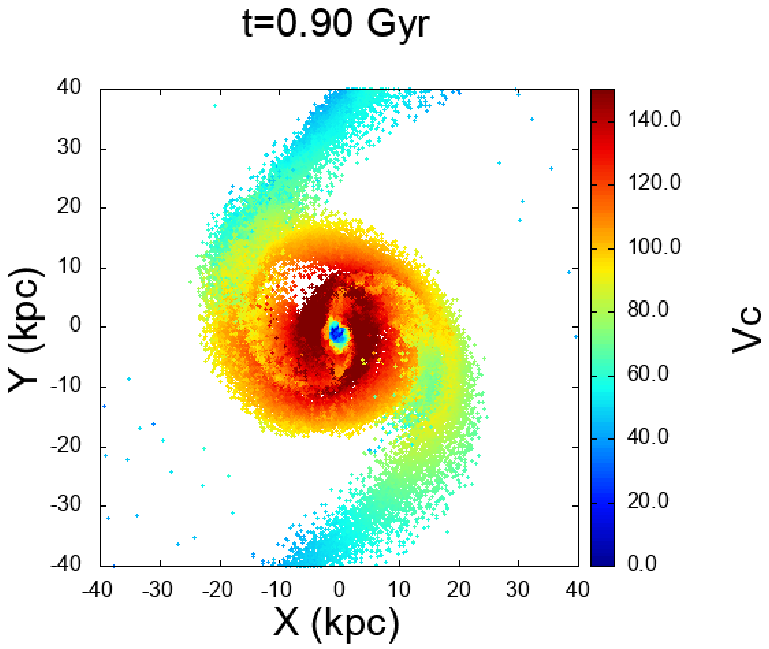}}
{\includegraphics[width = 2in]{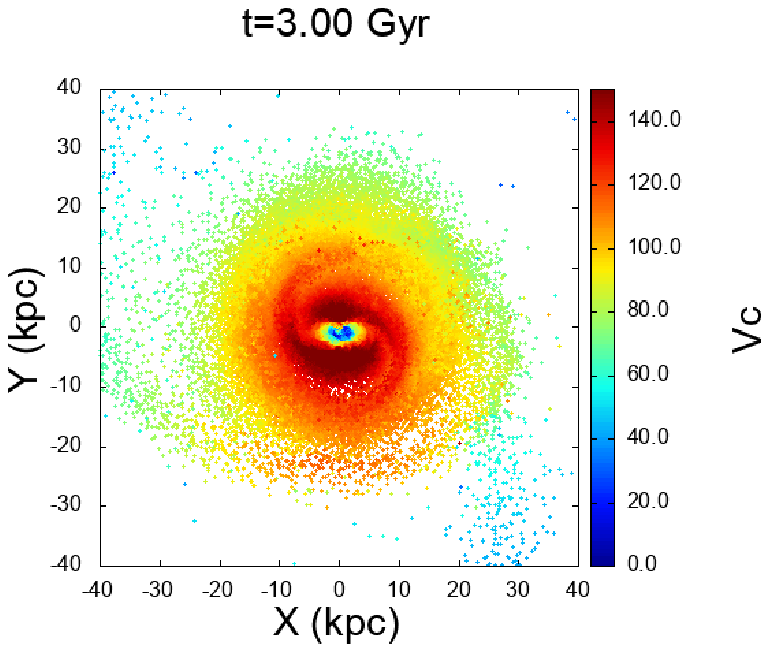}}
{\includegraphics[width = 2in]{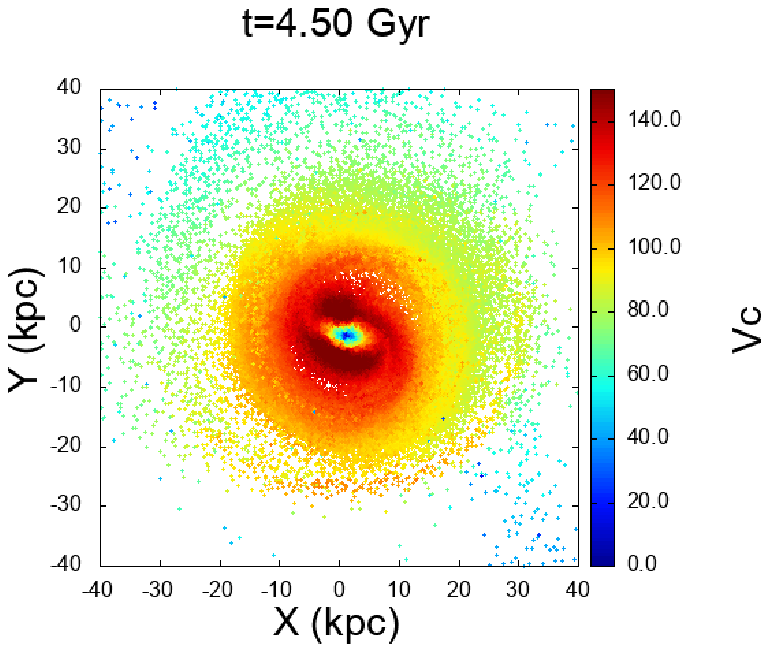}}\\
{\includegraphics[width = 2in]{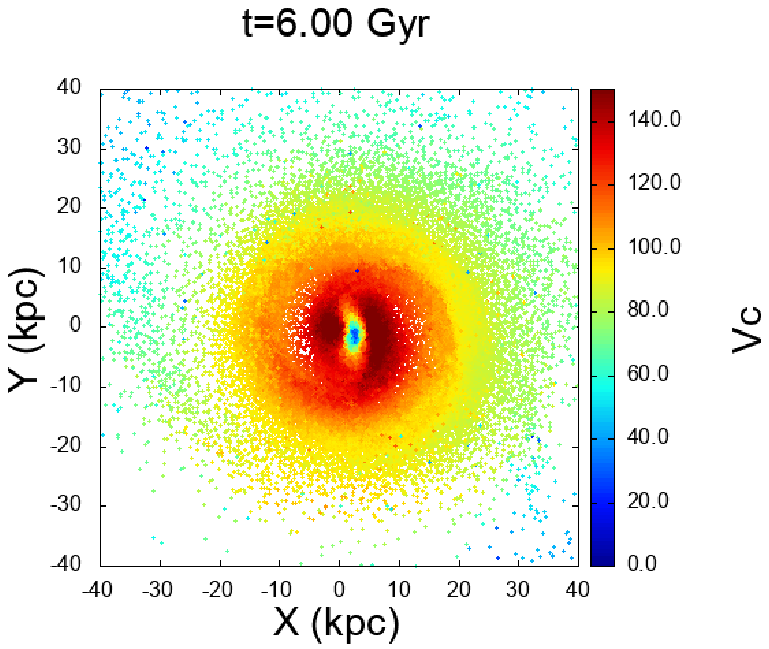}}
{\includegraphics[width = 2in]{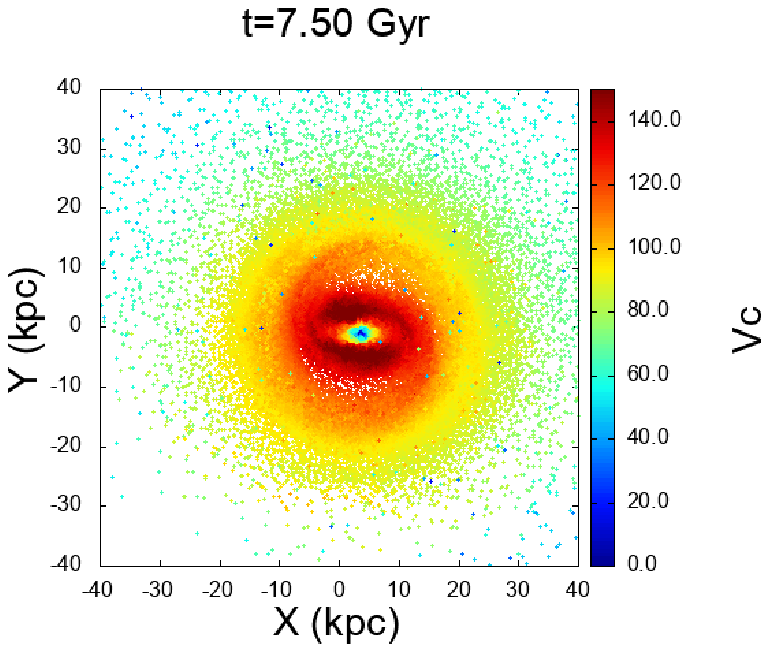}}
{\includegraphics[width = 2in]{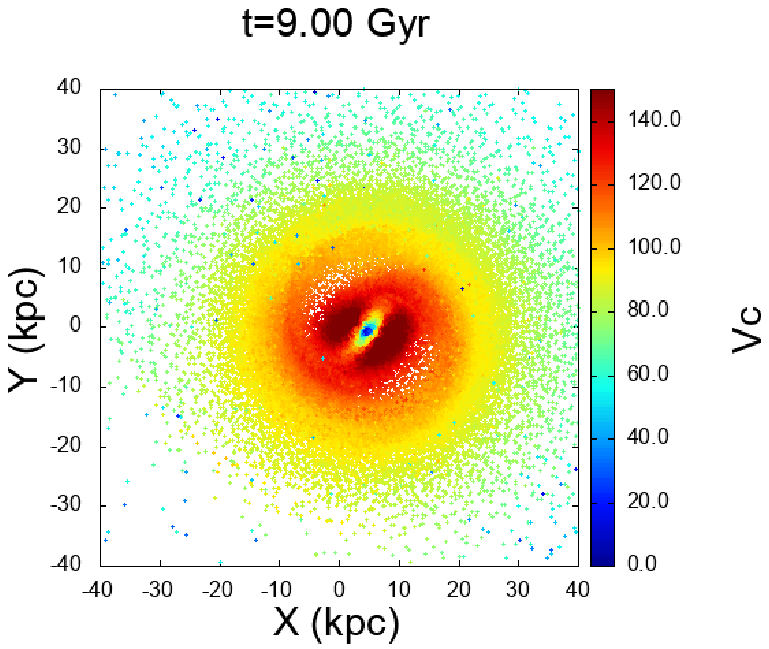}}
\caption{
Projection of various times slices of the GP component on the 
$XY$ plane: the color code corresponds to the modulus of
 the azimuthal  velocity. 
The corresponding time is reported at the top of each panel. 
}
\label{GP-Vphi} 
\end{figure*}

Figures \ref{GP-XY}-\ref{GP-Vphi} show the projection (on the $XY$ plane) of several snapshots, with a color code corresponding respectively to the logarithm of the number density integrated over the $Z$ axis and the radial and the azimuthal velocity component of the gas distribution \footnote{Movies of these runs can be found at the URL: {\tt https://tinyurl.com/rvyg5br}}. 
One may note a rapid initial change of shape and then the relaxation to a QSS, i.e., a phase in which the system inner disk becomes almost stable.
In particular, for $0.1$ Gyr  $< t <0.3$ Gyr the GP forms almost 1D filaments that get later warped forming a sort of spiral arms. 
{Note that the core and
the  spiral structures do not rotate with synchronized rotation
as the  tangential velocity is not constant at different distances. }
Such a disk has major axis $ \sim 14$ kpc and minor axis $\sim 6$ kpc so that its ellipticity is $e \approx 0.4$, i.e., its size is of the same order of magnitude of the initial size of the system.
Instead, the ellipticity is related both to the major to minor axis ratio of the initial system and to the amplitude of the initial angular momentum.
These parameters control how violent the collapse is and thus how large is the particle energy gain as a function of direction.
The smaller the angular momentum, the stronger the collapse and the larger is expected to be the ellipticity of the quasistationary disk formed by the gas.

From the visual inspection of these figures one may note that the compact and elliptical gaseous disk 
is surrounded by a sparser region in which there are long-lived but changing in time (i.e., nonstationary) spiral arms.
There is then, in the outermost regions of the systems 
(not visible in these figures; see below for a discussion), a fraction of particles that is evolving
in an out-of-equilibrium manner.
Let us 
now consider the time evolution of the structures present in these three regions.
We have identified particles belonging to the three mentioned regions 
in a snapshot at $t=6$ Gyr, where they can be easily disentangled, and we have traced backward and forward their evolution at the initial time and at $t=10$ Gyr.


\subsection{Energy and velocity probability distributions}

Both the velocity and energy PDF are rather different in the three regions. 
\begin{figure}
\includegraphics[width = 3.5in]{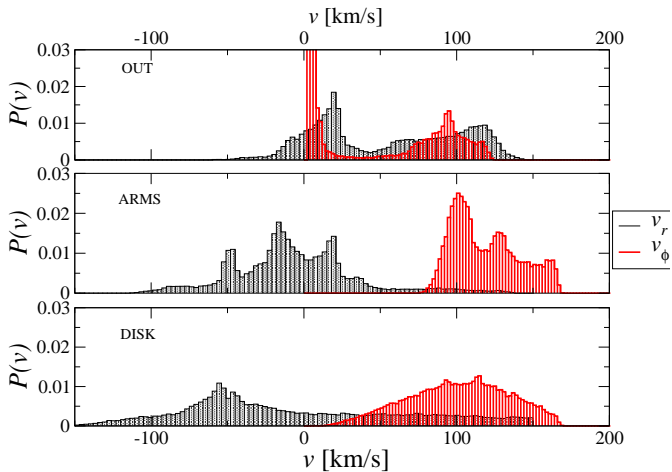}
\caption{PDF at $t=9$ Gyr 
of $v_R$ (black lines) and $v_\phi$ (red lines) respectively in the inner region (the disk, bottom panel), intermediate region 
(the arms, middle panel), and outermost region (upper panel).}
\label{figure_PvrPvt} 
\end{figure}
One may note (see Fig.\ref{figure_PvrPvt}) that particles in the inner region, i.e.,
 the elliptical disk, have a very spread PDF $P(v_R)$ with a variance of $\sigma_{v_R} \sim 50$ km/s.
In the intermediate region $P(v_R)$ is still peaked around zero but with a  smaller dispersion of about $\sigma_{v_R} \sim 10$ km/s. 
In the outermost regions of the system $P(v_R)$ develops a long tail toward large $v_R$ values, which however involves  a small fraction (i.e., $\sim 10\%$) of the gas matter.
Complementary to this tail the PDF of $v_\phi$ develops a peak for $v_\phi \rightarrow 0$ for the reasons we have already discussed above.
The PDF of the azimuthal velocity is peaked at high values of $v_\phi$ in the intermediate region, {  while in }  the inner region particles have smaller azimuthal velocities with a larger dispersion. 

\begin{figure}
\includegraphics[width = 3.5in]{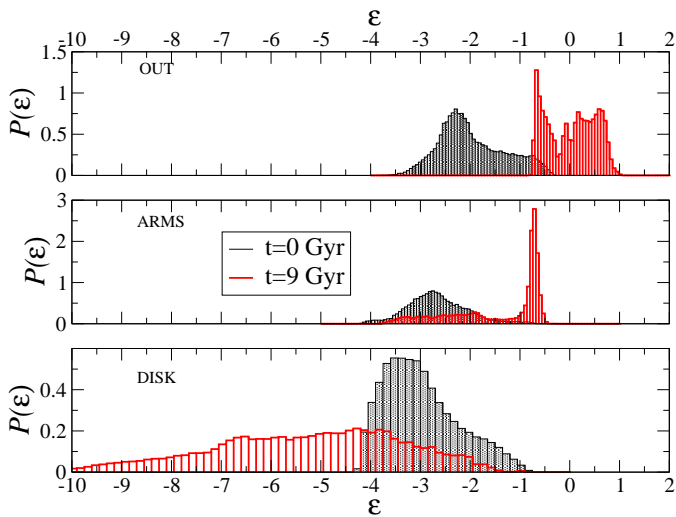}
\caption{
  Particle energy PDF of the 
GPs in the three different regions at $t=0$ and $t=9$ Gyr.
}
\label{figure_Pe-2} 
\end{figure}
By considering the behavior of the energy per unit mass distribution (see Fig.\ref{figure_Pe-2}) in the three different regions we can conclude that: particles  in the inner region (the disk) are strongly bound and have decreased their energy since the initial time and particles in the intermediate region (the arms) have energy close to, but smaller than, zero and have increased their energy from the initial state. 
Finally particles in the outermost regions are those which have mostly increased their energy, and some of them can even escape from the system.
Thus, there is a correlation between the energy gain or loss rate and the distance of the particles from the system center: the origin of such a correlation can be traced back to a particle's initial position.

\begin{figure}
{\includegraphics[width = 3.5in]{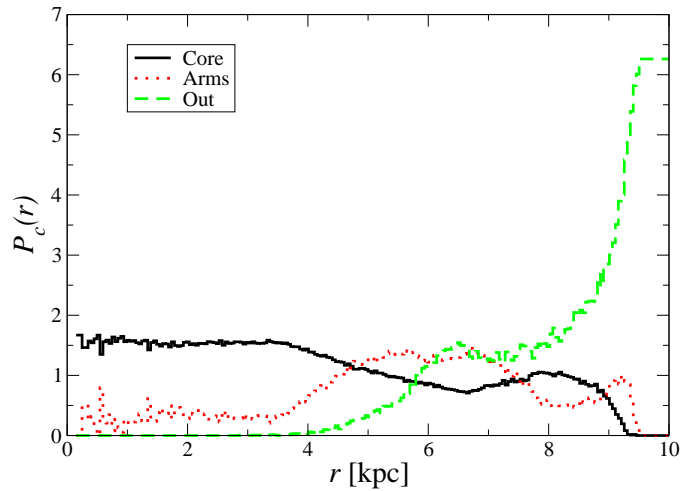}}
\caption{
Conditional probability as a function of the initial distance for a particle to be in the inner, intermediate or outer region of the system at $t=9$ Gyr.}
\label{figure-PR} 
\end{figure}

Indeed, Fig.\ref{figure-PR} shows the conditional probability for a particle of being member of a given group (i.e., inner disk, arms and outermost regions) as a function of its initial position.  
The conditional probability that a randomly chosen particle at distance $r$ in the initial configuration is in the outer region at $t=9$ Gyr is much larger if it was initially in the outermost shells. 
On the other hand if a particle was initially in the inner regions of the system,  the probably that it remains there is larger than for particles initially placed in the outer regions. 
Particles in the arms were initially placed in an intermediate region. 
\subsection{Origin of the spiral arms}

The formation of such a correlation implies that the change of energy and velocity during the collapse is correlated with the particle initial position. %
This implies that groups of particles coming from specific regions of the systems have a similar dynamical history and thus remain correlated during the evolution.
Such a correlation is thus the specific signature of the monolithic collapse. 
The underlying mechanism was outlined above: particles in the outer region of the initial distribution increase their energy because are still collapsing when the others (that will decrease their energy) are already re-expanding. 

\begin{figure*}
{\includegraphics[width = 2.2in]{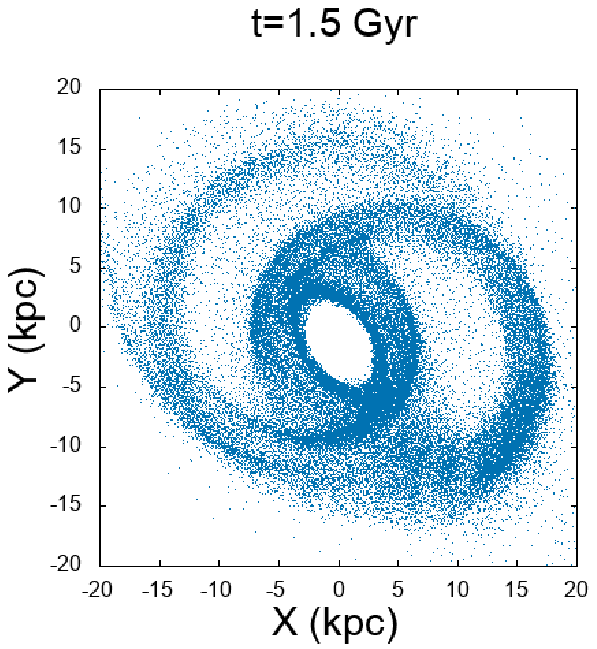}}
{\includegraphics[width = 2.2in]{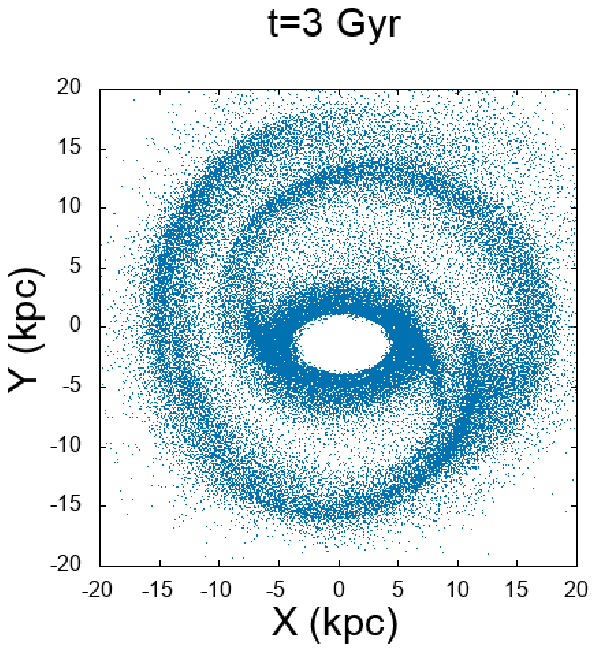}}
{\includegraphics[width = 2.2in]{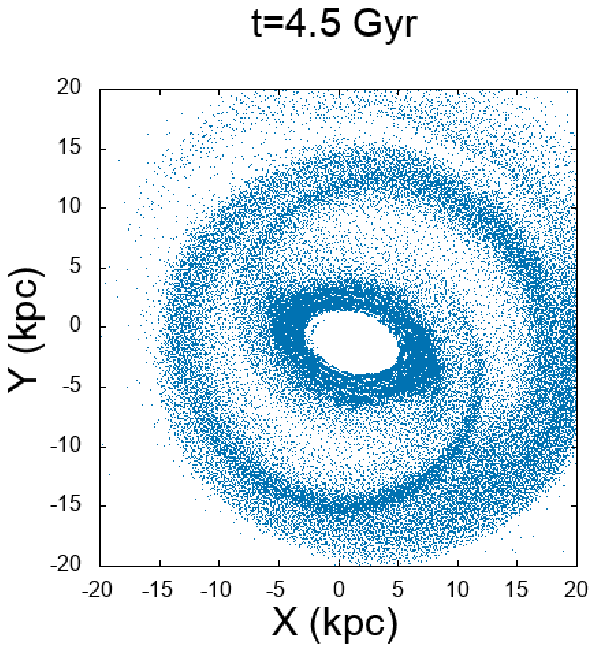}}\\
{\includegraphics[width = 2.2in]{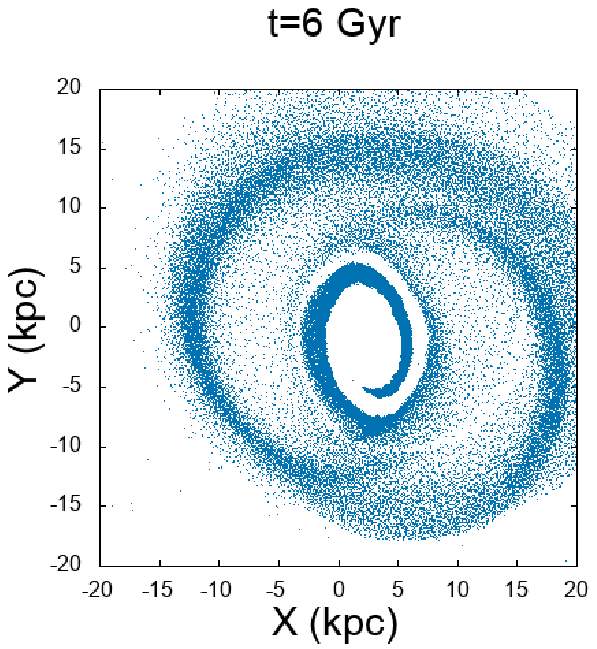}}
{\includegraphics[width = 2.2in]{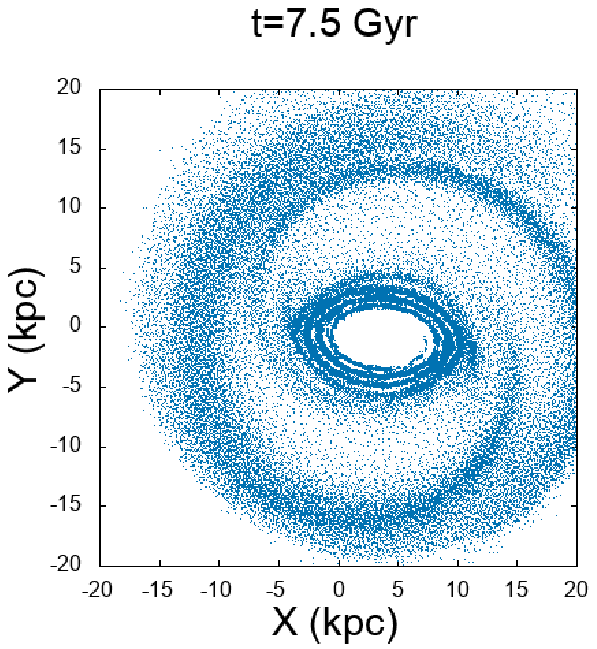}}
{\includegraphics[width = 2.2in]{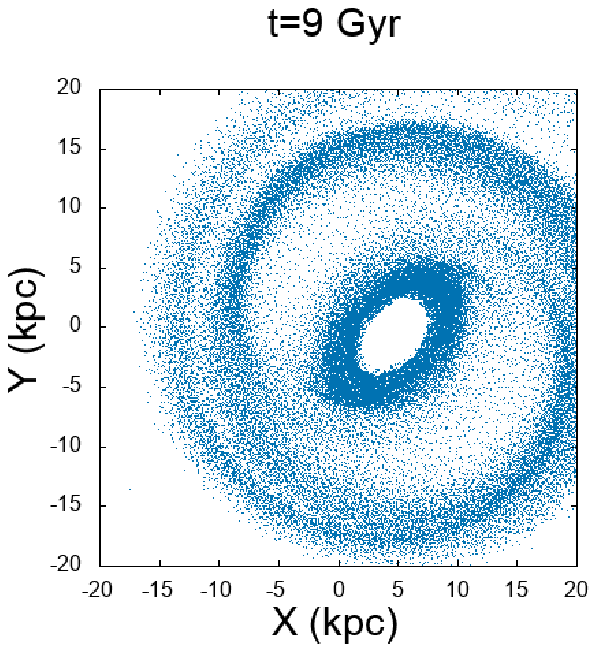}}
\caption{
Evolution of the spiral arms in the $XY$ plane: the particles belonging to the arms have been identified at $t=6$ Gyr and then traced backward or forward in time to $t=0$ Gyr and $t=9$ Gyr.}
\label{figure_arms-evol} 
\end{figure*}

Figure \ref{figure_arms-evol} shows the evolution of spiral arms  in the $XY$ plane. 
{The particles that form the arms are identified } 
 in an evolved snapshot, at $t=6$ Gyr: {the position 
 of these same particles is then tracked back to $t=0$ and forward to $t=10$ Gyr to reconstruct 
 the temporal evolution of the arms.}
One may see that particles move in a coherent way and that the majority of the particles in a given arm remain the same from its formation till the end of the simulation: that is the arms are not density waves as they involve the motions of particles.
The correlation  in both configuration and velocity space is developed during the collapse phase and persists over the whole run. 
On the very long timescales, i.e., $t \gg 10$ Gyr, the arms will be washed out for the effect of the velocity dispersion inside them.

Particles forming the arms were originally in the outer region of the system and thus their energy changes by a positive amount that is smaller than the absolute value of the energy variation of particles in the inner disk (see Fig.\ref{figure_Pe-2}). 
For this reason their radial velocity remains peaked at zero, while that of the inner disk particles  spreads and the PDF is close to uniform.
Given that the azimuthal velocity is larger than the radial velocity, the orbits are 
{  closer to} circular {  ones} than those of the inner disk particles. 
The symmetry of the arms is related to the symmetry of the initial conditions, as the two arms are formed by particles initially lying in a symmetric position along the system's major axis.  
During the evolution, the particles in the arms increase their spread in position and velocity and for this reason the arms are expected to be washed out in the long-run.
However, for what concerns the range of times we have considered (i.e., $\sim 10$ Gyr) the arms remain well formed.

It is interesting to note that 
in the 2D
phase space of 1D gravitational systems,
 spiral
structures are formed as a result of filamentary patterns that appear due
to differential rotation of an incompressible  fluid
\cite{Wright_etal_1982,Yan_Miller_1997,Tsuchiya_Gouda_2000,Joyce+Worrakitpoonpon_2011,Teles_etal_2011}. 
Such spiral
structures  are apparently similar to those observed in the present
work in the projected 2D plane occupied by the cooled compressed gas; 
whether the dynamical origin of these spiral arms is the same the two cases is, however, 
an open question and requires a complete analysis of the 6D phase-space: 
this goes beyond the scope of this work. In what follows
we will discuss only the properties of some projections of the 6D phase-space.

\subsection{Structures in  phase-space}

\begin{figure*}
{\includegraphics[width = 2.2in]{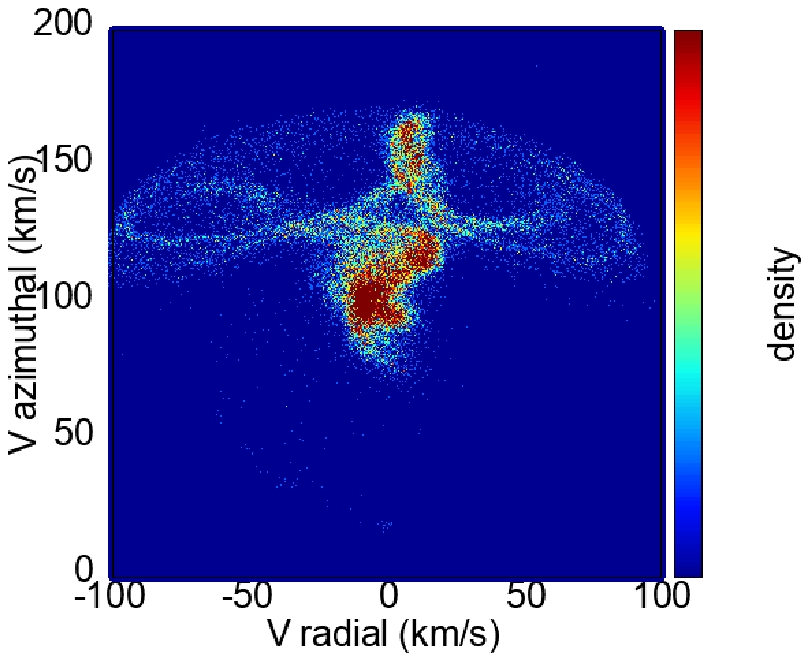}}
{\includegraphics[width = 2.2in]{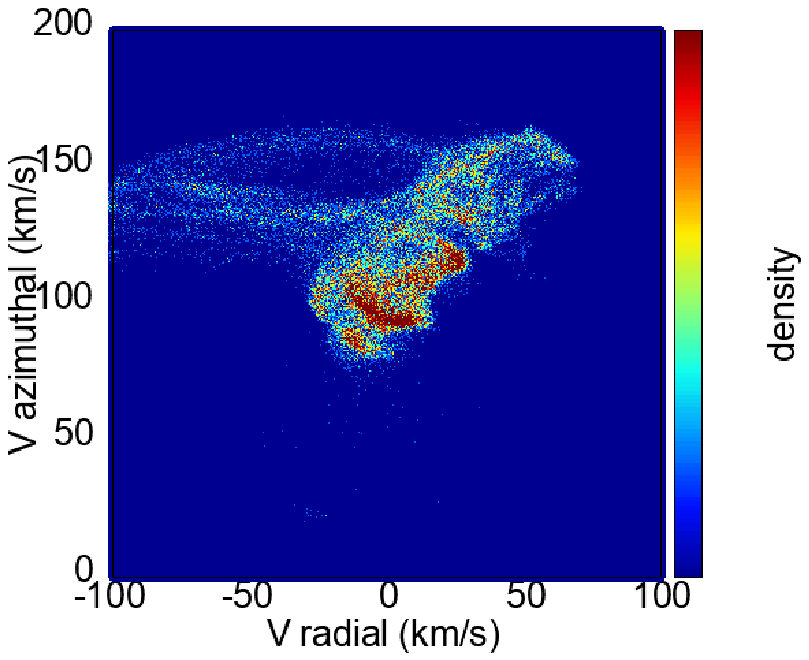}}
{\includegraphics[width = 2.2in]{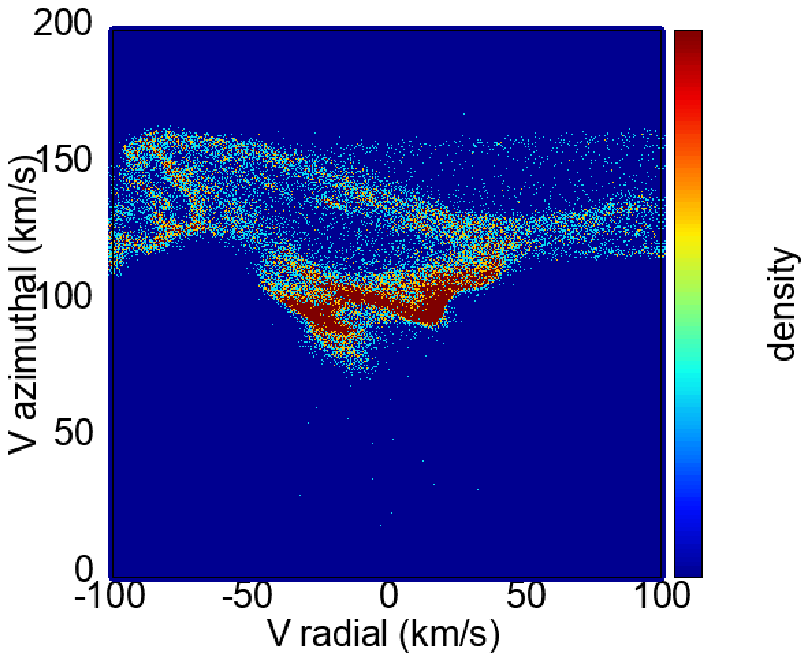}}\\
{\includegraphics[width = 2.2in]{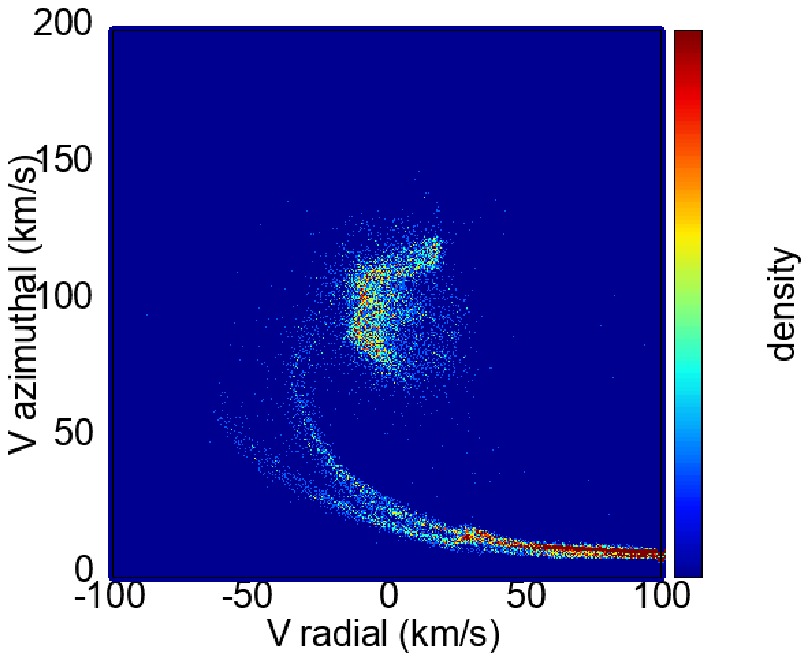}}
{\includegraphics[width = 2.2in]{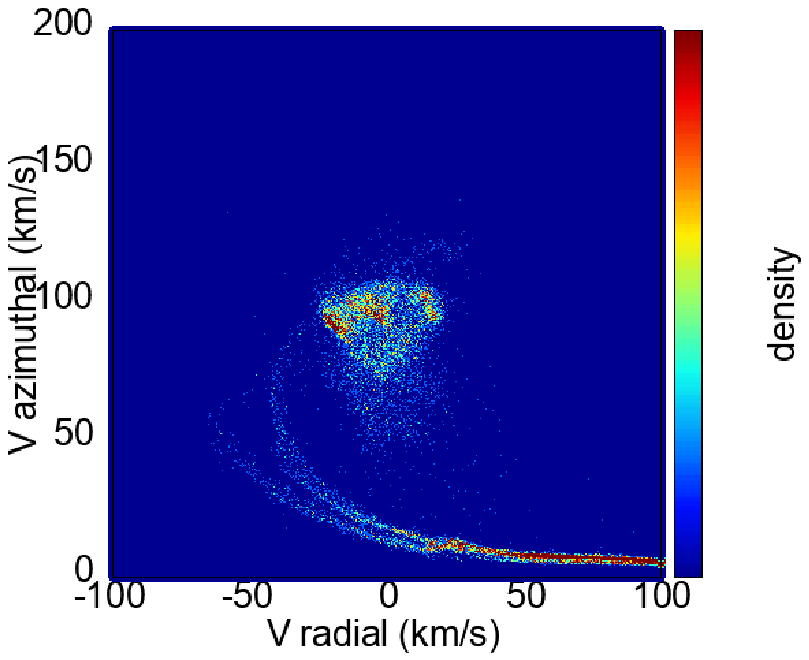}}
{\includegraphics[width = 2.2in]{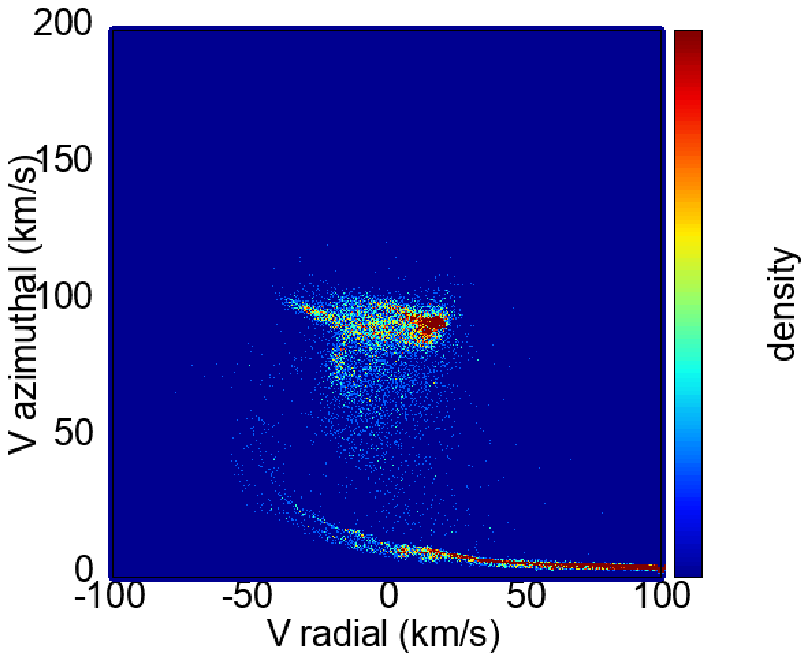}}\\
\caption{
Projection of the phase-space distribution of the GPs into the $v_R-v_\phi$ plane.
The upper row shows the spiral arms and the bottom row the outer regions and different times:
from left to right at $t=3, 4.5, 7.5$ Gyr.
}
\label{figure-phasespace} 
\end{figure*}

The projection of the phase space into the $v_R-v_\phi$ plane (see Fig.\ref{figure-phasespace}) reveals the presence of some nontrivial structures. It also reveals most 
notably that particles belonging to the inner disk have a spread distribution in both velocity components with some correlated structures which reflect the elliptical motions,
and particles in the arms occupying a very localized region in which $v_\phi$ is close to maximum and $v_R$ close to zero: such a region is, however, not symmetrical either with respect to $v_\phi$ or to $v_R$   and there are several sub-structures in it.
Such sub-structures in phase-space correspond to the presence of the real-space structures, i.e., the spiral arms.
Finally particles in the outer regions can be recognized by having a correlated stream corresponding to a decrease of $v_\phi$, when $v_R$ increases, that is, as a consequence of particles' angular momentum conservation. 

\begin{figure*}
{\includegraphics[width = 3in]{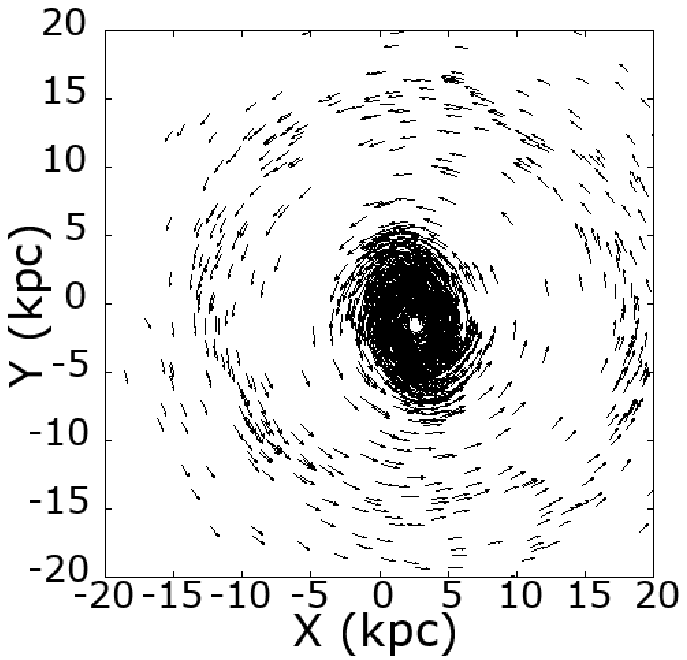}}
{\includegraphics[width = 3in]{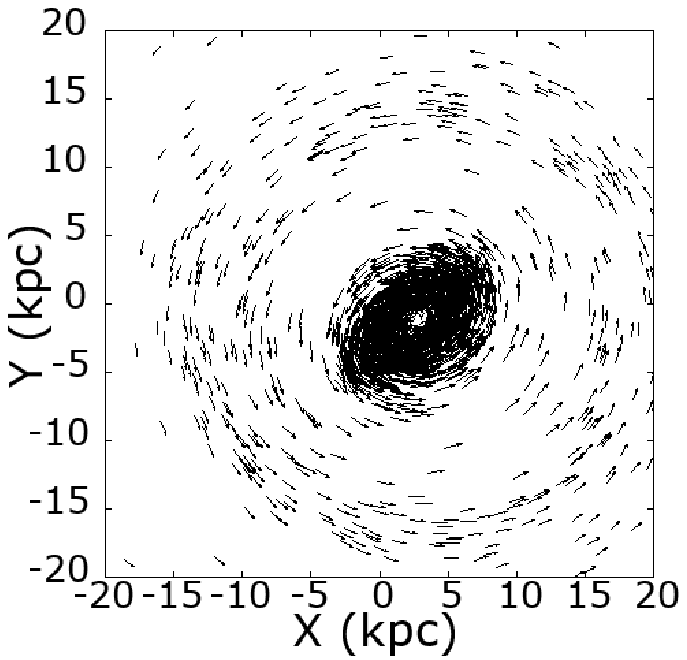}}
\caption{Projection of two snapshots of the GP component on the $XY$ plane at $t=6,6.15$ Gyr: the velocity and its direction for a sub-sample of the system's GPs has been shown with an arrow.}
\label{figure-elliptical-orbits} 
\end{figure*}

Let us now focus on the inner elliptical disk. 
It is not surprising that particles move on elliptical orbits in this region as shown by Fig. \ref{figure-elliptical-orbits}: one can see also that there is an overall precession of the whole disk.
These motions can be easily explained by considering that when a particle moves in an elliptical orbit from the perigee to the apogee it increases its distance from the center and thus it has a positive radial velocity component.
Clearly, the opposite occurs when a particle moves from the perigee to the apogee.  
It should be stressed that GPs move in the gravitational potential of the whole system that is dominated by the PP distribution that, as mentioned above, also form a disk (although with a larger thickness).
These particles are strongly bound and confined in phase space and their velocity 
is  predominantly oriented along the azimuthal direction, but a relatively large radial component was developed in the second phase of the collapse (i.e., $\tau < t < 5 \tau$).
For this reason, after the transient phase, they relax to elliptical orbits. 
%

\begin{figure*}
{\includegraphics[width = 3in]{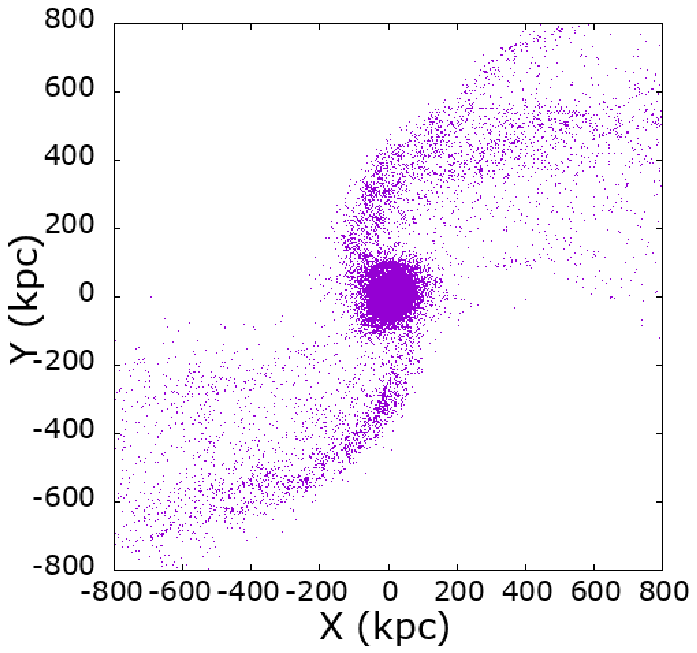}}
{\includegraphics[width = 3in]{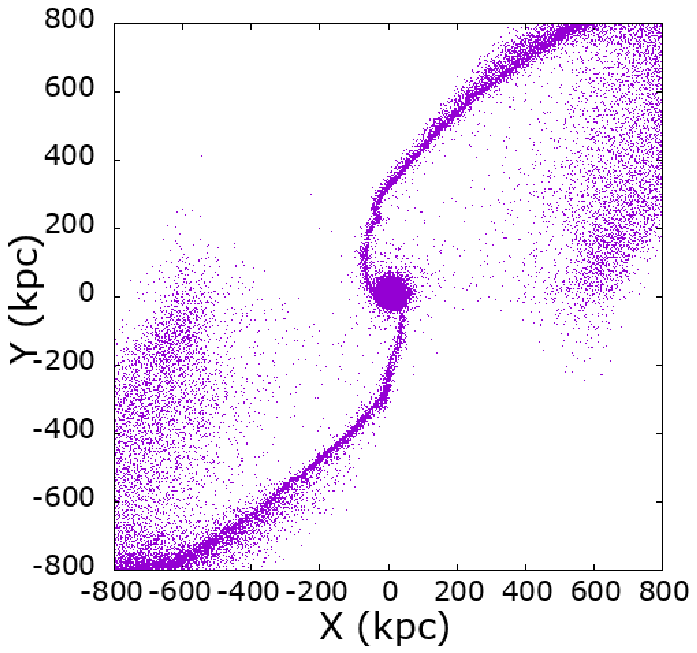}}
\caption{
Projection  of the GPs (left panel) and PPs (right panel) on the $XY$ plane at $t=6$ Gyr.}
\label{XY-LARGEDIST} 
\end{figure*}

Finally Fig. \ref{XY-LARGEDIST} shows the comparison of the GP and PP distributions at large distances from the center of mass of the system.
The spiral arms in the outermost regions are out-of-equilibrium and dominated by radial motions; GP trace the same structures formed by the heavier PP, although the latter component, having a larger velocity dispersion, traces more spread arms than the former one. 


\subsection{Jeans' equation}

The circular velocity is defined to be the velocity with which test particles would move on circular orbits at radius $R$ from the center of a self-gravitating disk with an axisymmetric gravitational potential $\Phi$ \citep{Binney_Tremaine_2008}:
\be
\label{vc1}
v_c^2(R) = \left( R \frac{\partial \Phi}{\partial R}\right)_{z\approx 0}\;.
\ee 
Let us assume that the system is in a steady state and that is axisymmetric so all derivatives with respect to $t$ and $\phi$ vanish.
Under these hypotheses and  by neglecting collisions, one may derive from the collisionless Boltzmann equation, the Jeans equation which links the density and the moments of the velocity distribution to the gravitational potential \citep{Binney_Tremaine_2008}. 
From the Jeans equation (in cylindrical coordinates) 
we can then derive the circular speed, by 
neglecting the cross-term $\langle v_\phi v_R \rangle$ that
represents a negligible correction \citep{Eilers_2019}, obtaining 
\be
\label{vc2}
v_{c,J}^2(R) = 
\langle v_\phi^2 \rangle  -\langle v_R^2 \rangle 
\left(1 
+\frac{\partial \ln(\nu) }{\partial \ln R }
+\frac{\partial \ln \langle v_R^2 \rangle  }{\partial \ln R }
\right)\;,
\ee
where $\nu=\nu(R,z)$ is the density and we have labeled the circular velocity as $v_{c,J}$ to recall that it is derived under the assumptions of the Jeans equation.

We have estimated $v_c$ in Eq.(\ref{vc1}) from the direct gravitational force summation and $v_{c,J}$ in Eq.(\ref{vc2}( by estimating both the radial behavior of the density $\nu$ and of the radial velocity dispersion $\langle v_R^2 \rangle$ 
on the plane (i.e., $z \approx 0$) and by computing numerically their logarithmic derivatives 
(for a more detailed discussion see Ref. \cite{Zofia_etal_2020}).
Results, as expected, show that $v_{c,J} \approx v_c$ in the inner disk while at large distances (i.e., $r>30$ kpc) $v_{c,J} > v_c$: the deviation of $v_{c,J}/v_c$ from unity correlates with the amplitude of the mean radial velocity $\langle v_R \rangle$ that describes the deviation from equilibrium of a system. 

\subsection{Discussion}   
\label{subsec:discussion}

The main result of our simulations is that a quasistationary rotating disk can be formed from the monolithic collapse of an isolated out-of-equilibrium overdensity of self-gravitating matter with a dissipational gas component. 
Around such a disk   long-lived but nonstationary spiral arms are formed 
whose velocity field is dominated by rotational motion but that also show  
large-scale gradients in all velocity components. 
At larger distances the whole system is surrounded by out-of-equilibrium spiral arms. 
The  physical mechanism  that gives rise to such an heterogeneous system is the variation of the system's mean-field potential energy in the short time interval around the global collapse. 
Such a variation amplifies any initial deviation from spherical symmetry and  causes a large change of the particle energy distribution.
{  There are two very different timescales:
(1)  the characteristic collapse timescale $\tau \sim 1/\sqrt{G\rho}$ (where $\rho$ is the initial density); 
and 
(2) the life-time  of the spiral arms $t_{arms}$ that we have shown to be $\gg \tau$. 
Because $\tau \ll t_{arms}$,  
 the formation of this kind of QSS can be compatible with 
astrophysical constraints both at small and high redshifts.
}

As a result of this process the purely nongaseous matter forms a 
thick disk that is dominated by the (azimuthal) rotational motion but in which there is a large velocity dispersion.
Instead, the gas forms a thin disk, almost 2D with a 
vertical height scale far shorter than the horizontal size, 
 where there are coherent rotational motions and a small 
velocity dispersion. 
Such a difference in the density configuration 
evolution between the nonvolatile matter and the gas occurs because when the system approaches 
its maximum contraction the latter component 
increases its density and thus can radiatively cool and sink to the nongaseous clumps. 
Since the gas is subjected to compression shock and radiative 
cooling with consequent kinetic and thermal energy dissipation, it develops a much flatter disk, where rotational 
motions are coherent and the velocity dispersion is smaller than that of the nongaseous matter. 
The quasistationary thin gaseous disk is 
thus embedded in the gravitational field of 
the thicker nongaseous disk that dominates the system mass and thus the potential energy.
The thin disk is in general elliptical, where the eccentricity depends on the 
violence of the collapse, i.e., on how much the system's gravitational radius has contracted, and thus on the shape of the ICs and on the initial angular momentum.

By analyzing the evolution of the spatial distribution 
of the SPH particles used to represent the gas, we have found that the they form long-lived but nonstationary spiral arms.
Such structures are formed by particles that have undergone to a similar dynamical history and that, consequently, remain correlated in both position and velocity. 
Their energy is larger than that possessed by the other particles forming the inner disk, even though they are still bound to the system.
They show a rough velocity field in which both radial and rotational motions are time-dependent and correlated.
The spiral arms are nonstationary mainly due to this latter characteristic.
On the other hand, the 
{  long-lived spiral nature of the spiral arms } arise 
from the correlation in phase space 
that the particles develop during the gravitational collapse. 
Finally both GPs and PPs form far-out-of equilibrium spiral arms in the very outermost regions of the system, where the velocity field is dominated by radial motions.
        

We have  discussed in detail the time evolution of a single numerical experiment in which the ICs were constituted by a prolate ellipsoid. 
We have, however, performed many other simulations and here we summarize our main results.
\begin{enumerate}
\item 
Systems with two-spiral arms,  in both the gas and nongaseous components, are formed provided that the starting configuration breaks spherical symmetry in the $XY$ plane as, for instance, the case of a prolate ellipsoid. 
Starting with an initial system consisting in an oblate ellipsoid, a multiple-arms system with a ring structure on the $XY$ plane can be formed, whose velocity field is characterized by a combination of rotational and radial motions.
When the initial system presents a configuration  more irregular than a simple ellipsoid it evolves acquiring more complex shapes:
however, as long as there is a major axis that is enough larger than the others, a two-arms spiral structure is formed.
In this regard, we performed some numerical experiments introducing some randomness in the shape of the ICs by considering uniform but irregular shapes.
Results do not change qualitatively but they are quantitatively different: for instance multiple-arm systems may arise and/or the systems can be characterized by subclumps that first collapse independently and then merge (some examples for systems constituted by only PP are reported in Ref. \cite{Benhaiem+SylosLabini+Joyce_2019}). 
It should be noticed that in various simulations  bars of different size-scale may be formed.
If the bar is as large as the system itself, then it presents a transient structure in which radial and rotational motions are of the same order. 
Instead, in some cases it can be formed a small bar in the inner regions that can typically survive for many dynamical times.
A more detailed characterization of these structures will be presented in a forthcoming work. 

\item 
If we give to the ICs some random motions in addition to solid-body rotational velocity,
the evolution notably 
changes only when the kinetic energy associated to 
the former becomes of the order of the kinetic energy associated to the latter. 
Figure \ref{Krandom} shows the comparison of the same 
initial configuration and initial virial ratio, but 
for one case the kinetic energy is fully rotational 
and for the other case the kinetic energy is half rotational and half random.
One may see that in the former case the phase-space 
correlation is broken, i.e., the projection of the
phase-space distribution in the $v_R-v_\phi$ plane
shows a much less structured shape when the random velocity is larger.
Correspondingly the spiral arms are washed out and 
the gas forms a structure-less disk.
\begin{figure*}
{\includegraphics[width = 3in]{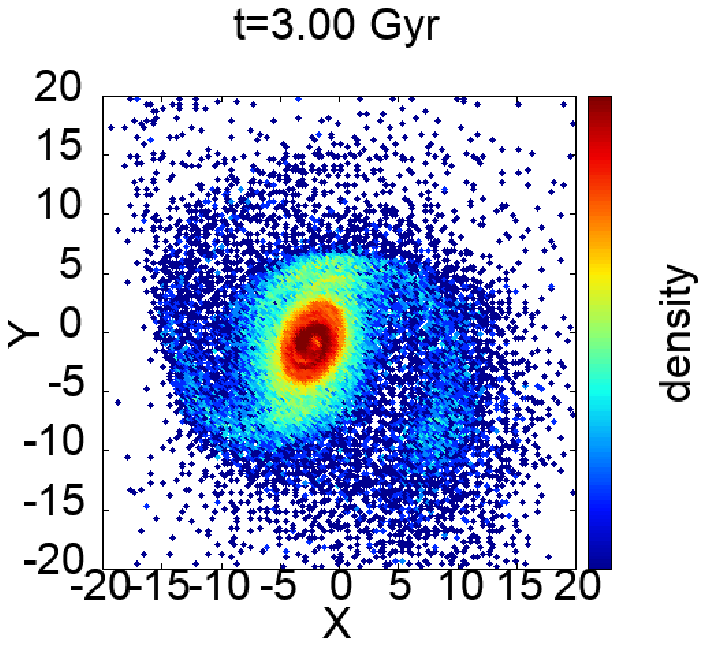}}
{\includegraphics[width = 3in]{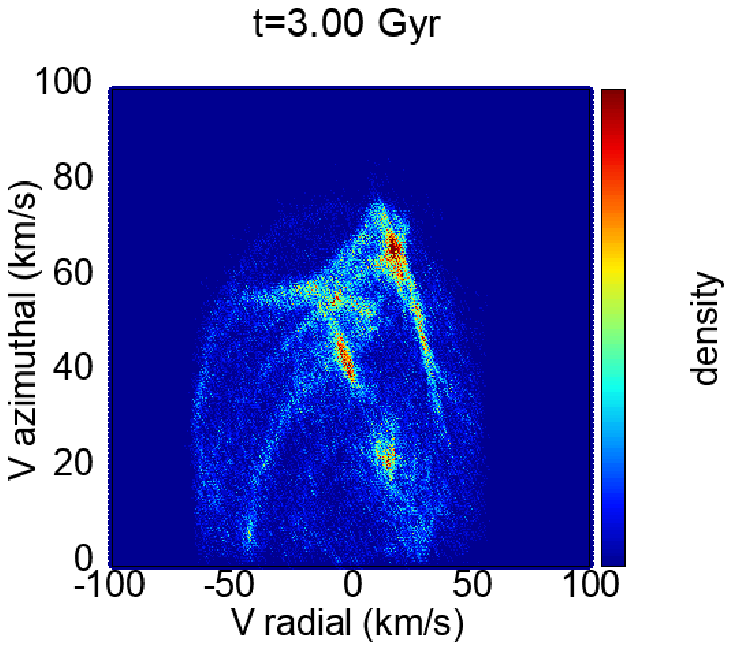}}\\
{\includegraphics[width = 3in]{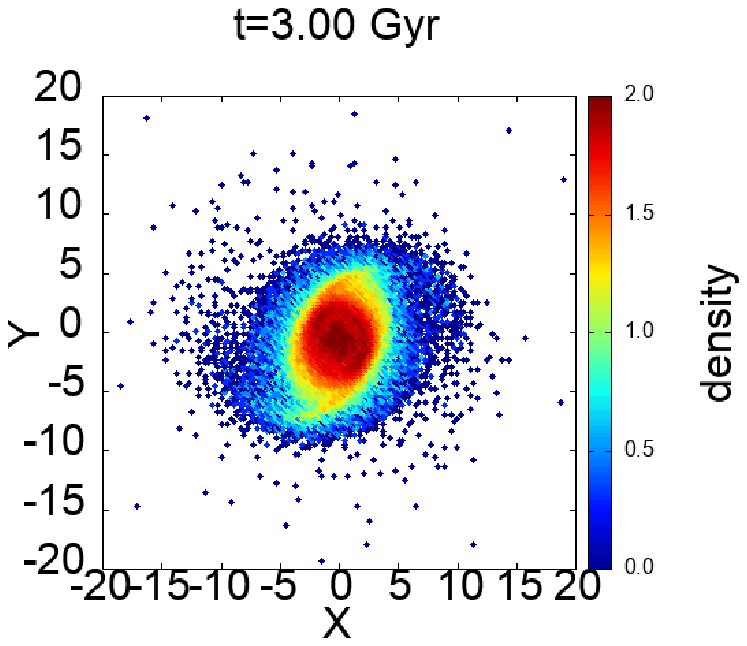}}
{\includegraphics[width = 3in]{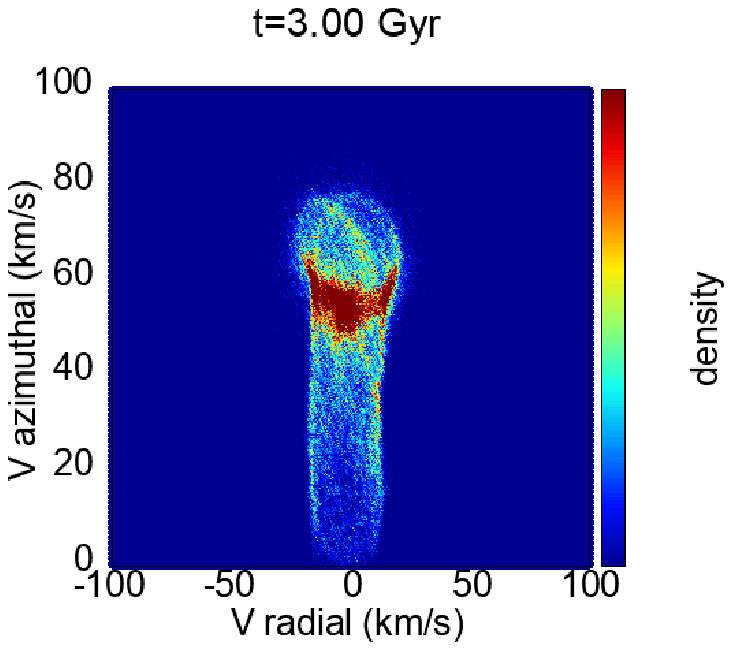}}
\caption{
Comparison of the same initial configuration and initial virial ratio but for one case the kinetic energy is fully rotational (upper panels) and for the other case the kinetic energy is half rotational and half random (bottom panels). 
In each row, the left panel shows the projection on the $XY$ plane and the right panel the phase-space distribution  projected on the $v_R-v_\phi$ plane at the same time (i.e., $t=3$ Gyr $\approx 10 \tau$).
}
\label{Krandom} 
\end{figure*}

\item 
If the initial ratio $Q$ tends to $1$ then the system is initially close to a stationary situation. 
In this condition the self-gravitating particles do not undergo  a strong collapse and the gravitational radius of the system remains close to its initial value.
Indeed, the system gently changes its phase-space configuration to reach a 
quasiequilibrium 
state and its mean field only slightly varies, so that the main source for the large changes of the phase-space properties is not active. 
In these conditions the system is closer to spherical symmetry but undergoes a small contraction around the rotation axis (see the upper panels of Fig.\ref{Kvirial1}): the velocity dispersion of the PP component remains almost isotropic.
On the other hand, the GP component, because of energy dissipation, forms a thin disk that is however structure-less as the mechanism originating structures like spiral arms is not active,  the variation of the system's mean field being too small (see the bottom panels of Fig.\ref{Kvirial1}).
Figure \ref{Kvirial2} shows the phase-space distribution projected on the $v_R-v_\phi$ plane: not surprisingly, this is structureless.
\begin{figure*}
{\includegraphics[width = 3in]{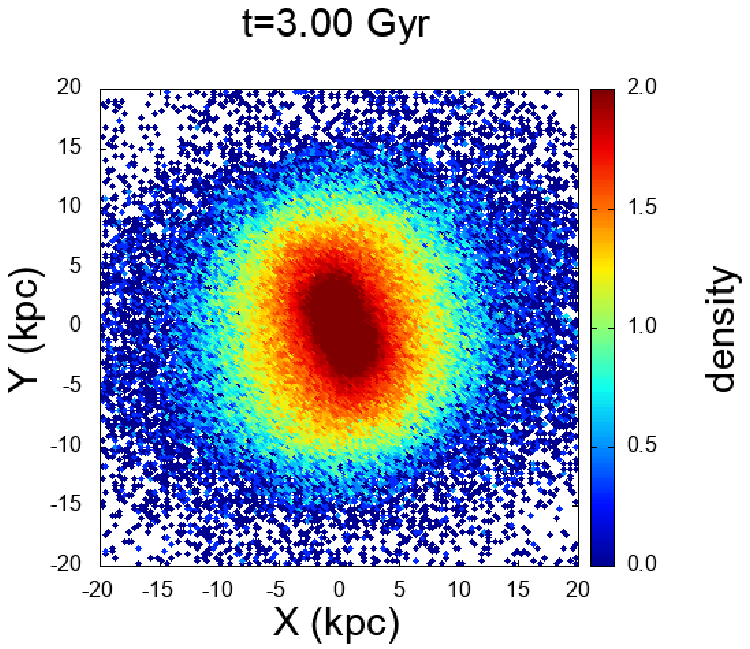}}
{\includegraphics[width = 3in]{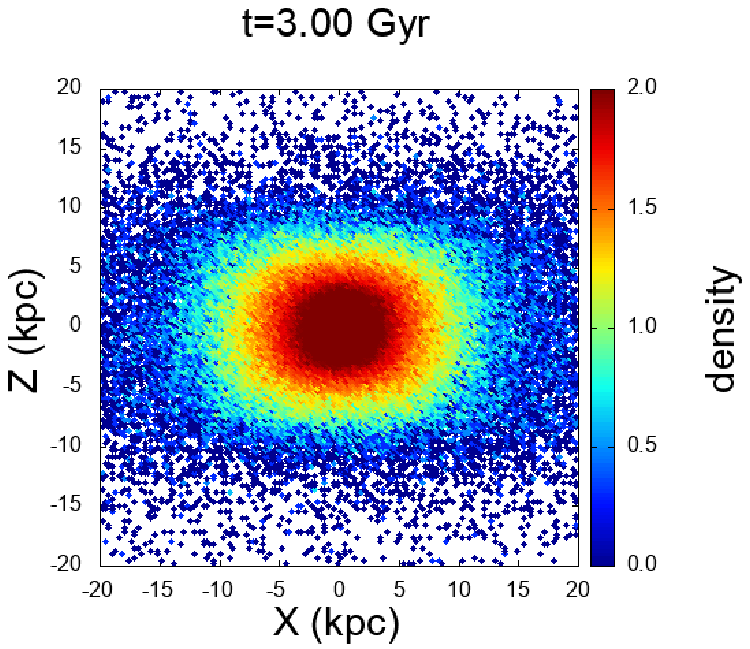}}\\
{\includegraphics[width = 3in]{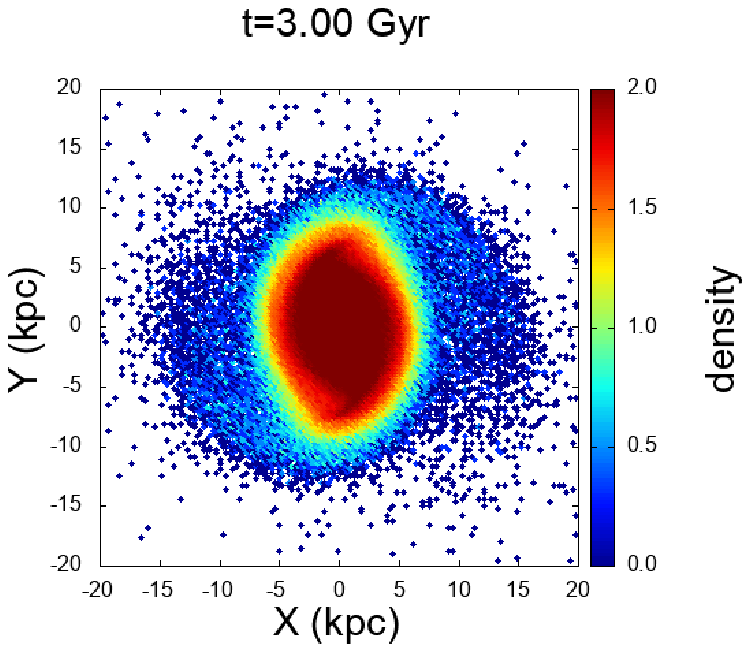}}
{\includegraphics[width = 3in]{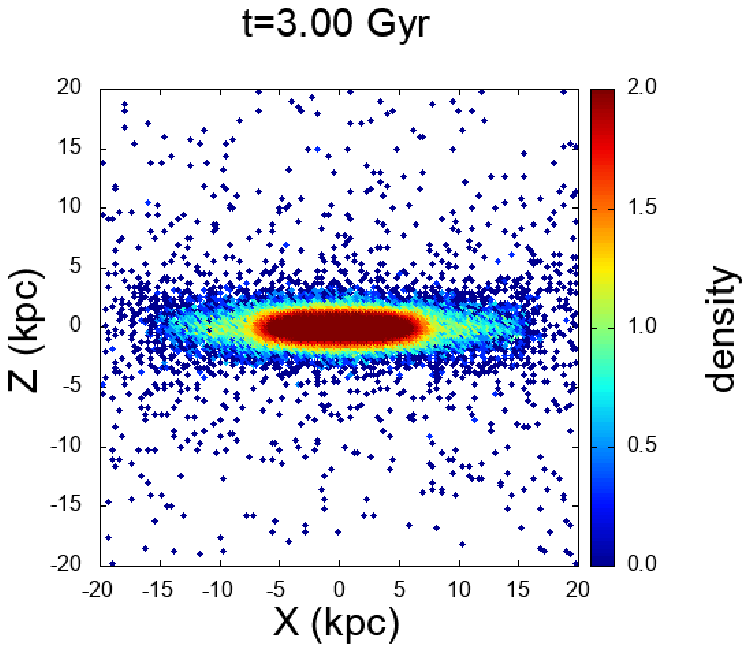}}
\caption{
In this case the initial condition is again an ellipsoid but with a configuration
close to quasistationary state, i.e., $Q=1$. 
Upper panels: projection of the self-gravitating particles in the $XY$ plane (left) and $XZ$ plane (right) at $t=3$ Gyr $\approx 10 \tau$.
Bottom panels: same for the gas.}
\label{Kvirial1} 
\end{figure*}

\begin{figure}
{\includegraphics[width = 3in]{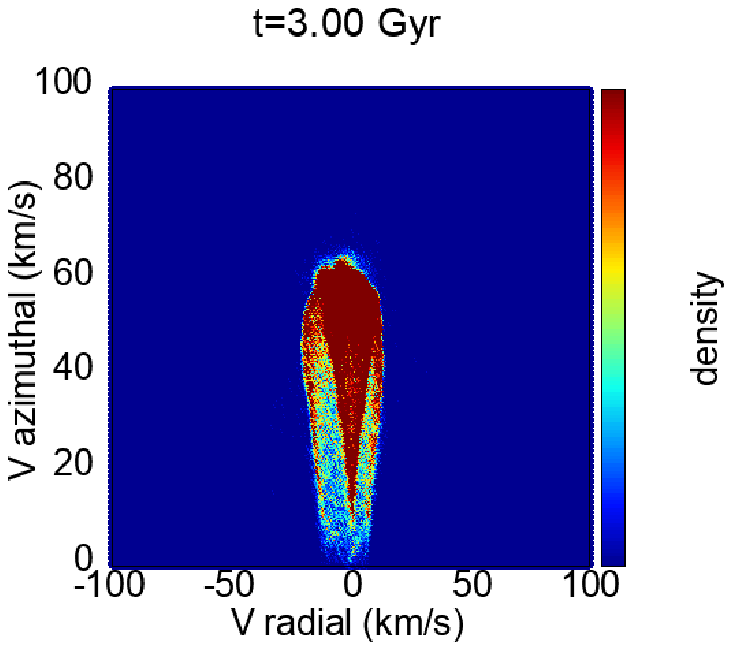}}
\caption{
Phase-space distribution 
of the gas for the simulation shown in Fig.\ref{Kvirial1} 
projected on the $v_R-v_\phi$ plane at $t=3$ Gyr $\approx 10 \tau$.}
\label{Kvirial2} 
\end{figure}

\item 
If the initial temperature of the gas is too high so that the internal energy $u$ becomes of the order of the particles potential energy per unit mass (see Eq.(\ref{u_to_e0})) then the gas diffuses without clustering. 
On the other hand, when the temperature is lower than $\sim 10^4$ K the gas behaves almost like the nonvolatile component, since the cooling is not efficient in irradiating away the thermal energy.
In this regard, w.e have performed some simulations with solely self-gravitating gas dynamics (i.e., no gaseous matter). 
In such a situation, for a typical temperature of the gas of $T \sim 4 \times 10^4$ K, i.e., such that cooling is very effective, the formation of spiral arms is inhibited and the system becomes isotropic even when the initial density distribution assumes an ellipsoidal shape. 
These tests show that the nongaseous matter plays the key role in the determination of the dynamical evolution of the system.
\end{enumerate}

%


\section{Conclusions}   
\label{sec:concl}

Understanding the origin and evolution of spiral structures has proved to be one of the harder problems in astrophysics. 
In this work we have studied 
{  the  formation} of long-lived, but nonstationary, spiral arms as 
a consequence of the rapid and violent collapse of an isolated system.
{  This physical mechanism is different from}
the slow and soft dynamical evolution that 
{  takes place} when a bottom-up aggregation process is at work or when a disk at equilibrium is softly {  perturbed.}
The key physical mechanism is the rapid variation of the system size during the collapse, corresponding to a large change of the mean-field potential which, in turn, causes a substantial variation of the 
particle energy distribution and thus of the system phase-space properties.

Typical cosmological scenarios of structure formations, like CDM models, 
assume that density fluctuations are strongly correlated and give rise to a
soft and slow bottom-up clustering mechanism: for this reason the formation
of a disk and spiral arms, via a rapid and violent mechanism, 
investigated in this work has been commonly overlooked in the literature.
Halo structures formed in CDM models are almost spherically
symmetric and present a quasi-isotropic velocity dispersion,
while disks are characterized by close to rotational and very quiet velocity fields.
On the other hand, quasistationary disks formed from a fast 
and violent dynamical mechanism are characterized by having 
large streaming motions in all velocity components.
In standard CDM type models  the distribution of gas and 
nongas matter are completely different, 
while in the present case they are correlated.

{  It is interesting to note that the paradigmatic model for nonlinear structure formation in cosmology, 
i.e., the spherical collapse model (SCM), says that an overdensity that detaches from the cosmological expansions
starts behaving like an isolated system when its density contrast is of order unity \citep{peebles_1983}. This purely
gravitational system thus undergoes a monolithic collapse of the type 
we have discussed in this work. 
In CDM scenarios, instead, because density fluctuations are too strongly correlated, 
clustering proceeds through a bottom-up hierarchical aggregation mechanism.  
Initial fluctuations must be highly suppressed below some scale as occurs in warm dark-matter-type scenarios, 
for a SCM-type scenario taking place in a cosmological setting. It is worth noticing that  
Peebles \cite{Peebles_2020} 
 has recently advocated precisely a monolithic top-down scenario for galaxy formation
 to overcome the difficulties of standard CDM-like models in explaining the main observations concerning galaxies. 
 In particular, Peebles 
 has proposed a warm dark matter 
 initial mass fluctuations power spectrum that has a sharp cut-off suppressing fluctuations at small enough scales. 
 Such scales are not probed by the
cosmic microwave background radiation, 
and thus do not have to satisfy strong observational constraints, 
but may be significant for galaxy
formation. 
}

{  Numerical experiments in which 
the initial conditions represent an isolated overdensity of massive particles with a 
dissipative gas component
represent a suitable playground to explore the combined effects  of gravitational
and gas dynamics in a system that undergoes to a monolithic collapse.
We have identified three essential features for the ICs to form a disk with long-lived spiral arms,
otherwise the system forms an ellipsoidal quasistationary configuration without the rich morphological structures observed in the 
present case.}

The first is that they have to be almost uniform, i.e.,
 internal perturbations must be efficiently suppressed.
Indeed, density perturbations inside the overdensity grow through gravitational clustering during the collapse, 
and they form substructures by a hierarchical aggregation process.
If the amplitude of the initial density fluctuations is too large and/or their spatial correlations too strong, 
{  then 
they go nonlinear on a scale comparable to the system's size} 
on a time-scale shorter than the collapse time $\tau$. 
{  At $t \sim \tau$ the system is made of large sub-clumps}, the collapse is halted and the evolution proceeds through an aggregation of the subclumps: 
{  in this condition} the system mean-field potential is only perturbed but does not undergo a rapid change.

The second condition is that the ICs must break spherical symmetry.
Indeed, the variation of the system mean-field potential triggers a large change of the particle 
energy distribution if a fraction of the particles have a collapse time longer than the bulk of the system mass.
In this condition those particles move for a short time interval in a 
rapidly varying potential field and thus gain some kinetic energy while all others become more bound.
The amount of energy gain depends on the time-lag of a particle to arrive at 
the center, with respect to the largest fraction of the system mass.
For an initially uniform system, such a lag is developed because of
the coupling of the system finite size with the growth of internal density fluctuations during the collapse phase.
As a result of this complex dynamical mechanism, 
particles that initially were in the outermost regions arrive 
later than the others at the center and thus gain the largest amount of energy. 
For this reason any anisotropy initially characterizing such a 
particle distribution contributes to the developing of the time lag and 
it is thus amplified by the collapse dynamics.

Finally, the third condition is that the system has a sufficient 
initial amount of angular momentum, otherwise rotation is inhibited and a disk cannot form.  

Given these three conditions, the dissipationless component of the system gives rise to a thick disk, in which rotational motions are predominant but still with  a large velocity dispersion.
Such a disk is surrounded by large size out-of-equilibrium spiral arms, with or without bars and/or rings, which are expanding on a secular timescale as the radial motion predominates on the transversal one \citep{Benhaiem+SylosLabini_2015,Benhaiem+SylosLabini_2017}.
When considering a two-phase system where gas is coupled to the dissipationless component, then the post-collapse configuration shows a more heterogeneous and richer phenomenology.
At small distances from the center, gas particles form a quasistationary thin disk in which rotation dominates but where orbits are generically elliptical and thus 
where radial motions are also present.

Around such a thin disk long-lived but nonstationary spiral arms are formed: they arise from the coherent motions of groups of particles that have undergone a very similar dynamical history. 
The coherence in the motion of these particles, which maintains the spiral pattern, is originated because they were initially close (in the outer regions of the system) so that they could gain a similar amount of energy during the gravitational collapse remaining correlated  in both position and velocity in their subsequent evolution. 
The life-time of the spiral arms for the system we have considered is 
{  much larger than the characteristic collapse timescale $\tau$  and}
 it is related to the velocity dispersion of the particles in the arms:
 the larger is the dispersion, the shorter is the life-time.
{  In the numerical experiments we have presented in this work, we have considered the  mass and size of 
a typical spiral galaxy, getting a characteristic collapse timescale of $\tau \sim 0.5$ Gyr, while the 
life-time of the spiral arms is $> 10$ Gyr.}

{ 
We notice that if the fraction of the system gas is marginal (like is the case of a very high star formation rate leading to a rapid gas to star phase transition), then the ﬁnal QSS has the shape of an ellipsoid whose ﬂatness parameter, in general, depends on both its initial value and on the amount of initial angular momentum 
 \cite{Benhaiem+SylosLabini_2015}. 
Thus the monolithic collapse process investigated in this 
work represents a viable evolutionary path for the
formation of elliptical galaxies (or of the  almost spherical globular clusters that also have 
no gas component) as it was firstly argued by 
\cite{vanalbada_1982}.
As noticed in  \cite{syloslabini_2012,syloslabini_2013}  
the signature of the violent origin of such ellipsoidal QSS is represented
by a characteristic $\sim r^{-4}$ density profile in the external regions and 
by an almost flat core. Such behaviors are clearly different from the density profile 
of the halo structures, that are ellipsoidal too,  
formed 
trough a bottom-up aggregation process 
in the context of CDM-type cosmological simulations
\citep{Navarro_etal_1997}: such a density profile is characterized by a cusp in the inner 
regions of the type $\sim r^{-1}$ and by a slower decay 
in the external regions, i.e $\sim r^{-3}$.
A further support to a monolithic collapse origin for elliptical galaxies is that in the case of a merger origin of ellipticals from spirals the phase space density, being the merger collisionless, should remain constant, respecting the Boltzmann's collisionless equation, while there are strong hints that in ellipticals it is higher
\cite{Carlberg_1987}. 
A more detailed study of the difference between 
purely
gravitational  
QSS 
formed through a top-down monolithic collapse 
and through a bottom-up aggregation mechanism will be presented 
in a forthcoming work.  
}



Different mechanisms to produce spiral arms and possibly bars have been proposed in the literature and they all assume that the galactic disk is already formed \cite{Dobbs_Baba_2014}. 
A model in which the spiral structure is interpreted as a stationary density wave was introduced by \cite{Lin+Shu_1964} (see \cite{Bertin+Lin_1996} for more details): this hypothesizes that 
the spiral arms arise from a periodic compression and rarefaction of the disk surface density that propagates through the disk and remains stationary over many orbital periods. The spiral arms characterizing the systems that we have discussed are very different from quasi-stationary density waves as they involve the motion of matter and not only of energy. 
A second mechanism hypothesized to produce spiral arms is given by the effect of local instabilities, or of external perturbations, in a rotating disk. Indeed, self-gravitating disks close to stationary equilibrium and dominated by rotational motions are remarkably responsive to small disturbances so that spiral arms 
can be transient, recurrent and initiated by swing amplified instabilities in the disk (see, e.g., \cite{Sellwood+Carlberg_1984,Sellwood+Carlberg_2014,Dobbs_Baba_2014,Sellwood+Carlberg_2019} and references therein).
In this context, the continuously changing recurrent transient patterns formed in simulations of isolated disk galaxy models, embedded in rigid halos, have a rather quiet velocity field, i.e. there are not  present nor large amplitude streaming motions neither a net radial velocity component and the system is dominated by rotational motions.
Indeed, the kind of perturbations considered do not sensibly change the system's mean field and give rise to a soft dynamical mechanism that is not able to change the particle energy PDF
as it occurs in the violent dynamics we have described in this work. As a result of the violence of the collapse, correlations in phase-space are stronger, and thus the life-time is longer, in the former case than in the latter one. 
Such correlations correspond to well-defined phase-space structures: in particular, we have highlighted the properties of the phase-space projection into the plane defined by the radial and azimuthal velocity components.

The main difference with the standard CDM scenario is that in that case the disk is embedded in the gravitational potential field of an halo structure, i.e. a system 
{  close to spherical} with an almost isotropic velocity dispersion.
In that situation the gaseous matter forms a disk where rotational motions dominate and whose dynamical properties are determined by the more massive halo structure. 
On the contrary, in the system we have studied in this paper, the rotating disk is embedded in the gravitational field of a more massive thicker disk that is also rotation dominated in its inner regions.
Then, in its outer regions, such a system is not yet relaxed and presents out-of-equilibrium features.
{  Such a situation} implies that in those regions it is not possible to simply recover a mass from the measurement of a velocity.
This represents an important warning that must be 
considered in detail when analyzing a given object as 
in general the assumption of stationary equilibrium 
with maximal rotational motions is taken for granted 
in the determination of galaxy masses from observed 
line-of-sight velocities or velocity dispersion. 
Indeed, the estimates of the quantity of dark matter, 
both in the Milky Way and in external galaxies, are generally based on the assumption that emitters motion is maximally rotational and/or that systems are relaxed into a quasi-stationary equilibrium state. 
Distance-dependent deviations from these assumptions 
naturally arise in the systems we have discussed: in such a 
situation the estimation of the amount of dark matter must
be revised taking into account more complex velocity fields
and dynamical mechanisms. 
Of course, from an observational point of view the problem 
is to detect the presence of nonrotational motions: this is not at all a simple task given the degeneracy between a radial and a rotational velocity field for nonaxisymmtric objects  \citep{SylosLabini_etal_2019}.

{  In this respect, it is worth noticing that }
the number of revolutions 
{  completed by} an object orbiting
with a velocity of $\sim 200$ km/s around a galaxy at a 
distance larger than $\sim 20$ kpc  in a Hubble's time 
$\sim 10$ Gyr is of the order of 10 or smaller
\citep{Benhaiem+SylosLabini+Joyce_2019}. 
If it is not at present possible to theoretically constraint 
the number of revolutions needed to reach a relaxed configuration
from a qualitative point of view, a reasonable requirement is that 
they must be $\gg 1$.
This simple 
observation raises  a serious warning about the 
possibility of considering the outermost regions of a galaxy 
in a relaxed equilibrium configuration, the hypothesis 
usually adopted to estimate the amount of  dark matter. 

There are three main directions that will be pursued in forthcoming works. 
First, we aim at studying the violent collapse in a full cosmological context, 
including other astrophysical effects beyond gas dynamics. 
In this respect, as discussed above, it is necessary to consider density 
fluctuations with suitable correlations that must be suppressed at small enough scales. 
Second, we will explore in more details the whole phase-space structure of these systems, comparing it to both other kind of quasistationary disks originated by different dynamical models (e.g., by the slow and soft dynamics acting in perturbed self-gravitating disks or in disks formed in the framework of CDM models) and to observations of the Milky Way, by considering the forthcoming data provided by the {\it Gaia } mission \citep{Gaia_2016}, and of external galaxies  \citep{SylosLabini_etal_2019}. 
Such a study is complementary to a wider investigation of different and more complex initial conditions, which include both irregular shapes and nonuniform and correlated matter density fluctuations.
Finally we plan a detailed study of external galaxy line-of-sight velocity maps with the aim of developing an alternative way to fit the data than the usually adopted one in which rotational motions are taken to be maximal at all scales. 

{  As a concluding remark it is worth mentioning
that 
the Milky Way velocity field
was recently found to 
exhibit several phase-space structures \cite{Antoja_etal_2018} together with 
velocity gradients in all  three components \cite{Katz_etal_2018,Wang_2018,MLC_FSL_2019,lopezcorredoira2020gaiadr2,Zofia_etal_2020}. 
Such findings can be understood in a model in which the galaxy has a nonstationary nature of 
the type we discussed in this work, but can also be explained as due to the effect of external perturbers. 
The forthcoming data of the {\it Gaia} mission 
\cite{Gaia_2016}
will eventually clarify the situation. }

\begin{acknowledgments}
We thank Volker Springel for making available to us his updated version of the {\tt Gadget-3} code.
FSL thanks Lapo Casetti, Edvige Corbelli, Martin L{\'o}pez-Corredoira, Michael Joyce, Yuri Baryshev and Zofia Chrob\'akov\'a for very useful discussions and comments.
Finally we thank the anonymous referees of this work 
who helped us to improve the presentation and discussion of our results.
\end{acknowledgments}. 


\begin{thebibliography}{79}%
\makeatletter
\providecommand \@ifxundefined [1]{%
 \@ifx{#1\undefined}
}%
\providecommand \@ifnum [1]{%
 \ifnum #1\expandafter \@firstoftwo
 \else \expandafter \@secondoftwo
 \fi
}%
\providecommand \@ifx [1]{%
 \ifx #1\expandafter \@firstoftwo
 \else \expandafter \@secondoftwo
 \fi
}%
\providecommand \natexlab [1]{#1}%
\providecommand \enquote  [1]{``#1''}%
\providecommand \bibnamefont  [1]{#1}%
\providecommand \bibfnamefont [1]{#1}%
\providecommand \citenamefont [1]{#1}%
\providecommand \href@noop [0]{\@secondoftwo}%
\providecommand \href [0]{\begingroup \@sanitize@url \@href}%
\providecommand \@href[1]{\@@startlink{#1}\@@href}%
\providecommand \@@href[1]{\endgroup#1\@@endlink}%
\providecommand \@sanitize@url [0]{\catcode `\\12\catcode `\$12\catcode
  `\&12\catcode `\#12\catcode `\^12\catcode `\_12\catcode `\%12\relax}%
\providecommand \@@startlink[1]{}%
\providecommand \@@endlink[0]{}%
\providecommand \url  [0]{\begingroup\@sanitize@url \@url }%
\providecommand \@url [1]{\endgroup\@href {#1}{\urlprefix }}%
\providecommand \urlprefix  [0]{URL }%
\providecommand \Eprint [0]{\href }%
\providecommand \doibase [0]{http://dx.doi.org/}%
\providecommand \selectlanguage [0]{\@gobble}%
\providecommand \bibinfo  [0]{\@secondoftwo}%
\providecommand \bibfield  [0]{\@secondoftwo}%
\providecommand \translation [1]{[#1]}%
\providecommand \BibitemOpen [0]{}%
\providecommand \bibitemStop [0]{}%
\providecommand \bibitemNoStop [0]{.\EOS\space}%
\providecommand \EOS [0]{\spacefactor3000\relax}%
\providecommand \BibitemShut  [1]{\csname bibitem#1\endcsname}%
\let\auto@bib@innerbib\@empty
\bibitem [{\citenamefont {Padmanabhan}(1990)}]{Padmanabhan_1989}%
  \BibitemOpen
  \bibfield  {author} {\bibinfo {author} {\bibfnamefont {T.}~\bibnamefont
  {Padmanabhan}},\ }\href@noop {} {\bibfield  {journal} {\bibinfo  {journal}
  {Phys. Rept.}\ }\textbf {\bibinfo {volume} {188}},\ \bibinfo {pages} {285}
  (\bibinfo {year} {1990})}\BibitemShut {NoStop}%
\bibitem [{\citenamefont {Dauxois}\ \emph {et~al.}(2002)\citenamefont
  {Dauxois}, \citenamefont {Ruffo}, \citenamefont {Arimondo},\ and\
  \citenamefont {Wilkens}}]{Dauxois_etal_2002}%
  \BibitemOpen
  \bibfield  {author} {\bibinfo {author} {\bibfnamefont {T.}~\bibnamefont
  {Dauxois}}, \bibinfo {author} {\bibfnamefont {S.}~\bibnamefont {Ruffo}},
  \bibinfo {author} {\bibfnamefont {E.}~\bibnamefont {Arimondo}}, \ and\
  \bibinfo {author} {\bibfnamefont {M.}~\bibnamefont {Wilkens}},\ }\enquote
  {\bibinfo {title} {Dynamics and thermodynamics of systems with long-range
  interactions: An introduction},}\ in\ \href {\doibase
  10.1007/3-540-45835-2_1} {\emph {\bibinfo {booktitle} {Dynamics and
  Thermodynamics of Systems with Long-Range Interactions}}}\ (\bibinfo
  {publisher} {Springer Berlin Heidelberg},\ \bibinfo {address} {Berlin,
  Heidelberg},\ \bibinfo {year} {2002})\ pp.\ \bibinfo {pages}
  {1--19}\BibitemShut {NoStop}%
\bibitem [{\citenamefont {Campa}\ \emph {et~al.}(2008)\citenamefont {Campa},
  \citenamefont {Giansanti}, \citenamefont {Morigi},\ and\ \citenamefont
  {Sylos~Labini}}]{Assisi}%
  \BibitemOpen
  \bibfield  {author} {\bibinfo {author} {\bibfnamefont {A.}~\bibnamefont
  {Campa}}, \bibinfo {author} {\bibfnamefont {A.}~\bibnamefont {Giansanti}},
  \bibinfo {author} {\bibfnamefont {G.}~\bibnamefont {Morigi}}, \ and\ \bibinfo
  {author} {\bibfnamefont {F.}~\bibnamefont {Sylos~Labini}},\ }\href@noop {}
  {\emph {\bibinfo {title} {Dynamics and Thermodynamics of Systems with Long
  Range Interactions: Theory and experiments}}}\ (\bibinfo  {publisher} {AIP
  Conference Proceedings},\ \bibinfo {year} {2008})\BibitemShut {NoStop}%
\bibitem [{\citenamefont {{Campa}}\ \emph {et~al.}(2009)\citenamefont
  {{Campa}}, \citenamefont {{Dauxois}},\ and\ \citenamefont
  {{Ruffo}}}]{Campa_etal_2009}%
  \BibitemOpen
  \bibfield  {author} {\bibinfo {author} {\bibfnamefont {A.}~\bibnamefont
  {{Campa}}}, \bibinfo {author} {\bibfnamefont {T.}~\bibnamefont {{Dauxois}}},
  \ and\ \bibinfo {author} {\bibfnamefont {S.}~\bibnamefont {{Ruffo}}},\
  }\href@noop {} {\bibfield  {journal} {\bibinfo  {journal} {Phys. Reports}\
  }\textbf {\bibinfo {volume} {480}},\ \bibinfo {pages} {57} (\bibinfo {year}
  {2009})}\BibitemShut {NoStop}%
\bibitem [{\citenamefont {Gabrielli}\ \emph {et~al.}(2010)\citenamefont
  {Gabrielli}, \citenamefont {Jouyce}, \citenamefont {Marcos},\ and\
  \citenamefont {Sicard}}]{gabrielli_etal_2010a}%
  \BibitemOpen
  \bibfield  {author} {\bibinfo {author} {\bibfnamefont {A.}~\bibnamefont
  {Gabrielli}}, \bibinfo {author} {\bibfnamefont {M.}~\bibnamefont {Jouyce}},
  \bibinfo {author} {\bibfnamefont {B.}~\bibnamefont {Marcos}}, \ and\ \bibinfo
  {author} {\bibfnamefont {F.}~\bibnamefont {Sicard}},\ }\href@noop {}
  {\bibfield  {journal} {\bibinfo  {journal} {J. Stat. Phys.}\ }\textbf
  {\bibinfo {volume} {141}},\ \bibinfo {pages} {970} (\bibinfo {year}
  {2010})}\BibitemShut {NoStop}%
\bibitem [{\citenamefont {Campa}\ \emph {et~al.}(2014)\citenamefont {Campa},
  \citenamefont {Dauxois}, \citenamefont {Fanelli},\ and\ \citenamefont
  {Ruffo}}]{Campa_etal_2014}%
  \BibitemOpen
  \bibfield  {author} {\bibinfo {author} {\bibfnamefont {A.}~\bibnamefont
  {Campa}}, \bibinfo {author} {\bibfnamefont {T.}~\bibnamefont {Dauxois}},
  \bibinfo {author} {\bibfnamefont {D.}~\bibnamefont {Fanelli}}, \ and\
  \bibinfo {author} {\bibfnamefont {S.}~\bibnamefont {Ruffo}},\ }\href@noop {}
  {\emph {\bibinfo {title} {Physics of Long-Range Interacting Systems}}}\
  (\bibinfo  {publisher} {Oxford},\ \bibinfo {year} {2014})\BibitemShut
  {NoStop}%
\bibitem [{\citenamefont {{Chavanis}}(2010)}]{chavanis_kEqns_2010}%
  \BibitemOpen
  \bibfield  {author} {\bibinfo {author} {\bibfnamefont {P.}~\bibnamefont
  {{Chavanis}}},\ }\href {\doibase 10.1088/1742-5468/2010/05/P05019} {\bibfield
   {journal} {\bibinfo  {journal} {Journal of Statistical Mechanics: Theory and
  Experiment}\ }\textbf {\bibinfo {volume} {5}},\ \bibinfo {pages} {19}
  (\bibinfo {year} {2010})}\BibitemShut {NoStop}%
\bibitem [{\citenamefont {Marcos}(2013)}]{Marcos_2013}%
  \BibitemOpen
  \bibfield  {author} {\bibinfo {author} {\bibfnamefont {B.}~\bibnamefont
  {Marcos}},\ }\href {\doibase 10.1103/PhysRevE.88.032112} {\bibfield
  {journal} {\bibinfo  {journal} {Phys. Rev. E}\ }\textbf {\bibinfo {volume}
  {88}},\ \bibinfo {pages} {032112} (\bibinfo {year} {2013})}\BibitemShut
  {NoStop}%
\bibitem [{\citenamefont {Levin}\ \emph {et~al.}(2014)\citenamefont {Levin},
  \citenamefont {Pakter}, \citenamefont {Rizzato}, \citenamefont {Teles},\ and\
  \citenamefont {Benetti}}]{Levin_etal_2014}%
  \BibitemOpen
  \bibfield  {author} {\bibinfo {author} {\bibfnamefont {Y.}~\bibnamefont
  {Levin}}, \bibinfo {author} {\bibfnamefont {R.}~\bibnamefont {Pakter}},
  \bibinfo {author} {\bibfnamefont {F.~B.}\ \bibnamefont {Rizzato}}, \bibinfo
  {author} {\bibfnamefont {T.~N.}\ \bibnamefont {Teles}}, \ and\ \bibinfo
  {author} {\bibfnamefont {F.~P.}\ \bibnamefont {Benetti}},\ }\href@noop {}
  {\bibfield  {journal} {\bibinfo  {journal} {Physics Reports}\ }\textbf
  {\bibinfo {volume} {535}},\ \bibinfo {pages} {1 } (\bibinfo {year}
  {2014})}\BibitemShut {NoStop}%
\bibitem [{\citenamefont {{Marcos}}\ \emph {et~al.}(2017)\citenamefont
  {{Marcos}}, \citenamefont {{Gabrielli}},\ and\ \citenamefont
  {{Joyce}}}]{Marcos_etal_2017}%
  \BibitemOpen
  \bibfield  {author} {\bibinfo {author} {\bibfnamefont {B.}~\bibnamefont
  {{Marcos}}}, \bibinfo {author} {\bibfnamefont {A.}~\bibnamefont
  {{Gabrielli}}}, \ and\ \bibinfo {author} {\bibfnamefont {M.}~\bibnamefont
  {{Joyce}}},\ }\href {\doibase 10.1103/PhysRevE.96.032102} {\bibfield
  {journal} {\bibinfo  {journal} {Phys.Rev.E}\ }\textbf {\bibinfo {volume}
  {96}},\ \bibinfo {eid} {032102} (\bibinfo {year} {2017})}\BibitemShut
  {NoStop}%
\bibitem [{\citenamefont {{Di Cintio}}\ \emph {et~al.}(2018)\citenamefont {{Di
  Cintio}}, \citenamefont {{Gupta}},\ and\ \citenamefont
  {{Casetti}}}]{diCintio_etal_2018}%
  \BibitemOpen
  \bibfield  {author} {\bibinfo {author} {\bibfnamefont {P.}~\bibnamefont {{Di
  Cintio}}}, \bibinfo {author} {\bibfnamefont {S.}~\bibnamefont {{Gupta}}}, \
  and\ \bibinfo {author} {\bibfnamefont {L.}~\bibnamefont {{Casetti}}},\ }\href
  {\doibase 10.1093/mnras/stx3244} {\bibfield  {journal} {\bibinfo  {journal}
  {Mon. Not. R. Astron. Soc.}\ }\textbf {\bibinfo {volume} {475}},\ \bibinfo
  {pages} {1137} (\bibinfo {year} {2018})}\BibitemShut {NoStop}%
\bibitem [{\citenamefont {{Capuzzo Dolcetta}}(2019)}]{RCDbook}%
  \BibitemOpen
  \bibfield  {author} {\bibinfo {author} {\bibfnamefont {R.~A.}\ \bibnamefont
  {{Capuzzo Dolcetta}}},\ }\href {\doibase 10.1007/978-3-030-25846-7} {\emph
  {\bibinfo {title} {{Classical Newtonian Gravity}}}}\ (\bibinfo  {publisher}
  {Springer International Publishing},\ \bibinfo {year} {2019})\BibitemShut
  {NoStop}%
\bibitem [{\citenamefont {{Lynden-Bell}}(1967)}]{lyndenbell}%
  \BibitemOpen
  \bibfield  {author} {\bibinfo {author} {\bibfnamefont {D.}~\bibnamefont
  {{Lynden-Bell}}},\ }\href {\doibase 10.1093/mnras/136.1.101} {\bibfield
  {journal} {\bibinfo  {journal} {Mon. Not. R. Astr. Soc.}\ }\textbf {\bibinfo
  {volume} {136}},\ \bibinfo {pages} {101} (\bibinfo {year}
  {1967})}\BibitemShut {NoStop}%
\bibitem [{\citenamefont {{Dauxois}}\ \emph {et~al.}(2002)\citenamefont
  {{Dauxois}}, \citenamefont {{Ruffo}}, \citenamefont {{Arimondo}},\ and\
  \citenamefont {{Wilkens}}}]{Dauxois:2002pv}%
  \BibitemOpen
  \bibfield  {author} {\bibinfo {author} {\bibfnamefont {T.}~\bibnamefont
  {{Dauxois}}}, \bibinfo {author} {\bibfnamefont {S.}~\bibnamefont {{Ruffo}}},
  \bibinfo {author} {\bibfnamefont {E.}~\bibnamefont {{Arimondo}}}, \ and\
  \bibinfo {author} {\bibfnamefont {M.}~\bibnamefont {{Wilkens}}},\ }\enquote
  {\bibinfo {title} {{Dynamics and Thermodynamics of Systems with Long-Range
  Interactions: An Introduction}},}\ in\ \href@noop {} {\emph {\bibinfo
  {booktitle} {Lecture Notes in Physics}}},\ Vol.\ \bibinfo {volume} {602}\
  (\bibinfo  {publisher} {Springer},\ \bibinfo {year} {2002})\ pp.\ \bibinfo
  {pages} {1--19}\BibitemShut {NoStop}%
\bibitem [{\citenamefont {{Chandrasekhar}}(1943)}]{chandra43}%
  \BibitemOpen
  \bibfield  {author} {\bibinfo {author} {\bibfnamefont {S.}~\bibnamefont
  {{Chandrasekhar}}},\ }\href {\doibase 10.1103/RevModPhys.15.1} {\bibfield
  {journal} {\bibinfo  {journal} {Reviews of Modern Physics}\ }\textbf
  {\bibinfo {volume} {15}},\ \bibinfo {pages} {1} (\bibinfo {year}
  {1943})}\BibitemShut {NoStop}%
\bibitem [{\citenamefont {Peebles}(1980)}]{peebles_1983}%
  \BibitemOpen
  \bibfield  {author} {\bibinfo {author} {\bibfnamefont {P.~J.~E.}\
  \bibnamefont {Peebles}},\ }\href@noop {} {\emph {\bibinfo {title} {{The
  Large-Scale Structure of the Universe}}}}\ (\bibinfo  {publisher} {Princeton
  University Press},\ \bibinfo {year} {1980})\BibitemShut {NoStop}%
\bibitem [{\citenamefont {{Wright}}\ \emph {et~al.}(1982)\citenamefont
  {{Wright}}, \citenamefont {{Miller}},\ and\ \citenamefont
  {{Stein}}}]{Wright_etal_1982}%
  \BibitemOpen
  \bibfield  {author} {\bibinfo {author} {\bibfnamefont {H.~L.}\ \bibnamefont
  {{Wright}}}, \bibinfo {author} {\bibfnamefont {B.~N.}\ \bibnamefont
  {{Miller}}}, \ and\ \bibinfo {author} {\bibfnamefont {W.~E.}\ \bibnamefont
  {{Stein}}},\ }\href {\doibase 10.1007/BF00651321} {\bibfield  {journal}
  {\bibinfo  {journal} {Astrophy.Spac.Science.}\ }\textbf {\bibinfo {volume}
  {84}},\ \bibinfo {pages} {421} (\bibinfo {year} {1982})}\BibitemShut
  {NoStop}%
\bibitem [{\citenamefont {{Yawn}}\ and\ \citenamefont
  {{Miller}}(1997)}]{Yan_Miller_1997}%
  \BibitemOpen
  \bibfield  {author} {\bibinfo {author} {\bibfnamefont {K.~R.}\ \bibnamefont
  {{Yawn}}}\ and\ \bibinfo {author} {\bibfnamefont {B.~N.}\ \bibnamefont
  {{Miller}}},\ }\href {\doibase 10.1103/PhysRevLett.79.3561} {\bibfield
  {journal} {\bibinfo  {journal} {Phys.Rev.Lett.}\ }\textbf {\bibinfo {volume}
  {79}},\ \bibinfo {pages} {3561} (\bibinfo {year} {1997})}\BibitemShut
  {NoStop}%
\bibitem [{\citenamefont {{Tsuchiya}}\ and\ \citenamefont
  {{Gouda}}(2000)}]{Tsuchiya_Gouda_2000}%
  \BibitemOpen
  \bibfield  {author} {\bibinfo {author} {\bibfnamefont {T.}~\bibnamefont
  {{Tsuchiya}}}\ and\ \bibinfo {author} {\bibfnamefont {N.}~\bibnamefont
  {{Gouda}}},\ }\href {\doibase 10.1103/PhysRevE.61.948} {\bibfield  {journal}
  {\bibinfo  {journal} {Phys.Rev.E}\ }\textbf {\bibinfo {volume} {61}},\
  \bibinfo {pages} {948} (\bibinfo {year} {2000})}\BibitemShut {NoStop}%
\bibitem [{\citenamefont {{Joyce}}\ and\ \citenamefont
  {{Worrakitpoonpon}}(2011)}]{Joyce+Worrakitpoonpon_2011}%
  \BibitemOpen
  \bibfield  {author} {\bibinfo {author} {\bibfnamefont {M.}~\bibnamefont
  {{Joyce}}}\ and\ \bibinfo {author} {\bibfnamefont {T.}~\bibnamefont
  {{Worrakitpoonpon}}},\ }\href {\doibase 10.1103/PhysRevE.84.011139}
  {\bibfield  {journal} {\bibinfo  {journal} {Phys.Rev.E}\ }\textbf {\bibinfo
  {volume} {84}},\ \bibinfo {eid} {011139} (\bibinfo {year}
  {2011})}\BibitemShut {NoStop}%
\bibitem [{\citenamefont {{Teles}}\ \emph {et~al.}(2011)\citenamefont
  {{Teles}}, \citenamefont {{Levin}},\ and\ \citenamefont
  {{Pakter}}}]{Teles_etal_2011}%
  \BibitemOpen
  \bibfield  {author} {\bibinfo {author} {\bibfnamefont {T.~N.}\ \bibnamefont
  {{Teles}}}, \bibinfo {author} {\bibfnamefont {Y.}~\bibnamefont {{Levin}}}, \
  and\ \bibinfo {author} {\bibfnamefont {R.}~\bibnamefont {{Pakter}}},\ }\href
  {\doibase 10.1111/j.1745-3933.2011.01112.x} {\bibfield  {journal} {\bibinfo
  {journal} {Mon.Not.R.Astron.Soc.}\ }\textbf {\bibinfo {volume} {417}},\
  \bibinfo {pages} {L21} (\bibinfo {year} {2011})}\BibitemShut {NoStop}%
\bibitem [{\citenamefont {Henon}(1964)}]{henon_1964}%
  \BibitemOpen
  \bibfield  {author} {\bibinfo {author} {\bibfnamefont {M.}~\bibnamefont
  {Henon}},\ }\href@noop {} {\bibfield  {journal} {\bibinfo  {journal} {Ann.
  Astrophys.}\ }\textbf {\bibinfo {volume} {27}},\ \bibinfo {pages} {1}
  (\bibinfo {year} {1964})}\BibitemShut {NoStop}%
\bibitem [{\citenamefont {van Albada}(1982)}]{vanalbada_1982}%
  \BibitemOpen
  \bibfield  {author} {\bibinfo {author} {\bibfnamefont {T.}~\bibnamefont {van
  Albada}},\ }\href@noop {} {\bibfield  {journal} {\bibinfo  {journal} {Mon.
  Not. R. Astr. Soc.}\ }\textbf {\bibinfo {volume} {201}},\ \bibinfo {pages}
  {939} (\bibinfo {year} {1982})}\BibitemShut {NoStop}%
\bibitem [{\citenamefont {Aarseth}\ \emph {et~al.}(1988)\citenamefont
  {Aarseth}, \citenamefont {Lin},\ and\ \citenamefont
  {Papaloizou}}]{aarseth_etal_1988}%
  \BibitemOpen
  \bibfield  {author} {\bibinfo {author} {\bibfnamefont {S.}~\bibnamefont
  {Aarseth}}, \bibinfo {author} {\bibfnamefont {D.}~\bibnamefont {Lin}}, \ and\
  \bibinfo {author} {\bibfnamefont {J.}~\bibnamefont {Papaloizou}},\
  }\href@noop {} {\bibfield  {journal} {\bibinfo  {journal} {Astrophys. J.}\
  }\textbf {\bibinfo {volume} {324}},\ \bibinfo {pages} {288} (\bibinfo {year}
  {1988})}\BibitemShut {NoStop}%
\bibitem [{\citenamefont {Aguilar}\ and\ \citenamefont
  {Merritt}(1990)}]{aguilar+merritt_1990}%
  \BibitemOpen
  \bibfield  {author} {\bibinfo {author} {\bibfnamefont {L.}~\bibnamefont
  {Aguilar}}\ and\ \bibinfo {author} {\bibfnamefont {D.}~\bibnamefont
  {Merritt}},\ }\href@noop {} {\bibfield  {journal} {\bibinfo  {journal}
  {Astrophys. J.}\ }\textbf {\bibinfo {volume} {354}},\ \bibinfo {pages} {73}
  (\bibinfo {year} {1990})}\BibitemShut {NoStop}%
\bibitem [{\citenamefont {Theis}\ and\ \citenamefont
  {Spurzem}(1999)}]{theis+spurzem_1999}%
  \BibitemOpen
  \bibfield  {author} {\bibinfo {author} {\bibfnamefont {C.}~\bibnamefont
  {Theis}}\ and\ \bibinfo {author} {\bibfnamefont {R.}~\bibnamefont
  {Spurzem}},\ }\href@noop {} {\bibfield  {journal} {\bibinfo  {journal}
  {Astron. Astrophys.}\ }\textbf {\bibinfo {volume} {341}},\ \bibinfo {pages}
  {361} (\bibinfo {year} {1999})}\BibitemShut {NoStop}%
\bibitem [{\citenamefont {Boily}\ \emph {et~al.}(2002)\citenamefont {Boily},
  \citenamefont {Athanassoula},\ and\ \citenamefont
  {Kroupa}}]{boily_etal_2002}%
  \BibitemOpen
  \bibfield  {author} {\bibinfo {author} {\bibfnamefont {C.}~\bibnamefont
  {Boily}}, \bibinfo {author} {\bibfnamefont {E.}~\bibnamefont {Athanassoula}},
  \ and\ \bibinfo {author} {\bibfnamefont {P.}~\bibnamefont {Kroupa}},\
  }\href@noop {} {\bibfield  {journal} {\bibinfo  {journal} {Mon. Not. R. Astr.
  Soc.}\ }\textbf {\bibinfo {volume} {332}},\ \bibinfo {pages} {971} (\bibinfo
  {year} {2002})}\BibitemShut {NoStop}%
\bibitem [{\citenamefont {Roy}\ and\ \citenamefont
  {Perez}(2004)}]{roy+perez_2004}%
  \BibitemOpen
  \bibfield  {author} {\bibinfo {author} {\bibfnamefont {F.}~\bibnamefont
  {Roy}}\ and\ \bibinfo {author} {\bibfnamefont {J.}~\bibnamefont {Perez}},\
  }\href@noop {} {\bibfield  {journal} {\bibinfo  {journal} {Mon. Not. R. Astr.
  Soc.}\ }\textbf {\bibinfo {volume} {348}},\ \bibinfo {pages} {62} (\bibinfo
  {year} {2004})}\BibitemShut {NoStop}%
\bibitem [{\citenamefont {{Boily}}\ and\ \citenamefont
  {{Athanassoula}}(2006)}]{boily+athanassoula_2006}%
  \BibitemOpen
  \bibfield  {author} {\bibinfo {author} {\bibfnamefont {C.~M.}\ \bibnamefont
  {{Boily}}}\ and\ \bibinfo {author} {\bibfnamefont {E.}~\bibnamefont
  {{Athanassoula}}},\ }\href {\doibase 10.1111/j.1365-2966.2006.10365.x}
  {\bibfield  {journal} {\bibinfo  {journal} {Mon. Not. R. Astr. Soc.}\
  }\textbf {\bibinfo {volume} {369}},\ \bibinfo {pages} {608} (\bibinfo {year}
  {2006})}\BibitemShut {NoStop}%
\bibitem [{\citenamefont {{Barnes}}\ \emph {et~al.}(2009)\citenamefont
  {{Barnes}}, \citenamefont {{Lanzel}},\ and\ \citenamefont
  {{Williams}}}]{barnes_etal_2009}%
  \BibitemOpen
  \bibfield  {author} {\bibinfo {author} {\bibfnamefont {E.~I.}\ \bibnamefont
  {{Barnes}}}, \bibinfo {author} {\bibfnamefont {P.~A.}\ \bibnamefont
  {{Lanzel}}}, \ and\ \bibinfo {author} {\bibfnamefont {L.~L.~R.}\ \bibnamefont
  {{Williams}}},\ }\href {\doibase 10.1088/0004-637X/704/1/372} {\bibfield
  {journal} {\bibinfo  {journal} {Astrophys. J.}\ }\textbf {\bibinfo {volume}
  {704}},\ \bibinfo {pages} {372} (\bibinfo {year} {2009})}\BibitemShut
  {NoStop}%
\bibitem [{\citenamefont {Joyce}\ \emph {et~al.}(2009)\citenamefont {Joyce},
  \citenamefont {Marcos},\ and\ \citenamefont
  {Sylos~Labini}}]{ejection_mjbmfsl}%
  \BibitemOpen
  \bibfield  {author} {\bibinfo {author} {\bibfnamefont {M.}~\bibnamefont
  {Joyce}}, \bibinfo {author} {\bibfnamefont {B.}~\bibnamefont {Marcos}}, \
  and\ \bibinfo {author} {\bibfnamefont {F.}~\bibnamefont {Sylos~Labini}},\
  }\href@noop {} {\bibfield  {journal} {\bibinfo  {journal} {Mon. Not. R.
  Astron. Soc.}\ }\textbf {\bibinfo {volume} {397}},\ \bibinfo {pages} {775}
  (\bibinfo {year} {2009})}\BibitemShut {NoStop}%
\bibitem [{\citenamefont {{Sylos Labini}}(2012)}]{syloslabini_2012}%
  \BibitemOpen
  \bibfield  {author} {\bibinfo {author} {\bibfnamefont {F.}~\bibnamefont
  {{Sylos Labini}}},\ }\href {\doibase 10.1111/j.1365-2966.2012.21019.x}
  {\bibfield  {journal} {\bibinfo  {journal} {Mon. Not. R. Astron. Soc.}\
  }\textbf {\bibinfo {volume} {423}},\ \bibinfo {pages} {1610} (\bibinfo {year}
  {2012})}\BibitemShut {NoStop}%
\bibitem [{\citenamefont {{Worrakitpoonpon}}(2015)}]{worrakitpoonpon_2014}%
  \BibitemOpen
  \bibfield  {author} {\bibinfo {author} {\bibfnamefont {T.}~\bibnamefont
  {{Worrakitpoonpon}}},\ }\href@noop {} {\bibfield  {journal} {\bibinfo
  {journal} {Mon. Not. R. Astr. Soc.}\ }\textbf {\bibinfo {volume} {466}},\
  \bibinfo {pages} {1335} (\bibinfo {year} {2015})}\BibitemShut {NoStop}%
\bibitem [{\citenamefont {{Merritt}}\ and\ \citenamefont
  {{Aguilar}}(1985)}]{merritt+aguilar_1985}%
  \BibitemOpen
  \bibfield  {author} {\bibinfo {author} {\bibfnamefont {D.}~\bibnamefont
  {{Merritt}}}\ and\ \bibinfo {author} {\bibfnamefont {L.~A.}\ \bibnamefont
  {{Aguilar}}},\ }\href@noop {} {\bibfield  {journal} {\bibinfo  {journal}
  {Mon. Not. R. Ast. Soc}\ }\textbf {\bibinfo {volume} {217}},\ \bibinfo
  {pages} {787} (\bibinfo {year} {1985})}\BibitemShut {NoStop}%
\bibitem [{\citenamefont {{Sylos Labini}}(2013)}]{syloslabini_2013}%
  \BibitemOpen
  \bibfield  {author} {\bibinfo {author} {\bibfnamefont {F.}~\bibnamefont
  {{Sylos Labini}}},\ }\href {\doibase 10.1093/mnras/sts365} {\bibfield
  {journal} {\bibinfo  {journal} {Mon. Not. R. Astron. Soc.}\ }\textbf
  {\bibinfo {volume} {429}},\ \bibinfo {pages} {679} (\bibinfo {year}
  {2013})}\BibitemShut {NoStop}%
\bibitem [{\citenamefont {{Sylos Labini}}\ \emph {et~al.}(2015)\citenamefont
  {{Sylos Labini}}, \citenamefont {{Benhaiem}},\ and\ \citenamefont
  {{Joyce}}}]{SylosLabini+Benhaiem+Joyce_2015}%
  \BibitemOpen
  \bibfield  {author} {\bibinfo {author} {\bibfnamefont {F.}~\bibnamefont
  {{Sylos Labini}}}, \bibinfo {author} {\bibfnamefont {D.}~\bibnamefont
  {{Benhaiem}}}, \ and\ \bibinfo {author} {\bibfnamefont {M.}~\bibnamefont
  {{Joyce}}},\ }\href {\doibase 10.1093/mnras/stv581} {\bibfield  {journal}
  {\bibinfo  {journal} {Mon.Not.R.Astron.Soc.}\ }\textbf {\bibinfo {volume}
  {449}},\ \bibinfo {pages} {4458} (\bibinfo {year} {2015})}\BibitemShut
  {NoStop}%
\bibitem [{\citenamefont {{Benhaiem}}\ and\ \citenamefont {{Sylos
  Labini}}(2015)}]{Benhaiem+SylosLabini_2015}%
  \BibitemOpen
  \bibfield  {author} {\bibinfo {author} {\bibfnamefont {D.}~\bibnamefont
  {{Benhaiem}}}\ and\ \bibinfo {author} {\bibfnamefont {F.}~\bibnamefont
  {{Sylos Labini}}},\ }\href {\doibase 10.1093/mnras/stv075} {\bibfield
  {journal} {\bibinfo  {journal} {Mon.Not.R.Astron.Soc.}\ }\textbf {\bibinfo
  {volume} {448}},\ \bibinfo {pages} {2634} (\bibinfo {year}
  {2015})}\BibitemShut {NoStop}%
\bibitem [{\citenamefont {{Benhaiem}}\ and\ \citenamefont {{Sylos
  Labini}}(2017)}]{Benhaiem+SylosLabini_2017}%
  \BibitemOpen
  \bibfield  {author} {\bibinfo {author} {\bibfnamefont {D.}~\bibnamefont
  {{Benhaiem}}}\ and\ \bibinfo {author} {\bibfnamefont {F.}~\bibnamefont
  {{Sylos Labini}}},\ }\href {\doibase 10.1051/0004-6361/201628698} {\bibfield
  {journal} {\bibinfo  {journal} {Astron.Astrophys.}\ }\textbf {\bibinfo
  {volume} {598}},\ \bibinfo {eid} {A95} (\bibinfo {year} {2017})}\BibitemShut
  {NoStop}%
\bibitem [{\citenamefont {{Benhaiem}}\ \emph {et~al.}(2017)\citenamefont
  {{Benhaiem}}, \citenamefont {{Joyce}},\ and\ \citenamefont {{Sylos
  Labini}}}]{Benhaiem+Joyce+SylosLabini_2017}%
  \BibitemOpen
  \bibfield  {author} {\bibinfo {author} {\bibfnamefont {D.}~\bibnamefont
  {{Benhaiem}}}, \bibinfo {author} {\bibfnamefont {M.}~\bibnamefont {{Joyce}}},
  \ and\ \bibinfo {author} {\bibfnamefont {F.}~\bibnamefont {{Sylos Labini}}},\
  }\href {\doibase 10.3847/1538-4357/aa96a7} {\bibfield  {journal} {\bibinfo
  {journal} {Astrophys.J.}\ }\textbf {\bibinfo {volume} {851}},\ \bibinfo {eid}
  {19} (\bibinfo {year} {2017})}\BibitemShut {NoStop}%
\bibitem [{\citenamefont {{Benhaiem}}\ \emph {et~al.}(2019)\citenamefont
  {{Benhaiem}}, \citenamefont {{Sylos Labini}},\ and\ \citenamefont
  {{Joyce}}}]{Benhaiem+SylosLabini+Joyce_2019}%
  \BibitemOpen
  \bibfield  {author} {\bibinfo {author} {\bibfnamefont {D.}~\bibnamefont
  {{Benhaiem}}}, \bibinfo {author} {\bibfnamefont {F.}~\bibnamefont {{Sylos
  Labini}}}, \ and\ \bibinfo {author} {\bibfnamefont {M.}~\bibnamefont
  {{Joyce}}},\ }\href {\doibase 10.1103/PhysRevE.99.022125} {\bibfield
  {journal} {\bibinfo  {journal} {Phys.Rev.E}\ }\textbf {\bibinfo {volume}
  {99}},\ \bibinfo {eid} {022125} (\bibinfo {year} {2019})}\BibitemShut
  {NoStop}%
\bibitem [{\citenamefont {{Goldreich}}\ and\ \citenamefont
  {{Lynden-Bell}}(1965{\natexlab{a}})}]{Goldreich+Lynden-Bell_1965a}%
  \BibitemOpen
  \bibfield  {author} {\bibinfo {author} {\bibfnamefont {P.}~\bibnamefont
  {{Goldreich}}}\ and\ \bibinfo {author} {\bibfnamefont {D.}~\bibnamefont
  {{Lynden-Bell}}},\ }\href {\doibase 10.1093/mnras/130.2.97} {\bibfield
  {journal} {\bibinfo  {journal} {Mon.Not.R.Astron.Soc.}\ }\textbf {\bibinfo
  {volume} {130}},\ \bibinfo {pages} {97} (\bibinfo {year}
  {1965}{\natexlab{a}})}\BibitemShut {NoStop}%
\bibitem [{\citenamefont {{Goldreich}}\ and\ \citenamefont
  {{Lynden-Bell}}(1965{\natexlab{b}})}]{Goldreich+Lynden-Bell_1965b}%
  \BibitemOpen
  \bibfield  {author} {\bibinfo {author} {\bibfnamefont {P.}~\bibnamefont
  {{Goldreich}}}\ and\ \bibinfo {author} {\bibfnamefont {D.}~\bibnamefont
  {{Lynden-Bell}}},\ }\href {\doibase 10.1093/mnras/130.2.125} {\bibfield
  {journal} {\bibinfo  {journal} {Mon.Not.R.Astron.Soc.}\ }\textbf {\bibinfo
  {volume} {130}},\ \bibinfo {pages} {125} (\bibinfo {year}
  {1965}{\natexlab{b}})}\BibitemShut {NoStop}%
\bibitem [{\citenamefont {{Sellwood}}\ and\ \citenamefont
  {{Carlberg}}(1984)}]{Sellwood+Carlberg_1984}%
  \BibitemOpen
  \bibfield  {author} {\bibinfo {author} {\bibfnamefont {J.~A.}\ \bibnamefont
  {{Sellwood}}}\ and\ \bibinfo {author} {\bibfnamefont {R.~G.}\ \bibnamefont
  {{Carlberg}}},\ }\href {\doibase 10.1086/162176} {\bibfield  {journal}
  {\bibinfo  {journal} {Astrophys.J.}\ }\textbf {\bibinfo {volume} {282}},\
  \bibinfo {pages} {61} (\bibinfo {year} {1984})}\BibitemShut {NoStop}%
\bibitem [{\citenamefont {{Sellwood}}\ and\ \citenamefont
  {{Carlberg}}(2014)}]{Sellwood+Carlberg_2014}%
  \BibitemOpen
  \bibfield  {author} {\bibinfo {author} {\bibfnamefont {J.~A.}\ \bibnamefont
  {{Sellwood}}}\ and\ \bibinfo {author} {\bibfnamefont {R.~G.}\ \bibnamefont
  {{Carlberg}}},\ }\href {\doibase 10.1088/0004-637X/785/2/137} {\bibfield
  {journal} {\bibinfo  {journal} {Astrophys.J.}\ }\textbf {\bibinfo {volume}
  {785}},\ \bibinfo {eid} {137} (\bibinfo {year} {2014})}\BibitemShut {NoStop}%
\bibitem [{\citenamefont {{Sellwood}}\ and\ \citenamefont
  {{Carlberg}}(2019)}]{Sellwood+Carlberg_2019}%
  \BibitemOpen
  \bibfield  {author} {\bibinfo {author} {\bibfnamefont {J.~A.}\ \bibnamefont
  {{Sellwood}}}\ and\ \bibinfo {author} {\bibfnamefont {R.~G.}\ \bibnamefont
  {{Carlberg}}},\ }\href {\doibase 10.1093/mnras/stz2132} {\bibfield  {journal}
  {\bibinfo  {journal} {Mon.Not.R.Astron.Soc.}\ }\textbf {\bibinfo {volume}
  {489}},\ \bibinfo {pages} {116} (\bibinfo {year} {2019})}\BibitemShut
  {NoStop}%
\bibitem [{\citenamefont {Binney}\ and\ \citenamefont
  {Tremaine}(2008)}]{Binney_Tremaine_2008}%
  \BibitemOpen
  \bibfield  {author} {\bibinfo {author} {\bibfnamefont {J.}~\bibnamefont
  {Binney}}\ and\ \bibinfo {author} {\bibfnamefont {S.}~\bibnamefont
  {Tremaine}},\ }\href@noop {} {\emph {\bibinfo {title} {Galactic Dynamics}}}\
  (\bibinfo  {publisher} {Princeton University Press},\ \bibinfo {year}
  {2008})\BibitemShut {NoStop}%
\bibitem [{\citenamefont {{Dobbs}}\ and\ \citenamefont
  {{Baba}}(2014)}]{Dobbs_Baba_2014}%
  \BibitemOpen
  \bibfield  {author} {\bibinfo {author} {\bibfnamefont {C.}~\bibnamefont
  {{Dobbs}}}\ and\ \bibinfo {author} {\bibfnamefont {J.}~\bibnamefont
  {{Baba}}},\ }\href {\doibase 10.1017/pasa.2014.31} {\bibfield  {journal}
  {\bibinfo  {journal} {Pub.Astron.Soc.Austr.}\ }\textbf {\bibinfo {volume}
  {31}},\ \bibinfo {eid} {e035} (\bibinfo {year} {2014})}\BibitemShut {NoStop}%
\bibitem [{\citenamefont {{Navarro}}\ \emph {et~al.}(1997)\citenamefont
  {{Navarro}}, \citenamefont {{Frenk}},\ and\ \citenamefont
  {{White}}}]{Navarro_etal_1997}%
  \BibitemOpen
  \bibfield  {author} {\bibinfo {author} {\bibfnamefont {J.~F.}\ \bibnamefont
  {{Navarro}}}, \bibinfo {author} {\bibfnamefont {C.~S.}\ \bibnamefont
  {{Frenk}}}, \ and\ \bibinfo {author} {\bibfnamefont {S.~D.~M.}\ \bibnamefont
  {{White}}},\ }\href {\doibase 10.1086/304888} {\bibfield  {journal} {\bibinfo
   {journal} {Astrophys. J.}\ }\textbf {\bibinfo {volume} {490}},\ \bibinfo
  {pages} {493} (\bibinfo {year} {1997})}\BibitemShut {NoStop}%
\bibitem [{\citenamefont {{Eggen}}\ \emph {et~al.}(1962)\citenamefont
  {{Eggen}}, \citenamefont {{Lynden-Bell}},\ and\ \citenamefont
  {{Sandage}}}]{ELS_1962}%
  \BibitemOpen
  \bibfield  {author} {\bibinfo {author} {\bibfnamefont {O.~J.}\ \bibnamefont
  {{Eggen}}}, \bibinfo {author} {\bibfnamefont {D.}~\bibnamefont
  {{Lynden-Bell}}}, \ and\ \bibinfo {author} {\bibfnamefont {A.~R.}\
  \bibnamefont {{Sandage}}},\ }\href {\doibase 10.1086/147433} {\bibfield
  {journal} {\bibinfo  {journal} {Astrophys.J.}\ }\textbf {\bibinfo {volume}
  {136}},\ \bibinfo {pages} {748} (\bibinfo {year} {1962})}\BibitemShut
  {NoStop}%
\bibitem [{\citenamefont {{Katz}}(1991)}]{Katz_1991}%
  \BibitemOpen
  \bibfield  {author} {\bibinfo {author} {\bibfnamefont {N.}~\bibnamefont
  {{Katz}}},\ }\href {\doibase 10.1086/169696} {\bibfield  {journal} {\bibinfo
  {journal} {Astrophys.J.}\ }\textbf {\bibinfo {volume} {368}},\ \bibinfo
  {pages} {325} (\bibinfo {year} {1991})}\BibitemShut {NoStop}%
\bibitem [{\citenamefont {{Katz}}\ and\ \citenamefont
  {{Gunn}}(1991)}]{Katz+Gunn_1991}%
  \BibitemOpen
  \bibfield  {author} {\bibinfo {author} {\bibfnamefont {N.}~\bibnamefont
  {{Katz}}}\ and\ \bibinfo {author} {\bibfnamefont {J.~E.}\ \bibnamefont
  {{Gunn}}},\ }\href {\doibase 10.1086/170367} {\bibfield  {journal} {\bibinfo
  {journal} {Astrophys.J.}\ }\textbf {\bibinfo {volume} {377}},\ \bibinfo
  {pages} {365} (\bibinfo {year} {1991})}\BibitemShut {NoStop}%
\bibitem [{\citenamefont {{Katz}}(1992)}]{Katz_1992}%
  \BibitemOpen
  \bibfield  {author} {\bibinfo {author} {\bibfnamefont {N.}~\bibnamefont
  {{Katz}}},\ }\href {\doibase 10.1086/171366} {\bibfield  {journal} {\bibinfo
  {journal} {Astrophys.J.}\ }\textbf {\bibinfo {volume} {391}},\ \bibinfo
  {pages} {502} (\bibinfo {year} {1992})}\BibitemShut {NoStop}%
\bibitem [{\citenamefont {{Springel}}(2005)}]{Springel_2005}%
  \BibitemOpen
  \bibfield  {author} {\bibinfo {author} {\bibfnamefont {V.}~\bibnamefont
  {{Springel}}},\ }\href {\doibase 10.1111/j.1365-2966.2005.09655.x} {\bibfield
   {journal} {\bibinfo  {journal} {Mon.Not.R.Astron.Soc.}\ }\textbf {\bibinfo
  {volume} {364}},\ \bibinfo {pages} {1105} (\bibinfo {year}
  {2005})}\BibitemShut {NoStop}%
\bibitem [{\citenamefont {{Katz}}\ \emph {et~al.}(1996)\citenamefont {{Katz}},
  \citenamefont {{Weinberg}},\ and\ \citenamefont {{Hernquist}}}]{katz&al1996}%
  \BibitemOpen
  \bibfield  {author} {\bibinfo {author} {\bibfnamefont {N.}~\bibnamefont
  {{Katz}}}, \bibinfo {author} {\bibfnamefont {D.~H.}\ \bibnamefont
  {{Weinberg}}}, \ and\ \bibinfo {author} {\bibfnamefont {L.}~\bibnamefont
  {{Hernquist}}},\ }\href {\doibase 10.1086/192305} {\bibfield  {journal}
  {\bibinfo  {journal} {Astrophy.J.Suppl.}\ }\textbf {\bibinfo {volume}
  {105}},\ \bibinfo {pages} {19} (\bibinfo {year} {1996})}\BibitemShut
  {NoStop}%
\bibitem [{\citenamefont {Abel}\ \emph {et~al.}(1997)\citenamefont {Abel},
  \citenamefont {Anninos}, \citenamefont {Zhang},\ and\ \citenamefont
  {Norman}}]{abel&al97}%
  \BibitemOpen
  \bibfield  {author} {\bibinfo {author} {\bibfnamefont {T.}~\bibnamefont
  {Abel}}, \bibinfo {author} {\bibfnamefont {P.}~\bibnamefont {Anninos}},
  \bibinfo {author} {\bibfnamefont {Y.}~\bibnamefont {Zhang}}, \ and\ \bibinfo
  {author} {\bibfnamefont {M.~L.}\ \bibnamefont {Norman}},\ }\href {\doibase
  https://doi.org/10.1016/S1384-1076(97)00010-9} {\bibfield  {journal}
  {\bibinfo  {journal} {New Astronomy}\ }\textbf {\bibinfo {volume} {2}},\
  \bibinfo {pages} {181 } (\bibinfo {year} {1997})}\BibitemShut {NoStop}%
\bibitem [{\citenamefont {Yoshida}\ \emph {et~al.}(2003)\citenamefont
  {Yoshida}, \citenamefont {Abel}, \citenamefont {Hernquist},\ and\
  \citenamefont {Sugiyama}}]{yoshida&al03}%
  \BibitemOpen
  \bibfield  {author} {\bibinfo {author} {\bibfnamefont {N.}~\bibnamefont
  {Yoshida}}, \bibinfo {author} {\bibfnamefont {T.}~\bibnamefont {Abel}},
  \bibinfo {author} {\bibfnamefont {L.}~\bibnamefont {Hernquist}}, \ and\
  \bibinfo {author} {\bibfnamefont {N.}~\bibnamefont {Sugiyama}},\ }\href
  {\doibase 10.1086/375810} {\bibfield  {journal} {\bibinfo  {journal} {The
  Astrophysical Journal}\ }\textbf {\bibinfo {volume} {592}},\ \bibinfo {pages}
  {645} (\bibinfo {year} {2003})}\BibitemShut {NoStop}%
\bibitem [{\citenamefont {{Lucy}}(1977)}]{Lucy_1977}%
  \BibitemOpen
  \bibfield  {author} {\bibinfo {author} {\bibfnamefont {L.~B.}\ \bibnamefont
  {{Lucy}}},\ }\href {\doibase 10.1086/112164} {\bibfield  {journal} {\bibinfo
  {journal} {Astron.J.}\ }\textbf {\bibinfo {volume} {82}},\ \bibinfo {pages}
  {1013} (\bibinfo {year} {1977})}\BibitemShut {NoStop}%
\bibitem [{\citenamefont {{Gingold}}\ and\ \citenamefont
  {{Monaghan}}(1977)}]{Gingold+Monaghan_1977}%
  \BibitemOpen
  \bibfield  {author} {\bibinfo {author} {\bibfnamefont {R.~A.}\ \bibnamefont
  {{Gingold}}}\ and\ \bibinfo {author} {\bibfnamefont {J.~J.}\ \bibnamefont
  {{Monaghan}}},\ }\href {\doibase 10.1093/mnras/181.3.375} {\bibfield
  {journal} {\bibinfo  {journal} {Mon.Not.R.Astron.Soc.}\ }\textbf {\bibinfo
  {volume} {181}},\ \bibinfo {pages} {375} (\bibinfo {year}
  {1977})}\BibitemShut {NoStop}%
\bibitem [{\citenamefont {{Monaghan}}(1992)}]{Monaghan_1992}%
  \BibitemOpen
  \bibfield  {author} {\bibinfo {author} {\bibfnamefont {J.~J.}\ \bibnamefont
  {{Monaghan}}},\ }\href {\doibase 10.1146/annurev.aa.30.090192.002551}
  {\bibfield  {journal} {\bibinfo  {journal} {Ann.Rev.Astron.Astrophys.}\
  }\textbf {\bibinfo {volume} {30}},\ \bibinfo {pages} {543} (\bibinfo {year}
  {1992})}\BibitemShut {NoStop}%
\bibitem [{\citenamefont {{Monaghan}}(2005)}]{Monaghan_2005}%
  \BibitemOpen
  \bibfield  {author} {\bibinfo {author} {\bibfnamefont {J.~J.}\ \bibnamefont
  {{Monaghan}}},\ }\href {\doibase 10.1088/0034-4885/68/8/R01} {\bibfield
  {journal} {\bibinfo  {journal} {Reports on Progress in Physics}\ }\textbf
  {\bibinfo {volume} {68}},\ \bibinfo {pages} {1703} (\bibinfo {year}
  {2005})}\BibitemShut {NoStop}%
\bibitem [{\citenamefont {{Monaghan}}(1997)}]{Mon97}%
  \BibitemOpen
  \bibfield  {author} {\bibinfo {author} {\bibfnamefont {J.~J.}\ \bibnamefont
  {{Monaghan}}},\ }\href {\doibase 10.1006/jcph.1997.5846} {\bibfield
  {journal} {\bibinfo  {journal} {Journal of Computational Physics}\ }\textbf
  {\bibinfo {volume} {138}},\ \bibinfo {pages} {801} (\bibinfo {year}
  {1997})}\BibitemShut {NoStop}%
\bibitem [{\citenamefont {{Peebles}}(1969)}]{Peebles_1969}%
  \BibitemOpen
  \bibfield  {author} {\bibinfo {author} {\bibfnamefont {P.~J.~E.}\
  \bibnamefont {{Peebles}}},\ }\href {\doibase 10.1086/149876} {\bibfield
  {journal} {\bibinfo  {journal} {Astrophys.J.}\ }\textbf {\bibinfo {volume}
  {155}},\ \bibinfo {pages} {393} (\bibinfo {year} {1969})}\BibitemShut
  {NoStop}%
\bibitem [{\citenamefont {{Knebe}}\ and\ \citenamefont
  {{Power}}(2008)}]{knebe2008}%
  \BibitemOpen
  \bibfield  {author} {\bibinfo {author} {\bibfnamefont {A.}~\bibnamefont
  {{Knebe}}}\ and\ \bibinfo {author} {\bibfnamefont {C.}~\bibnamefont
  {{Power}}},\ }\href {\doibase 10.1086/586702} {\bibfield  {journal} {\bibinfo
   {journal} {Astrophys.J.}\ }\textbf {\bibinfo {volume} {678}},\ \bibinfo
  {pages} {621} (\bibinfo {year} {2008})}\BibitemShut {NoStop}%
\bibitem [{\citenamefont {{Kratter}}\ \emph {et~al.}(2008)\citenamefont
  {{Kratter}}, \citenamefont {{Matzner}},\ and\ \citenamefont
  {{Krumholz}}}]{kra08}%
  \BibitemOpen
  \bibfield  {author} {\bibinfo {author} {\bibfnamefont {K.~M.}\ \bibnamefont
  {{Kratter}}}, \bibinfo {author} {\bibfnamefont {C.~D.}\ \bibnamefont
  {{Matzner}}}, \ and\ \bibinfo {author} {\bibfnamefont {M.~R.}\ \bibnamefont
  {{Krumholz}}},\ }\href {\doibase 10.1086/587543} {\bibfield  {journal}
  {\bibinfo  {journal} {The Astrophysical Journal}\ }\textbf {\bibinfo {volume}
  {681}},\ \bibinfo {pages} {375} (\bibinfo {year} {2008})}\BibitemShut
  {NoStop}%
\bibitem [{\citenamefont {{Wan}}\ \emph {et~al.}(2017)\citenamefont {{Wan}},
  \citenamefont {{Liu}},\ and\ \citenamefont {{Deng}}}]{liu17}%
  \BibitemOpen
  \bibfield  {author} {\bibinfo {author} {\bibfnamefont {J.-C.}\ \bibnamefont
  {{Wan}}}, \bibinfo {author} {\bibfnamefont {C.}~\bibnamefont {{Liu}}}, \ and\
  \bibinfo {author} {\bibfnamefont {L.-C.}\ \bibnamefont {{Deng}}},\ }\href
  {\doibase 10.1088/1674-4527/17/8/79} {\bibfield  {journal} {\bibinfo
  {journal} {Research in Astronomy and Astrophysics}\ }\textbf {\bibinfo
  {volume} {17}},\ \bibinfo {eid} {079} (\bibinfo {year} {2017})}\BibitemShut
  {NoStop}%
\bibitem [{\citenamefont {Levin}\ \emph {et~al.}(2008)\citenamefont {Levin},
  \citenamefont {Pakter},\ and\ \citenamefont {Rizzato}}]{levin_etal_2008}%
  \BibitemOpen
  \bibfield  {author} {\bibinfo {author} {\bibfnamefont {Y.}~\bibnamefont
  {Levin}}, \bibinfo {author} {\bibfnamefont {R.}~\bibnamefont {Pakter}}, \
  and\ \bibinfo {author} {\bibfnamefont {F.}~\bibnamefont {Rizzato}},\
  }\href@noop {} {\bibfield  {journal} {\bibinfo  {journal} {Phys. Rev.}\
  }\textbf {\bibinfo {volume} {E78}},\ \bibinfo {pages} {021130} (\bibinfo
  {year} {2008})}\BibitemShut {NoStop}%
\bibitem [{\citenamefont {Eilers}\ \emph {et~al.}(2019)\citenamefont {Eilers},
  \citenamefont {Hogg}, \citenamefont {Rix},\ and\ \citenamefont
  {Ness}}]{Eilers_2019}%
  \BibitemOpen
  \bibfield  {author} {\bibinfo {author} {\bibfnamefont {A.-C.}\ \bibnamefont
  {Eilers}}, \bibinfo {author} {\bibfnamefont {D.~W.}\ \bibnamefont {Hogg}},
  \bibinfo {author} {\bibfnamefont {H.-W.}\ \bibnamefont {Rix}}, \ and\
  \bibinfo {author} {\bibfnamefont {M.~K.}\ \bibnamefont {Ness}},\ }\href
  {\doibase 10.3847/1538-4357/aaf648} {\bibfield  {journal} {\bibinfo
  {journal} {The Astrophysical Journal}\ }\textbf {\bibinfo {volume} {871}},\
  \bibinfo {pages} {120} (\bibinfo {year} {2019})}\BibitemShut {NoStop}%
\bibitem [{\citenamefont {\v{Z}. Chrob\'akov\'a}\ \emph
  {et~al.}(2020)\citenamefont {\v{Z}. Chrob\'akov\'a}, \citenamefont
  {L\'opez-Corredoira}, \citenamefont {Labini}, \citenamefont {Wang},\ and\
  \citenamefont {Nagy}}]{Zofia_etal_2020}%
  \BibitemOpen
  \bibfield  {author} {\bibinfo {author} {\bibnamefont {\v{Z}.
  Chrob\'akov\'a}}, \bibinfo {author} {\bibfnamefont {M.}~\bibnamefont
  {L\'opez-Corredoira}}, \bibinfo {author} {\bibfnamefont {F.~S.}\ \bibnamefont
  {Labini}}, \bibinfo {author} {\bibfnamefont {H.-F.}\ \bibnamefont {Wang}}, \
  and\ \bibinfo {author} {\bibfnamefont {R.}~\bibnamefont {Nagy}},\ }\href@noop
  {} {\bibfield  {journal} {\bibinfo  {journal} {Astron.Astrophys}\ } (\bibinfo
  {year} {in the press 2020})},\ \Eprint
  {http://arxiv.org/abs/arXiv:2007.14825} {arXiv:2007.14825} \BibitemShut
  {NoStop}%
\bibitem [{\citenamefont {Peebles}(2020)}]{Peebles_2020}%
  \BibitemOpen
  \bibfield  {author} {\bibinfo {author} {\bibfnamefont {P.~J.~E.}\
  \bibnamefont {Peebles}},\ }\href@noop {} {\enquote {\bibinfo {title}
  {Formation of the large nearby galaxies},}\ } (\bibinfo {year} {2020}),\
  \Eprint {http://arxiv.org/abs/2005.07588} {arXiv:2005.07588}
  \BibitemShut {NoStop}%
\bibitem [{\citenamefont {{Carlberg}}(1987)}]{Carlberg_1987}%
  \BibitemOpen
  \bibfield  {author} {\bibinfo {author} {\bibfnamefont {R.~G.}\ \bibnamefont
  {{Carlberg}}},\ }in\ \href {\doibase 10.1007/978-94-009-3971-4_29} {\emph
  {\bibinfo {booktitle} {Structure and Dynamics of Elliptical Galaxies}}},\
  \bibinfo {series} {IAU Symposium}, Vol.\ \bibinfo {volume} {127},\ \bibinfo
  {editor} {edited by\ \bibinfo {editor} {\bibfnamefont {P.~T.}\ \bibnamefont
  {{de Zeeuw}}}}\ (\bibinfo {year} {1987})\ pp.\ \bibinfo {pages}
  {353--364}\BibitemShut {NoStop}%
\bibitem [{\citenamefont {{Lin}}\ and\ \citenamefont
  {{Shu}}(1964)}]{Lin+Shu_1964}%
  \BibitemOpen
  \bibfield  {author} {\bibinfo {author} {\bibfnamefont {C.~C.}\ \bibnamefont
  {{Lin}}}\ and\ \bibinfo {author} {\bibfnamefont {F.~H.}\ \bibnamefont
  {{Shu}}},\ }\href {\doibase 10.1086/147955} {\bibfield  {journal} {\bibinfo
  {journal} {Astrophys.J.}\ }\textbf {\bibinfo {volume} {140}},\ \bibinfo
  {pages} {646} (\bibinfo {year} {1964})}\BibitemShut {NoStop}%
\bibitem [{\citenamefont {{Bertin}}\ and\ \citenamefont
  {{Lin}}(1996)}]{Bertin+Lin_1996}%
  \BibitemOpen
  \bibfield  {author} {\bibinfo {author} {\bibfnamefont {G.}~\bibnamefont
  {{Bertin}}}\ and\ \bibinfo {author} {\bibfnamefont {C.~C.}\ \bibnamefont
  {{Lin}}},\ }\href@noop {} {\emph {\bibinfo {title} {{Spiral structure in
  galaxies a density wave theory}}}}\ (\bibinfo  {publisher} {Cambridge, MA MIT
  Press},\ \bibinfo {year} {1996})\BibitemShut {NoStop}%
\bibitem [{\citenamefont {{Sylos Labini}}\ \emph {et~al.}(2019)\citenamefont
  {{Sylos Labini}}, \citenamefont {{Benhaiem}}, \citenamefont {{Comer{\'o}n}},\
  and\ \citenamefont {{L{\'o}pez-Corredoira}}}]{SylosLabini_etal_2019}%
  \BibitemOpen
  \bibfield  {author} {\bibinfo {author} {\bibfnamefont {F.}~\bibnamefont
  {{Sylos Labini}}}, \bibinfo {author} {\bibfnamefont {D.}~\bibnamefont
  {{Benhaiem}}}, \bibinfo {author} {\bibfnamefont {S.}~\bibnamefont
  {{Comer{\'o}n}}}, \ and\ \bibinfo {author} {\bibfnamefont {M.}~\bibnamefont
  {{L{\'o}pez-Corredoira}}},\ }\href {\doibase 10.1051/0004-6361/201833834}
  {\bibfield  {journal} {\bibinfo  {journal} {Astron.Astrophys.}\ }\textbf
  {\bibinfo {volume} {622}},\ \bibinfo {eid} {A58} (\bibinfo {year}
  {2019})}\BibitemShut {NoStop}%
\bibitem [{\citenamefont {{Gaia Collaboration}}\ \emph
  {et~al.}(2016)\citenamefont {{Gaia Collaboration}}, \citenamefont {{Prusti}},
  \citenamefont {{de Bruijne}}, \citenamefont {{Brown}}, \citenamefont
  {{Vallenari}}, \citenamefont {{Babusiaux}}, \citenamefont {{Bailer-Jones}},
  \citenamefont {{Bastian}}, \citenamefont {{Biermann}}, \citenamefont
  {{Evans}},\ and\ \citenamefont {et~al.}}]{Gaia_2016}%
  \BibitemOpen
  \bibfield  {author} {\bibinfo {author} {\bibnamefont {{Gaia Collaboration}}},
  \bibinfo {author} {\bibfnamefont {T.}~\bibnamefont {{Prusti}}}, \bibinfo
  {author} {\bibfnamefont {J.~H.~J.}\ \bibnamefont {{de Bruijne}}}, \bibinfo
  {author} {\bibfnamefont {A.~G.~A.}\ \bibnamefont {{Brown}}}, \bibinfo
  {author} {\bibfnamefont {A.}~\bibnamefont {{Vallenari}}}, \bibinfo {author}
  {\bibfnamefont {C.}~\bibnamefont {{Babusiaux}}}, \bibinfo {author}
  {\bibfnamefont {C.~A.~L.}\ \bibnamefont {{Bailer-Jones}}}, \bibinfo {author}
  {\bibfnamefont {U.}~\bibnamefont {{Bastian}}}, \bibinfo {author}
  {\bibfnamefont {M.}~\bibnamefont {{Biermann}}}, \bibinfo {author}
  {\bibfnamefont {D.~W.}\ \bibnamefont {{Evans}}}, \ and\ \bibinfo {author}
  {\bibnamefont {et~al.}},\ }\href {\doibase 10.1051/0004-6361/201629272}
  {\bibfield  {journal} {\bibinfo  {journal} {Astron.Astrophys.}\ }\textbf
  {\bibinfo {volume} {595}},\ \bibinfo {eid} {A1} (\bibinfo {year}
  {2016})}\BibitemShut {NoStop}%
\bibitem [{\citenamefont {{Antoja}}\ \emph {et~al.}(2018)\citenamefont
  {{Antoja}}, \citenamefont {{Helmi}}, \citenamefont {{Romero-G{\'o}mez}},
  \citenamefont {{Katz}}, \citenamefont {{Babusiaux}}, \citenamefont
  {{Drimmel}}, \citenamefont {{Evans}}, \citenamefont {{Figueras}},
  \citenamefont {{Poggio}}, \citenamefont {{Reyl{\'e}}}, \citenamefont
  {{Robin}}, \citenamefont {{Seabroke}},\ and\ \citenamefont
  {{Soubiran}}}]{Antoja_etal_2018}%
  \BibitemOpen
  \bibfield  {author} {\bibinfo {author} {\bibfnamefont {T.}~\bibnamefont
  {{Antoja}}}, \bibinfo {author} {\bibfnamefont {A.}~\bibnamefont {{Helmi}}},
  \bibinfo {author} {\bibfnamefont {M.}~\bibnamefont {{Romero-G{\'o}mez}}},
  \bibinfo {author} {\bibfnamefont {D.}~\bibnamefont {{Katz}}}, \bibinfo
  {author} {\bibfnamefont {C.}~\bibnamefont {{Babusiaux}}}, \bibinfo {author}
  {\bibfnamefont {R.}~\bibnamefont {{Drimmel}}}, \bibinfo {author}
  {\bibfnamefont {D.~W.}\ \bibnamefont {{Evans}}}, \bibinfo {author}
  {\bibfnamefont {F.}~\bibnamefont {{Figueras}}}, \bibinfo {author}
  {\bibfnamefont {E.}~\bibnamefont {{Poggio}}}, \bibinfo {author}
  {\bibfnamefont {C.}~\bibnamefont {{Reyl{\'e}}}}, \bibinfo {author}
  {\bibfnamefont {A.~C.}\ \bibnamefont {{Robin}}}, \bibinfo {author}
  {\bibfnamefont {G.}~\bibnamefont {{Seabroke}}}, \ and\ \bibinfo {author}
  {\bibfnamefont {C.}~\bibnamefont {{Soubiran}}},\ }\href {\doibase
  10.1038/s41586-018-0510-7} {\bibfield  {journal} {\bibinfo  {journal}
  {Nature}\ }\textbf {\bibinfo {volume} {561}},\ \bibinfo {pages} {360}
  (\bibinfo {year} {2018})}\BibitemShut {NoStop}%
\bibitem [{\citenamefont {{Gaia Collaboration}}\ \emph
  {et~al.}(2018)\citenamefont {{Gaia Collaboration}}, \citenamefont {{Katz}},
  \citenamefont {{Antoja}}, \citenamefont {{Romero-G{\'o}mez}}, \citenamefont
  {{Drimmel}}, \citenamefont {{Reyl{\'e}}}, \citenamefont {{Seabroke}},
  \citenamefont {{Soubiran}}, \citenamefont {{Babusiaux}}, \citenamefont {{Di
  Matteo}},\ and\ \citenamefont {et~al.}}]{Katz_etal_2018}%
  \BibitemOpen
  \bibfield  {author} {\bibinfo {author} {\bibnamefont {{Gaia Collaboration}}},
  \bibinfo {author} {\bibfnamefont {D.}~\bibnamefont {{Katz}}}, \bibinfo
  {author} {\bibfnamefont {T.}~\bibnamefont {{Antoja}}}, \bibinfo {author}
  {\bibfnamefont {M.}~\bibnamefont {{Romero-G{\'o}mez}}}, \bibinfo {author}
  {\bibfnamefont {R.}~\bibnamefont {{Drimmel}}}, \bibinfo {author}
  {\bibfnamefont {C.}~\bibnamefont {{Reyl{\'e}}}}, \bibinfo {author}
  {\bibfnamefont {G.~M.}\ \bibnamefont {{Seabroke}}}, \bibinfo {author}
  {\bibfnamefont {C.}~\bibnamefont {{Soubiran}}}, \bibinfo {author}
  {\bibfnamefont {C.}~\bibnamefont {{Babusiaux}}}, \bibinfo {author}
  {\bibfnamefont {P.}~\bibnamefont {{Di Matteo}}}, \ and\ \bibinfo {author}
  {\bibnamefont {et~al.}},\ }\href {\doibase 10.1051/0004-6361/201832865}
  {\bibfield  {journal} {\bibinfo  {journal} {Astron.Astrophys.}\ }\textbf
  {\bibinfo {volume} {616}},\ \bibinfo {eid} {A11} (\bibinfo {year}
  {2018})}\BibitemShut {NoStop}%
\bibitem [{\citenamefont {Wang}\ \emph {et~al.}(2018)\citenamefont {Wang},
  \citenamefont {Liu}, \citenamefont {Xu}, \citenamefont {Wan},\ and\
  \citenamefont {Deng}}]{Wang_2018}%
  \BibitemOpen
  \bibfield  {author} {\bibinfo {author} {\bibfnamefont {H.-F.}\ \bibnamefont
  {Wang}}, \bibinfo {author} {\bibfnamefont {C.}~\bibnamefont {Liu}}, \bibinfo
  {author} {\bibfnamefont {Y.}~\bibnamefont {Xu}}, \bibinfo {author}
  {\bibfnamefont {J.-C.}\ \bibnamefont {Wan}}, \ and\ \bibinfo {author}
  {\bibfnamefont {L.}~\bibnamefont {Deng}},\ }\href {\doibase
  10.1093/mnras/sty1058} {\bibfield  {journal} {\bibinfo  {journal} {Monthly
  Notices of the Royal Astronomical Society}\ }\textbf {\bibinfo {volume}
  {478}},\ \bibinfo {pages} {3367-3379} (\bibinfo {year} {2018})}\BibitemShut
  {NoStop}%
\bibitem [{\citenamefont {{L{\'o}pez-Corredoira}}\ and\ \citenamefont {{Sylos
  Labini}}(2019)}]{MLC_FSL_2019}%
  \BibitemOpen
  \bibfield  {author} {\bibinfo {author} {\bibfnamefont {M.}~\bibnamefont
  {{L{\'o}pez-Corredoira}}}\ and\ \bibinfo {author} {\bibfnamefont
  {F.}~\bibnamefont {{Sylos Labini}}},\ }\href {\doibase
  10.1051/0004-6361/201833849} {\bibfield  {journal} {\bibinfo  {journal}
  {Astron.Astrophys.}\ }\textbf {\bibinfo {volume} {621}},\ \bibinfo {eid}
  {A48} (\bibinfo {year} {2019})}\BibitemShut {NoStop}%
\bibitem [{\citenamefont {{L{\'o}pez-Corredoira}}\ \emph
  {et~al.}(2020)\citenamefont {{L{\'o}pez-Corredoira}}, \citenamefont
  {{Garz{\'o}n}}, \citenamefont {{Wang}}, \citenamefont {{Sylos Labini}},
  \citenamefont {{Nagy}}, \citenamefont {{Chrob{\'a}kov{\'a}}}, \citenamefont
  {{Chang}},\ and\ \citenamefont {{Villarroel}}}]{lopezcorredoira2020gaiadr2}%
  \BibitemOpen
  \bibfield  {author} {\bibinfo {author} {\bibfnamefont {M.}~\bibnamefont
  {{L{\'o}pez-Corredoira}}}, \bibinfo {author} {\bibfnamefont {F.}~\bibnamefont
  {{Garz{\'o}n}}}, \bibinfo {author} {\bibfnamefont {H.~F.}\ \bibnamefont
  {{Wang}}}, \bibinfo {author} {\bibfnamefont {F.}~\bibnamefont {{Sylos
  Labini}}}, \bibinfo {author} {\bibfnamefont {R.}~\bibnamefont {{Nagy}}},
  \bibinfo {author} {\bibfnamefont {{\v{Z}}.}~\bibnamefont
  {{Chrob{\'a}kov{\'a}}}}, \bibinfo {author} {\bibfnamefont {J.}~\bibnamefont
  {{Chang}}}, \ and\ \bibinfo {author} {\bibfnamefont {B.}~\bibnamefont
  {{Villarroel}}},\ }\href {\doibase 10.1051/0004-6361/201936711} {\bibfield
  {journal} {\bibinfo  {journal} {Astronom.Astrophys.}\ }\textbf {\bibinfo
  {volume} {634}},\ \bibinfo {eid} {A66} (\bibinfo {year} {2020})}\BibitemShut
  {NoStop}%
\end{thebibliography}

%

\end{document}